\documentclass[10pt]{article}

\usepackage{amsmath}
\usepackage{amssymb}

\usepackage{graphicx}

\usepackage{cite}

\usepackage{color} 

\usepackage[labelfont=bf,labelsep=period,justification=raggedright]{caption}

\makeatletter
\renewcommand{\@biblabel}[1]{\quad#1.}
\makeatother

\date{}

\usepackage{latexsym,amsfonts,bbm} 

\usepackage[english]{babel} 
\usepackage[latin1]{inputenc} 
\usepackage{times} 
\usepackage[T1]{fontenc} 
\usepackage{curves} 
\usepackage{microtype}

\graphicspath{
{./Figures/Figure1/}
{./Figures/Figure2/}
{./Figures/Figure3/}
{./Figures/Figure4/}
{./Figures/Figure5/}
{./Figures/Figure6/}
{./Figures/Figure7/}
{./Figures/Figure8/}
{./Figures/Figure9/}
{./Figures/Figure10/}
{./Figures/Figure11/}
{./Figures/Figure12/}
{./Figures/Figure13/}
{./Figures/Figure14/}
{./Figures/Figure15/}
{./Figures/Figure16/}
{./Figures/Figure17/}
{./Figures/Figure18/}
{./Figures/Figure19/}
}

\renewcommand{\[}{\begin{equation}} 
\renewcommand{\]}{\end{equation}} 

\renewcommand{\O}{\operatorname{\mathbb{O}}} 
 
\newcommand{\E}{\operatorname{\mathbb{E}}} 
\newcommand{\F}{\operatorname{\mathbb{F}}}

\newcommand{\one}{\mathbbm{1}} 
\newcommand{\R}{\mathbb{R}} 
 
\renewcommand{\deg}{^{\circ}} 
\newcommand{\erf}{\operatorname{\mathrm{erf}}} 

\renewcommand{\vec}{\mathbf}

\usepackage{listings}
\usepackage{setspace}
\usepackage{textcomp}

\usepackage{booktabs}
\usepackage[table]{xcolor}

\begin{document}

% Title must be 150 characters or less
\begin{flushleft}
{\Large
\textbf{Mean-Field Analysis of Orientation Selectivity in Inhibition-Dominated Networks of Spiking Neurons}
}
% Insert Author names, affiliations and corresponding author email.
\\
Sadra Sadeh$^{1}$, 
Stefano Cardanobile$^{1}$, 
Stefan Rotter$^{1,\ast}$
\\
\bf{1} Bernstein Center Freiburg \& Faculty of Biology, University of Freiburg, Freiburg, Germany
%\\
%\bf{2} Bernstein Center Freiburg \& Faculty of Biology, University of Freiburg, Freiburg, Germany
%\\
%\bf{3} Bernstein Center Freiburg \& Faculty of Biology, University of Freiburg, Freiburg, Germany
\\
$\ast$ E-mail: stefan.rotter@biologie.uni-freiburg.de
\end{flushleft}

%%%%%%%%%%%%%%%%%%%%%%%%%%%%%%%%%%%%%%%%%%%%%%%%%%%%%%%%%%%%%%%%%%%%%%%%%%%%%%%% 
%%%%%%%%%%%%%%%%%%%%%%%%%%%%%% Abstract 
%%%%%%%%%%%%%%%%%%%%%%%%%%%%%%%%%%%%%%%%%%%%%%%%%%%%%%%%%%%%%%%%%%%%%%%%%%%%%%%%

% Please keep the abstract between 250 and 300 words
\section*{Abstract}

Mechanisms underlying the emergence of orientation selectivity in the primary visual cortex are highly debated. 
Here we study the contribution of inhibition-dominated random recurrent networks to orientation selectivity, and more generally to sensory processing. 
By simulating and analyzing large-scale networks of spiking neurons, we investigate tuning amplification and contrast invariance of orientation selectivity in these networks. 
In particular, we show how selective attenuation of the common mode and amplification of the modulation component take place in these networks. 
Selective attenuation of the baseline, which is governed by the exceptional eigenvalue of the connectivity matrix, removes the unspecific, redundant signal component and ensures the invariance of selectivity across different contrasts. 
Selective amplification of modulation, which is governed by the operating regime of the network and depends on the strength of coupling, amplifies the informative signal component and thus increases the signal-to-noise ratio. 
Here, we perform a mean-field analysis which accounts for this process.

% Please keep the Author Summary between 150 and 200 words
% Use first person. PLoS ONE authors please skip this step. 
% Author Summary not valid for PLoS ONE submissions.   
%\section*{Author Summary}

%%%%%%%%%%%%%%%%%%%%%%%%%%%%%%%%%%%%%%%%%%%%%%%%%%%%%%%%%%%%%%%%%%%%%%%%%%%%%%%% 
%%%%%%%%%%%%%%%%%%%%%%%%%%%%%% Introduction
%%%%%%%%%%%%%%%%%%%%%%%%%%%%%%%%%%%%%%%%%%%%%%%%%%%%%%%%%%%%%%%%%%%%%%%%%%%%%%%% 

\section{Introduction}
\label{Sec_Intro} 

Neurons in sensory cortices of mammals often respond selectively to certain features of a stimulus. 
A well-known example in the visual system is the orientation of an elongated bar \cite{Hubel1962, Hubel1968}. 
This specificity of neuronal responses is believed to be a fundamental building block of stimulus processing and perception in the mammalian brain. 
Although well studied for more than half a century now, it is not fully clear which neuronal mechanisms generate this selectivity. 
In particular, it is not clear which kind of network structure is necessary for its establishment, and whether different system architectures are being employed by different species. 

In the center of the current debate is the role of feedforward vs.\ recurrent connectivity in the initial establishment of selectivity, as well as its further properties like contrast invariance \cite{Sclar1982, Alitto2004, Niell2008} and tuning sharpening \cite{Ferster2000, Sompolinsky1997}. 
Models relying on a purely feedforward structure, as it was originally suggested by \cite{Hubel1962}, cannot explain why the orientation tuning of both neuronal spiking and membrane potentials is invariant with respect to stimulus contrast \cite{Anderson2000} (see \cite{Finn2007}, however, for a more elaborate feedforward model to account for contrast invariance). 
On the other hand, the prevailing recurrent network models for the intra-cortical origin of contrast-invariant orientation selectivity \cite{Ben-Yishai1995, Somers1995} cannot explain how highly selective neuronal responses emerge in mice \cite{Niell2008}, as they rely on feature specific connectivity which rodents seem to lack, at least at the onset of eye opening \cite{ko2013}.\footnote{Feature-specific connectivity, however, appears later during deveolpment \cite{ko2013}, consistent with previous reports \cite{ko2011, Jia2010}.}
Also, it is not clear whether the orderly arrangement of preferred features on the cortical surface, which has been described in most primates and carnivores \cite{Ohki2006, Bonhoeffer1991, Blasdel, Ts'o1990}, is necessary for the emergence of feature selectivity, or serves any function at all \cite{Horton2005}. 
Different answers to these questions would have radically different implications for understanding higher brain functions, and how to study them. 

Here we investigate the emergence of contrast invariant orientation selectivity in large-scale networks of spiking neurons with dominant inhibition. 
The biological motivation for studying such networks come from experimental studies which show functional dominance of inhibition \cite{Rudolph2007, Haider2013}. 
The dominance of inhibition is probably a consequence of the dense, local pattern of inhibitory connectivity, which has been reported for different cortices \cite{Fino2011, Packer2011, Hofer2011}. 
From a theoretical point of view, such networks are well-studied \cite{VanVreeswijk1996, Brunel2000} and they have been shown to exhibit asynchronous-irregular (AI) activity states that are in many respects resembling the spiking activity recorded in the mammalian neocortex. 

We first show that highly selective tuning curves, which are contrast invariant, can be obtained in these networks, even in absence of any feature-specific connectivity and any spatial network structure. 
We then analyze these networks by proposing a simplified mean-field description, which predicts the main properties of output orientation selectivity in the networks. 
The analysis identifies the mechanisms responsible for tuning amplification and contrast invariance. 
We show that the results hold for a wide range of parameters, and for networks operating in different recurrent regimes. 

%%%%%%%%%%%%%%%%%%%%%%%%%%%%%%%%%%%%%%%%%%%%%%%%%%%%%%%%%%%%%%%%%%%%%%%%%%%%%%%% 
%%%%%%%%%%%%%%%%%%%%%%%%%%%%%% Results 
%%%%%%%%%%%%%%%%%%%%%%%%%%%%%%%%%%%%%%%%%%%%%%%%%%%%%%%%%%%%%%%%%%%%%%%%%%%%%%%% 

% Results and Discussion can be combined.
\section{Results}
\label{Sec_Results}

%%%%%%%%%%%%%%%%%%%%%%%%%%%%%%%%%%%%%%%%%%%%%%%%%%%%%%%%%%%%%%%%%%%%%%%%%%%%%%%% 
\subsection{Contrast invariant orientation selectivity in random networks} 
\label{Sec_CIOS}

We consider a large-scale network model of spiking neurons, representing a small volume of cerebral cortex. 
Our model consists of a recurrent network of $12\,500$ leaky integrate-and-fire (LIF) neurons, $80\%$ excitatory and $20\%$ inhibitory \cite{Braitenberg1998}. 
Each neuron receives input from $10\,\%$ of the excitatory population and $10\,\%$ of the inhibitory population, sampled randomly. This is the same network configuration as the one considered by \cite{Brunel2000}. 
The network is strongly inhibition dominated, as individual inhibitory synapses are arranged to be $8$ times more effective than the excitatory ones (see Methods for details). 
We fix the random connectivity, and only change the strength of recurrent coupling, measured as the amplitude of the postsynaptic potential (PSP) at each synapse. 
The post-synaptic currents are modeled as alpha-functions, with time constant $\tau_\mathrm{syn}$ (see Methods). 
We refer to the peak value of PSP, $J_\alpha$, as $\mathrm{EPSP}$, and to the total PSP as $J = \tau_\mathrm{syn} e J_\alpha$. 

%%%%%%%%%%%%%%%%%%%%%%%%%%%%%%%%%%%%%%%%%%%%%%%%%%%%%%%%%%%%%%%%%%%%%%%%%%%%%%%%
%%%%%%%%%% Fig1 

\begin{figure}[h!] 
\centering\includegraphics[width=6.0in]{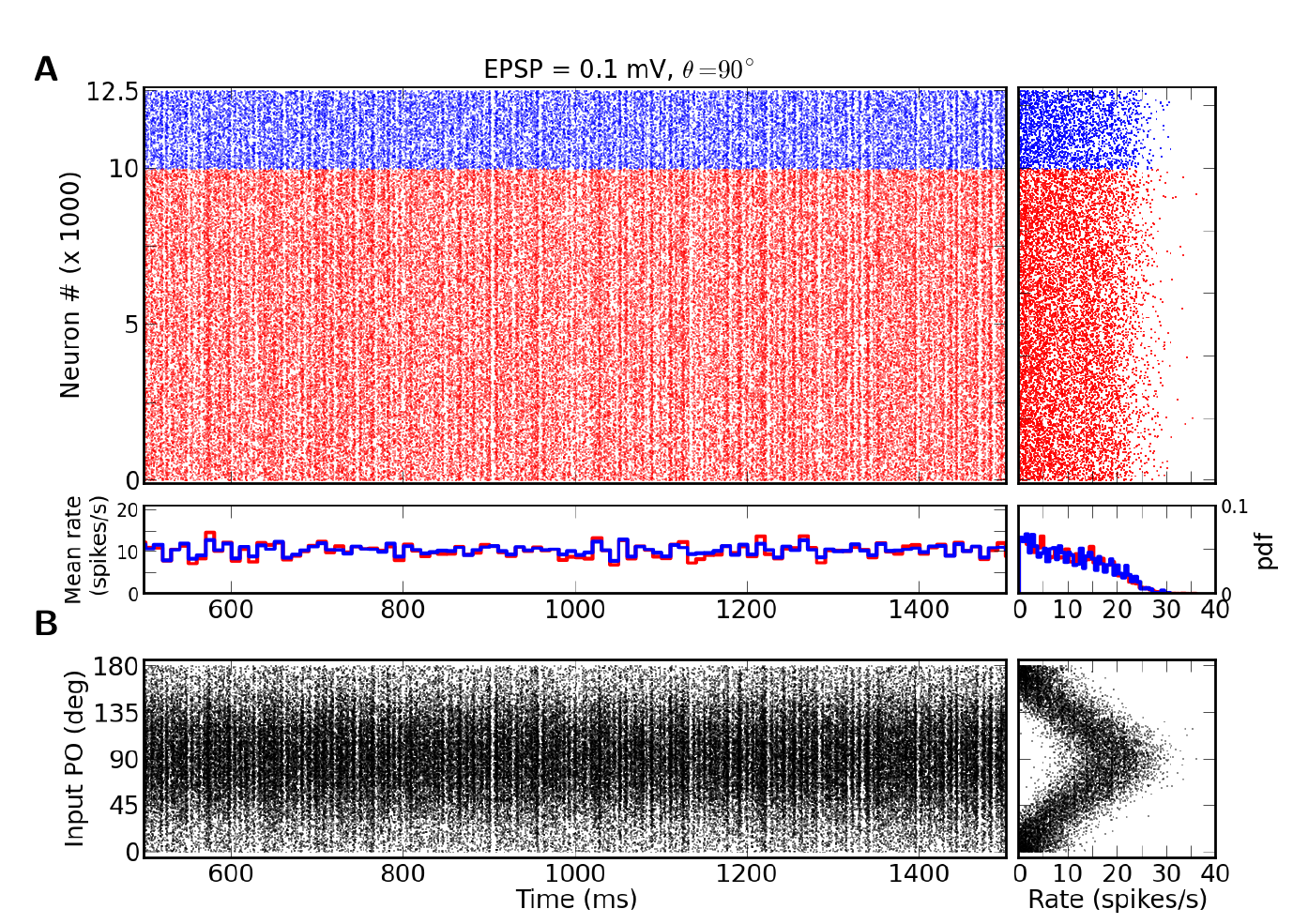} 
\caption{\textbf{Raster plot of network activity.} 
    (\textbf{A})~Typical raster plot, shown is $1$ second of the activity of a balanced network with recurrent synaptic couplings of amplitude $\mathrm{EPSP} = 0.1\,\mathrm{mV}$, in response to a stimulus of orientation $\theta = 90\deg$ at a medium contrast ($s_B = 16\,000~\mathrm{spikes}/\mathrm{s}$).     
    Spikes of excitatory and inhibitory neurons are plotted in red and blue, respectively. 
    The plot on the right shows the average firing rate of neurons, computed from the spike count during $6\,\mathrm{s}$ of simulation. 
    The histogram on the bottom depticts the time resolved firing rates of the excitatory (red) and the inhibitory (blue) populations, respectively, using a time window of $10\,\mathrm{ms}$. 
    The histogram on the bottom right shows the probability density function of time averaged firing rates of individual excitatory (red) and inhibitory (blue) neurons in the network, respectively. 
    (\textbf{B})~Same spike trains as above, with neurons sorted according to their input preferred orientations (indicated on the y-axis). 
    As above, the plot on the right indicates the firing rate of each neuron computed from the spikes emitted during the simulation.}    
    \label{Fig_RasPlt_0.1} 

\end{figure} 

%\begin{center}
%\fbox{Fig.~1 approximately here}
%\end{center}
%%%%%%%%%%%%%%%%%%%%%%%%%%%%%%%%%%%%%%%%%%%%%%%%%%%%%%%%%%%%%%%%%%%%%%%%%%%%%%%%

The response of the network with $\mathrm{EPSP} = 0.1\,\mathrm{mV}$ to the stimulus of certain orientation, $\theta$, is shown in Fig.~\ref{Fig_RasPlt_0.1}.
The external input, $s$, that a neuron receives is modeled as a homogeneous Poisson process, which is slightly ($m = 10\,\%$) modulated by the orientation of the stimulus according to a cosine function: 
$s(\theta) = s_B \bigl[ 1 + m \cos(2(\theta - \theta^*)) \bigr]$. 
A random preferred orientation (PO), $\theta^*$, of the external input is assigned to each neuron. 
The feedforward efficacy is fixed in all simulations to $\mathrm{EPSP}_\mathrm{ffw} = 0.1\,\mathrm{mV}$. 
The raster plot of the activity for half a second is shown in Fig.~\ref{Fig_RasPlt_0.1}A, which indicates that with these parameters the network is indeed operating in the asynchronous irregular (AI) state.

%%%%%%%%%%%%%%%%%%%%%%%%%%%%%%%%%%%%%%%%%%%%%%%%%%%%%%%%%%%%%%%%%%%%%%%%%%%%%%%%
%%%%%%%%%% Fig2 

\begin{figure}[h!] 
\centering\includegraphics[width=6.0in]{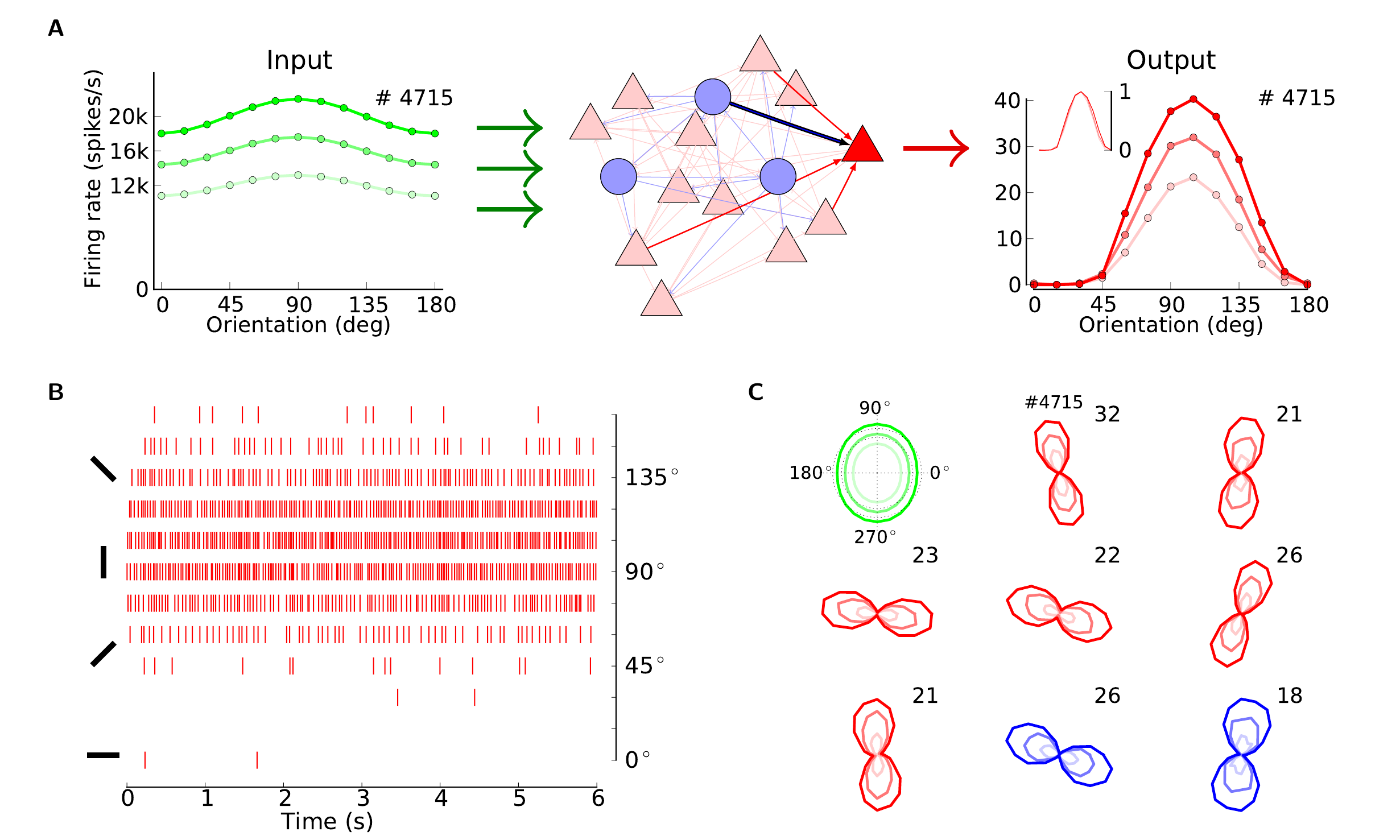} 
\caption{\textbf{Tuning amplification and contrast invariance of neuronal responses in the network.}    
    (\textbf{A})~The random network (center), composed of excitatory (red, $80\,\%$) and inhibitory (blue, $20\,\%$) neurons, transforms weakly tuned input (green) to highly selective responses (red). The sample tuning curves are shown for unit $\#~4715$, which is an excitatory neuron. 
    Neurons have very similar tuning curves for different contrasts, and they are indistinguishable after normalization by their respective peak value (see inset). $\mathrm{EPSP} = 0.1\,\mathrm{mV}$.     
    (\textbf{B})~Spikes elicited by the same excitatory neuron as in (A) in response to different orientations of a stimulus at a medium contrast.     
    (\textbf{C})~More samples of output tuning curves from the network, in polar representation. 
    Radial axis encodes the firing rate of the respective neuron at each orientation, indicated by the angle, with the maximum firing rate indicated as a number next to each plot. 
    Both excitatory (red) and inhibitory (blue) neurons are highly selective in their responses, despite their weakly selective inputs (green) and recurrent inputs of mixed selectivities. 
    Lighter colors correspond to tuning curves for lower contrasts, respectively.} 
    \label{Fig_TunCurves}    

\end{figure}

%\begin{center}
%\fbox{Fig.~2 approximately here} 
%\end{center}
%%%%%%%%%%%%%%%%%%%%%%%%%%%%%%%%%%%%%%%%%%%%%%%%%%%%%%%%%%%%%%%%%%%%%%%%%%%%%%%%

If neurons are sorted according to their preferred orientation, the differences in the firing rates become visible (Fig.~\ref{Fig_RasPlt_0.1}B). 
Neurons with a preferred orientation closer to the orientation of the stimulus on average respond with higher rates, while the neurons closer to the orthogonal orientation are mostly silent. 
The cosine tuning of the input is reflected by the cosine tuning of firing rates across the population (Fig.~\ref{Fig_RasPlt_0.1}B, right).

If we repeat the stimulation of the network with different orientations, the individual tuning curves for each neuron are obtained. 
For a sample neuron from the same network this is shown in Fig.~\ref{Fig_TunCurves}A-B. 
Although the neuron receives input that is only weakly tuned, the network is capable of amplifying the selectivity, and the output tuning is much more pronounced. 
This emerging selectivity is independent of stimulus contrast, $C$, reflected by the stimulus-specific change in the mean firing rate of the external input, $s_B$. 
Fig.~\ref{Fig_TunCurves}A shows, for a sample neuron, that the shape of tuning curves remains unchanged for different contrasts. 
Normalizing the output tuning curves by the peak value \cite{Anderson2000} yields exactly the same curve for all input intensities (Fig.~\ref{Fig_TunCurves}A, inset). 
The other neurons in the network show the same behavior, as the polar plots in Fig.~\ref{Fig_TunCurves}C demonstrate for a larger sample. 
Note that excitatory and inhibitory neurons are both highly selective.

\begin{figure}[h!] 
\centering\includegraphics[width=6.0in]{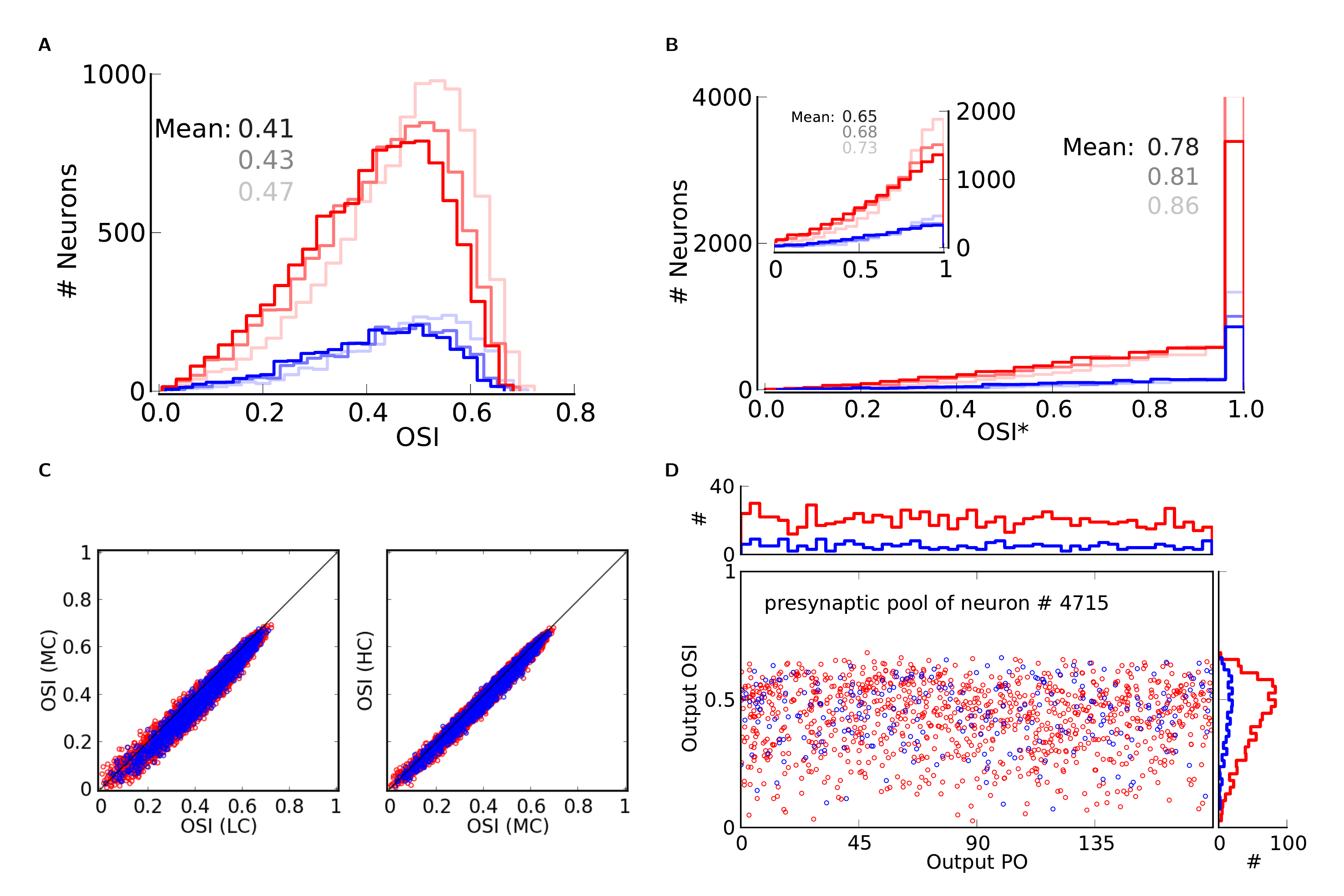} 
\caption{\textbf{Orientation selectivity across the population.}    
    (\textbf{A})~Distribution of a global measure of orientation selectivity, $\mathrm{OSI} = 1 - \text{Circular Variance}$ \cite{Ringach2002}, in the network. 
    Lighter colors show the distributions for lower contrasts, respectively. 
    All inputs have an OSI of $0.05$.    
    (\textbf{B})~Distribution of an alternative measure of orientation selectivity often used by experimentalists \cite{Niell2008}. 
    OSI* is the difference of activity at preferred and orthogonal orientations, normalized by their sum, $(r_\mathrm{{pref}} - r_\mathrm{{orth}})/(r_\mathrm{{pref}}+r_\mathrm{{orth}})$. 
    $(r_\mathrm{{pref}}$ and $r_\mathrm{{orth}})$ are obtained from the best fit of a cosine function to output tuning curves, evaluated at $\mathrm{Output~PO}$ and $\mathrm{Output~PO} + 90\deg$, respectively. 
    Alternatively, OSI* can be computed from a linear interpolation of data points (inset). 
    Lighter colors show the distributions for lower contrasts, respectively. 
    All inputs have an OSI* of $0.1$.     
    (\textbf{C})~The OSI of all neurons for medium contrast (MC) vs.\ low contrast (LC), and for high contrast (HC) vs.\ medium contrast are plotted in the left and right panels, respectively. 
    The diagonal line indicates a perfect contrast invariance of OSI.    
    (\textbf{D})~Output PO vs.\ Output OSI for all neurons in the presynaptic pool of the neuron shown in Fig.~2A. A stimulus of medium contrast has been applied. 
    The neuron receives input from presynaptic neurons that are themselves highly selective on average (OSI distribution on the right), and which uniformly cover the whole range of possible Output POs (distribution on top). 
    The Output OSI and Output PO of the target neuron are $0.65$ and $105\deg$, respectively. 
    Other neurons receive similarly heterogeneous inputs (not shown).   }     
    \label{Fig_OsiPop} 

\end{figure}

To quantify orientation selectivity across the population, we compute two orientation selectivity indexes, OSI \cite{Ringach2002, Niell2008}, for all tuning curves (Fig.~\ref{Fig_OsiPop}). 
Both measures (Fig.~\ref{Fig_OsiPop}A and B) show that the mean OSI is increased by the network, to a level that is compatible with the results reported in animal experiments \cite{Niell2008, Ringach2002, Dragoi2001, Chapman1993}. 
Moreover, the selectivity is maintained upon increasing the input intensity, as both the mean value and the shape of OSI distributions are very similar for different contrasts. 

%%%%%%%%%%%%%%%%%%%%%%%%%%%%%%%%%%%%%%%%%%%%%%%%%%%%%%%%%%%%%%%%%%%%%%%%%%%%%%%%
%%%%%%%%%% Fig3 

%\begin{center}
%\fbox{Fig.~3 approximately here} 
%\end{center} 
%%%%%%%%%%%%%%%%%%%%%%%%%%%%%%%%%%%%%%%%%%%%%%%%%%%%%%%%%%%%%%%%%%%%%%%%%%%%%%%%

To directly verify this invariance, we compare the OSI of all neurons at different contrasts. 
Plotting the OSIs at the medium contrast (MC) vs.\ the lowest contrast (LC), and at the highest contrast (HC) vs.\ the medium, indeed reveal that the majority of neurons show a remarkable robustness of their tuning curves upon a change in contrast (Fig.~\ref{Fig_OsiPop}C).

The high selectivity in the network emerges despite the fact that each neuron receives input from a large pool of neurons with heterogeneous selectivity and different preferred orientations (Fig.~\ref{Fig_OsiPop}D). 
In fact, the PO distribution of presynaptic neurons is essentially uniform (Fig.~\ref{Fig_OsiPop}D, top histogram), and the presynaptic OSI distribution (Fig.~\ref{Fig_OsiPop}D, right histogram) is very similar to the OSI distribution of the whole population (Fig.~\ref{Fig_OsiPop}A). 
Therefore, the output response is highly selective, despite the fact that the input is quite heterogeneous, as reported in experiments \cite{Jia2010, Varga2011, Chen2011}. 

This result is similar to a recent study of orientation selectivity in rodents \cite{Hansel2012}, in that both show random networks are capable of generating selective output responses.
In the following, we provide a detailed mathematical analysis of the mechanisms involved in this process.
Our mean-field analysis indeed enables us to compute the mean output responses of networks quite precisely.

%%%%%%%%%%%%%%%%%%%%%%%%%%%%%%%%%%%%%%%%%%%%%%%%%%%%%%%%%%%%%%%%%%%%%%%%%%%%%%%% 
\subsection{A reduced linear rate model of the network} 
\label{Sec_NtwrkAnls}

\begin{figure}[h!] 
\centering\includegraphics[width=6.0in]{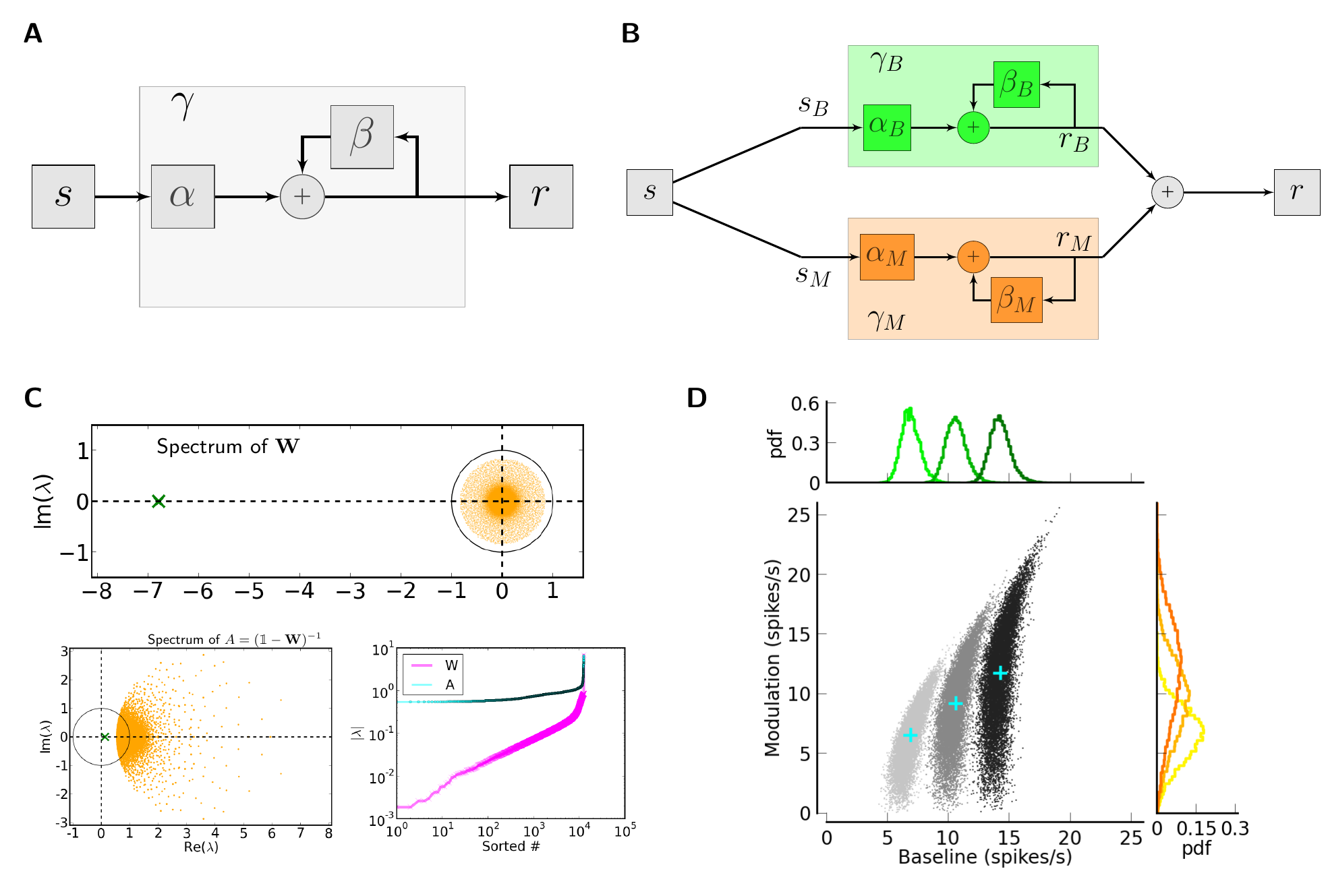} 
\caption{\textbf{Selective processing of baseline and modulation. }    
    (\textbf{A})~General reduced circuit model for the operation of a network on its inputs.     
    (\textbf{B})~Reduced circuit model for selective operation of the network on baseline and modulation components of an input vector.      
    (\textbf{C})~Top: Eigenvalue distribution of the weight matrix, $\vec{W}$, shown for $\mathrm{EPSP} = 0.1\,\mathrm{mV}$ ($J = \tau_\mathrm{syn} e \mathrm{EPSP} = 0.136\,\mathrm{mV}$). 
    For normalization, each entry is divided by the reset potential, $V_{\mathrm{reset}}=20\,\mathrm{mV}$. 
    The `exceptional eigenvalue' (green) corresponds to the uniform eigenmode, i.e.\ the baseline, and the bulk of the spectrum (orange) determines the response of the network to perturbations of a uniform input. 
    Bottom, left: Eigenvalue distribution for the matrix $(\vec{\one} - \vec{W})^{-1}$, which gives the stationary firing rates. 
    %Note that for illustration purposes eigenvalues outside the range chosen here have not been shown. 
    Bottom, right: Sorted magnitudes of eigenvalues of $\vec{W}$ and $(\vec{\one} - \vec{W})^{-1}$.    
    (\textbf{D})~Baseline and modulation components for individual neurons in the network with $\mathrm{EPSP} = 0.1\,\mathrm{mV}$. 
    Scatter plot (center) shows the modulation vs.\ baseline component of output tuning curves for all neurons of the network. 
    Baseline and modulation are taken as the mean (F0) and the second Fourier component (F2) of individual tuning curves, respectively. 
    The markers (cyan crosses) show the center of mass of each cloud. 
    The histograms indicate the marginal distributions of baseline (green, top) and modulation (orange, right) components, respectively.  } 
    \label{Fig_SepPath}        

\end{figure}

To illustrate the main network processing, we first recruit a linear rate model of the network.
We start by a reduced diagrammatic description of the network \cite{Douglas1995}.
This is equivalent to the description of the network in its stationary state, in terms of neuronal firing rates $\vec{r}$ that arise as a result of a stimulus $\vec{s}$ driving a recurrent network $\vec{W}$ (see Methods, Sect.~\ref{Sec_tempmean}): 
\[ 
\vec{v} = -\vec{r} + \vec{W} \vec{r} + J_s \vec{s}. 
\label{Eq1} 
\] 
The matrix $\vec{W}$ encodes the recurrent synaptic connections in a network comprising $N$ neurons, $\vec{s}$ and $\vec{r}$ are the $N$-dimensional vectors of firing rates of input/stimulus and output/response, respectively, $\vec{v}$ is the $N$-dimensional vector of time-averaged membrane potentials, and $J_s$ is the single-neuron gain. If $\vec{v} = \vec{0}$, the input-output relation is given by 
\[ 
0 = -\vec{r} + \vec{W} \vec{r} + J_s \vec{s}, 
\label{Eq2a} 
\] 
and if the matrix $\vec{\one} - \vec{W}$ is invertible, this readily implies 
\[ 
\vec{r} = J_s (\vec{\one}-\vec{W})^{-1}\vec{s} = \vec{A} \vec{s}. 
\label{Eq2b} 
\] 
The single-neuron gain $J_s$ in Eq.~(\ref{Eq2a}) subsumes the feedforward processing of input to all neurons, before recruiting any lateral interactions ($\alpha$ in Fig.~\ref{Fig_SepPath}A). 
The operation of the recurrent network on a rate vector $\vec{r}$ is given by the feedback $\vec{W} \vec{r}$ appearing in the same equation. 
If the feedback gain $\beta$ was the same for all activity configurations, $\gamma = \alpha/(1-\beta)$ would be the associated closed-loop gain of the system (Fig.~\ref{Fig_SepPath}A). 
In this case, any stimulus would be amplified or attenuated by the network with a uniform factor $\gamma$. 
As a consequence, if $\vec{\phi}$ is a stimulus feature systematically varied in an experiment, the output tuning curves $\vec{r}(\vec{\phi})$ would have the same shape as the input tuning curves $\vec{s}(\vec{\phi})$, as they would merely be rescaled by $\gamma$ as a whole. 
However, the amplification of orientation selectivity observed in our simulated networks suggests that the amplification factor of the untuned part (baseline) is considerably smaller than that of the tuned part (modulation). 
As a consequence, the selective amplification/attenuation of the networks considered here is reflected by an extended diagram (Fig.~\ref{Fig_SepPath}B) with two separate channels, and different overall gains for baseline and modulation, respectively. 
Even if the feedforward gain is identical for baseline and modulation, the feedback gain is different for each stimulus component, leading to different closed-loop gains $\gamma_B$ and $\gamma_M$ for each branch of the diagram. 

%%%%%%%%%%%%%%%%%%%%%%%%%%%%%%%%%%%%%%%%%%%%%%%%%%%%%%%%%%%%%%%%%%%%%%%%%%%%%%%%
%%%%%%%%%% Fig4 

%\begin{center} 
%\fbox{Fig.~4 approximately here} 
%\end{center} 
%%%%%%%%%%%%%%%%%%%%%%%%%%%%%%%%%%%%%%%%%%%%%%%%%%%%%%%%%%%%%%%%%%%%%%%%%%%%%%%%

The emergence of different processing pathways is a consequence of the linear recurrent dynamics: 
Any activity vector $\vec{x}$ (describing either input $\vec{s}$ to the network, or output $\vec{r}$ from the network) can be decomposed in terms of a sum $\vec{x} = \vec{x}_B + \vec{x}_M$. 
The part $\vec{x}_B = \langle \vec{x} \rangle$ is a pure baseline vector, representing the mean response rate of each neuron across all stimuli. 
The remaining part $\vec{x}_M = \vec{x} - \vec{x}_B$ is a pure modulation vector, with zero baseline. 
If the input is processed by a recurrent network that operates linearly on the input according to $\vec{r} = \vec{A} \vec{s}$ for some effective matrix $\vec{A}$ (cf.~Eq.~(\ref{Eq2b})) it is evident that $\vec{r}_B = \vec{A} \vec{s}_B$ and $\vec{r}_M = \vec{A} \vec{s}_M$ (see Sect.~\ref{Sec_BM} for further explanation). 
In particular, there is no cross-talk in the processing of baseline, $\vec{s}_B$, and modulation, $\vec{s}_M$, whatsoever. 
The independent, non-interfering processing of the baseline and the modulation component of the input is exactly corresponding to the two separate pathways depicted in Fig.~\ref{Fig_SepPath}B. 

For the networks considered in this work, mean and variance of all inputs are identical, and all neurons in the recurrent network have the same number of excitatory and inhibitory recurrent afferents. 
Therefore, the entries of the baseline vector $\vec{s}_B$ of the input are all identical, and it is mapped to a baseline output firing rate $\vec{r}_B$, which is again a uniform vector. 
This means that uniform vectors are eigenvectors of the matrices $\vec{W}$ and $\vec{A}$, respectively. 
The eigenvalue belonging to these eigenvectors is exactly the feedback gain $\beta_B$ of the baseline. 
In networks with dominant inhibition $\beta_B$ is negative. 
As a consequence, the corresponding closed-loop gain $\gamma_B$ is a positive number, smaller than $1$ (Fig.~\ref{Fig_SepPath}C). 
The effective enhancement of feature specificity mediated by the recurrent operation of the network is the result of different closed-loop gains for modulation, $\gamma_M$, and baseline, $\gamma_B$. 
For the example network of Fig.~1-3, this leads to comparable strengths of baseline and modulation in the output tuning curves (Fig.~\ref{Fig_SepPath}D), despite having much weaker modulation in the input. 

To see how these gains change with the strength of recurrent coupling, we fixed the network connectivity and only changed the post-synaptic amplitudes in the network.
We then computed the mean baseline and modulation gains in each network, corresponding to cross marks in Fig.~\ref{Fig_SepPath}D.
Note that the gains are now computed from individual tuning curves, $r_i(\theta)$. 
For each output tuning curve, the baseline and modulation component is computed as the zeroth and the second Fourier component, respectively, and then the average values are computed across the population (see Methods for details).
For the cosine tuning we are using here these gains are equal to population gains we described before (for rate vectors over the population).

\begin{figure}[h!] 
\centering\includegraphics[width=4.5in]{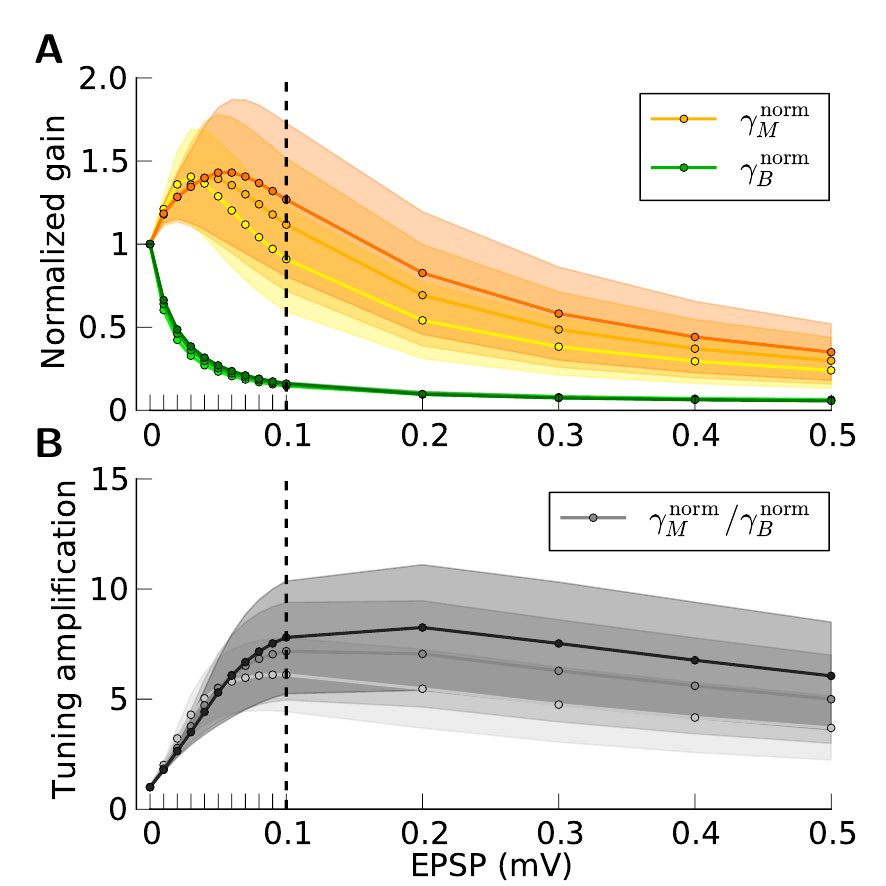} 
\caption{\textbf{Network gains for the selective processing of baseline and modulation.}    
    (\textbf{A})~Normalized baseline gain, $\gamma^\mathrm{norm}_B$ (green), and modulation gain, $\gamma^\mathrm{norm}_M$ (orange), for a network with a fixed connectivity, but different strengths of recurrent synaptic couplings. 
    Shaded regions represent $\mathrm{mean} \pm \mathrm{std}$. 
    Lighter colors correspond to lower contrasts.    
    (\textbf{B})~As a result of selective attenuation of the baseline, the normalized modulation-to-baseline ratio ($\gamma^\mathrm{norm}_M/\gamma^\mathrm{norm}_B$) is generally much larger than $1$. 
    The degree of recurrence of the network presented in Figs.~1-4 is marked by dashed lines.  } 
    \label{Fig_BasModGains} 
    
\end{figure}

The normalized gains, with respect to corresponding gains at zero recurrence, are plotted in Fig.~\ref{Fig_BasModGains}A.
Increasing the strength of recurrent couplings in the network results in a monotonic decrease of $\gamma_B$ (green curve), since the network is inhibition dominated.
The associated change of $\gamma_M$ (orange curve), however, is non-monotonic: 
It increases until a certain degree of recurrence is reached, and then it decreases slowly, while always remaining significantly larger than $\gamma_B$. 
As a result, the modulation ratio, $\gamma_M/\gamma_B$, of the network exhibits a peak for some degree of recurrence (Fig.~\ref{Fig_BasModGains}B), reflecting optimal performance with regard to tuning amplification.

%%%%%%%%%%%%%%%%%%%%%%%%%%%%%%%%%%%%%%%%%%%%%%%%%%%%%%%%%%%%%%%%%%%%%%%%%%%%%%%%
%%%%%%%%%% Fig5 

%\begin{center} 
%\fbox{Fig.~5 approximately here} 
%\end{center}

%%%%%%%%%%%%%%%%%%%%%%%%%%%%%%%%%%%%%%%%%%%%%%%%%%%%%%%%%%%%%%%%%%%%%%%%%%%%%%%%

%%%%%%%%%%%%%%%%%%%%%%%%%%%%%%%%%%%%%%%%%%%%%%%%%%%%%%%%%%%%%%%%%%%%%%%%%%%%%%%% 
%%%%%%%%%%%%%%%%%%%%%%%%%%%%%%%%%%%%%%%%%%%%%%%%%%%%%%%%%%%%%%%%%%%%%%%%%%%%%%%% 
\subsection{Stability of the network} 
\label{Sec_MFA}

\begin{figure}[h!] 
\centering\includegraphics[width=6.0in]{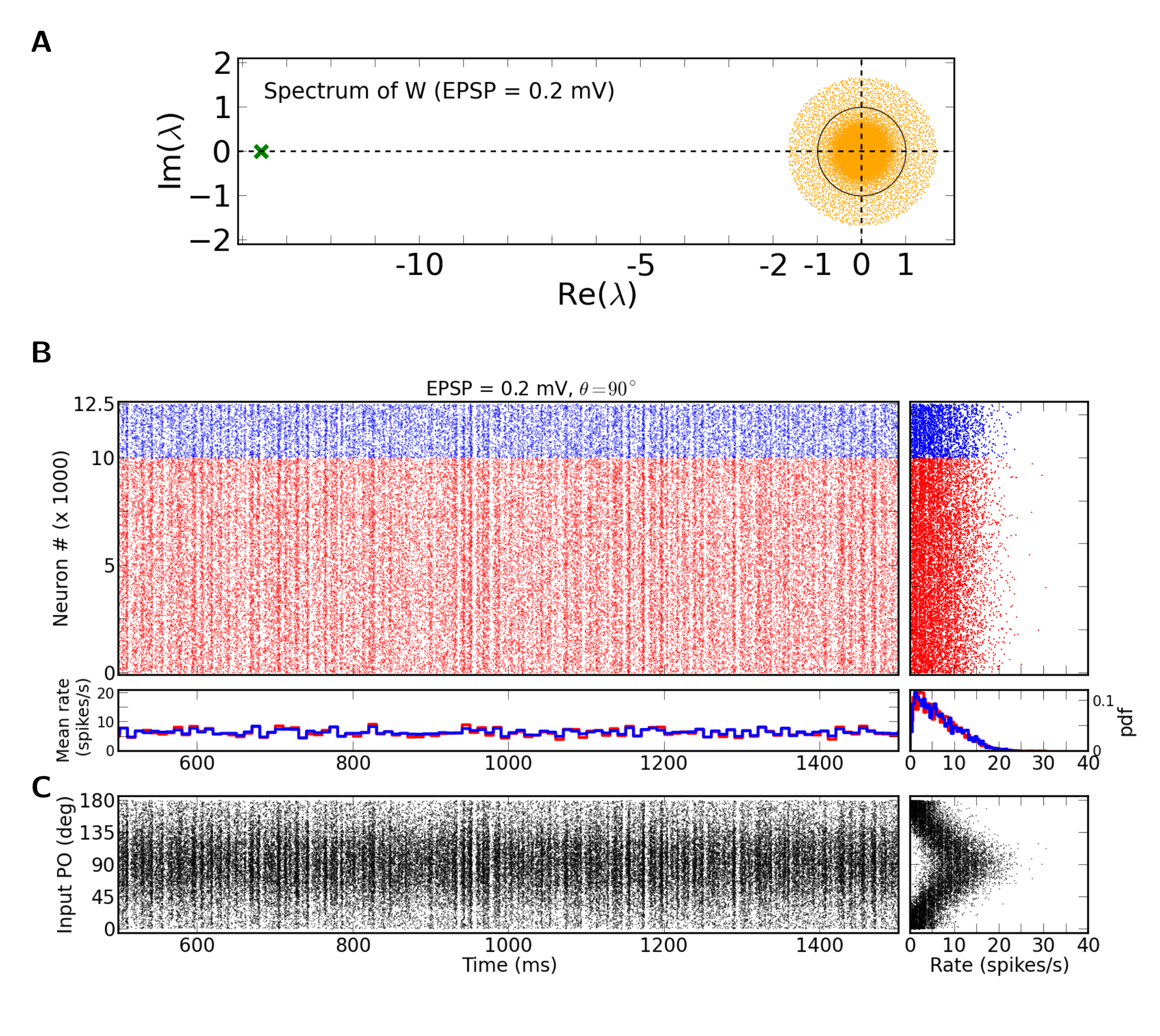} 
\caption{\textbf{Stable dynamics of a network with an unstable eingenvalue spectrum.}    
    (\textbf{A})~Eigenvalue distribution of the weight matrix, $\vec{W}$, shown for $\mathrm{EPSP} = 0.2\,\mathrm{mV}$ ($J = 0.27\,\mathrm{mV}$). 
    The same normalization as in Fig.~\ref{Fig_SepPath}C is employed, i.e.\ each entry is divided by the reset potential, $V_{\mathrm{reset}}=20\,\mathrm{mV}$.         
    (\textbf{B-C})~Same plots as in Fig.~\ref{Fig_RasPlt_0.1}A, B, for $\mathrm{EPSP} = 0.2\,\mathrm{mV}$.    } 
    \label{Fig_InstabDyn} 
    
\end{figure}

One puzzling observation here is that the behavior of networks does not exhibit any dynamic instability as the recurrent coupling in the network is increased. 
This is counter-intuitive, as the radius of the bulk spectrum of the network, $\rho$, increases linearly with the EPSP amplitude in the network (see Eq.~\ref{Eq_rho} in Methods). 
In fact, already at $\mathrm{EPSP} = 0.2\,\mathrm{mV}$ this radius is larger than $1$, as illustrated in Fig.~\ref{Fig_InstabDyn}A, which could result in instability upon stimulating the network with a modulated input. 
However, the network does not show such instability.

%%%%%%%%%%%%%%%%%%%%%%%%%%%%%%%%%%%%%%%%%%%%%%%%%%%%%%%%%%%%%%%%%%%%%%%%%%%%%%%% 
%%%%%%%%%% Fig6 

%\begin{center} 
%\fbox{Fig.~6 approximately here} 
%\end{center} 
%%%%%%%%%%%%%%%%%%%%%%%%%%%%%%%%%%%%%%%%%%%%%%%%%%%%%%%%%%%%%%%%%%%%%%%%%%%%%%%% 

\begin{figure}[h!] 
\centering\includegraphics[width=6.0in]{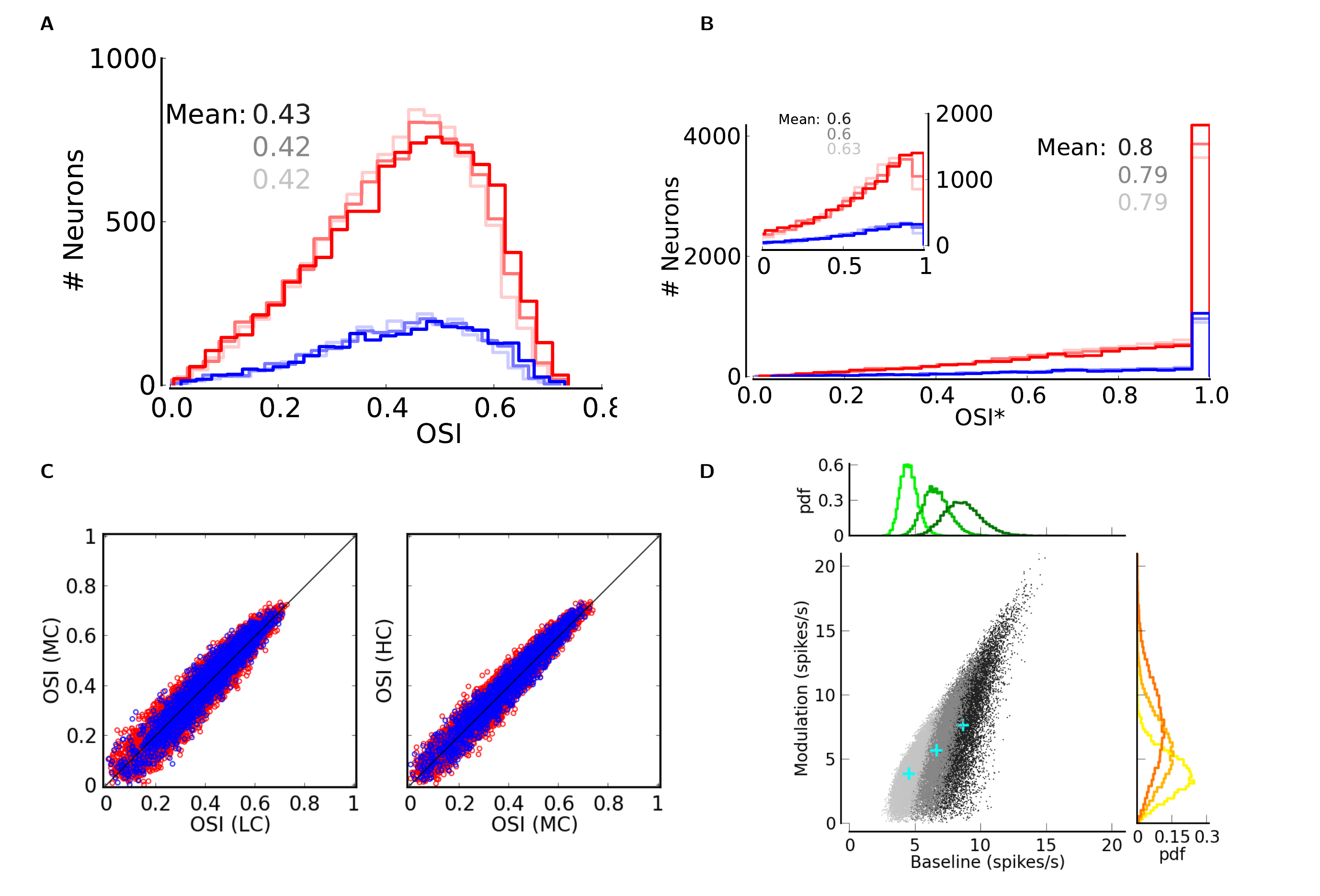} 
\caption{\textbf{Qualitativly similiar functional properties of a network with an unstable spectrum.}    
    (\textbf{A-B})~Same distributions of OSI and OSI* as in Fig.~\ref{Fig_OsiPop}A, B, for a network with $\mathrm{EPSP} = 0.2\,\mathrm{mV}$.        
    (\textbf{C})~Contrast invariance of the OSI measure, as in Fig.~\ref{Fig_OsiPop}C, for $\mathrm{EPSP} = 0.2\,\mathrm{mV}$. 
    (\textbf{D})~Distribution of baseline (F0) and modulation (F2) component in a network with $\mathrm{EPSP} = 0.2\,\mathrm{mV}$. 
    Labeling is the same as in Fig.~\ref{Fig_SepPath}D.  } 
    \label{Fig_InstabFunc} 
        
\end{figure}

First, the network dynamics does not change significantly, as demonstrated in Fig.~\ref{Fig_InstabDyn}B and \ref{Fig_InstabDyn}C. 
Moreover, the same functional properties are implied from the output tuning curves of neurons in the network (Fig.~\ref{Fig_InstabFunc}), without any sign of unstable operation. 
Altogether, it seems that the networks experience a smooth transition as EPSP amplitudes increase, and no abrupt change of the state. 
Instabilities are avoided by some dynamic mechanism. 
We come back to this point in Sect.~\ref{Sec_MGOPN}. 

%%%%%%%%%%%%%%%%%%%%%%%%%%%%%%%%%%%%%%%%%%%%%%%%%%%%%%%%%%%%%%%%%%%%%%%%%%%%%%%% 
%%%%%%%%%% Fig7 

%\begin{center} 
%\fbox{Fig.~7 approximately here} 
%\end{center}
%%%%%%%%%%%%%%%%%%%%%%%%%%%%%%%%%%%%%%%%%%%%%%%%%%%%%%%%%%%%%%%%%%%%%%%%%%%%%%%% 

%%%%%%%%%%%%%%%%%%%%%%%%%%%%%%%%%%%%%%%%%%%%%%%%%%%%%%%%%%%%%%%%%%%%%%%%%%%%%%%% 
%%%%%%%%%%%%%%%%%%%%%%%%%%%%%%%%%%%%%%%%%%%%%%%%%%%%%%%%%%%%%%%%%%%%%%%%%%%%%%%% 
\subsection{A simplified mean-field analysis of the network} 
\label{Sec_MFA}

To compute the gains for baseline and modulation of inputs, respectively, we employ a simplified mean-field approximation, which considers the corresponding average gains. 
For that, we need to compute the mean baseline and modulation rate of output tuning curves in the network. 

\begin{figure}[h!] 
\centering\includegraphics[width=4.5in]{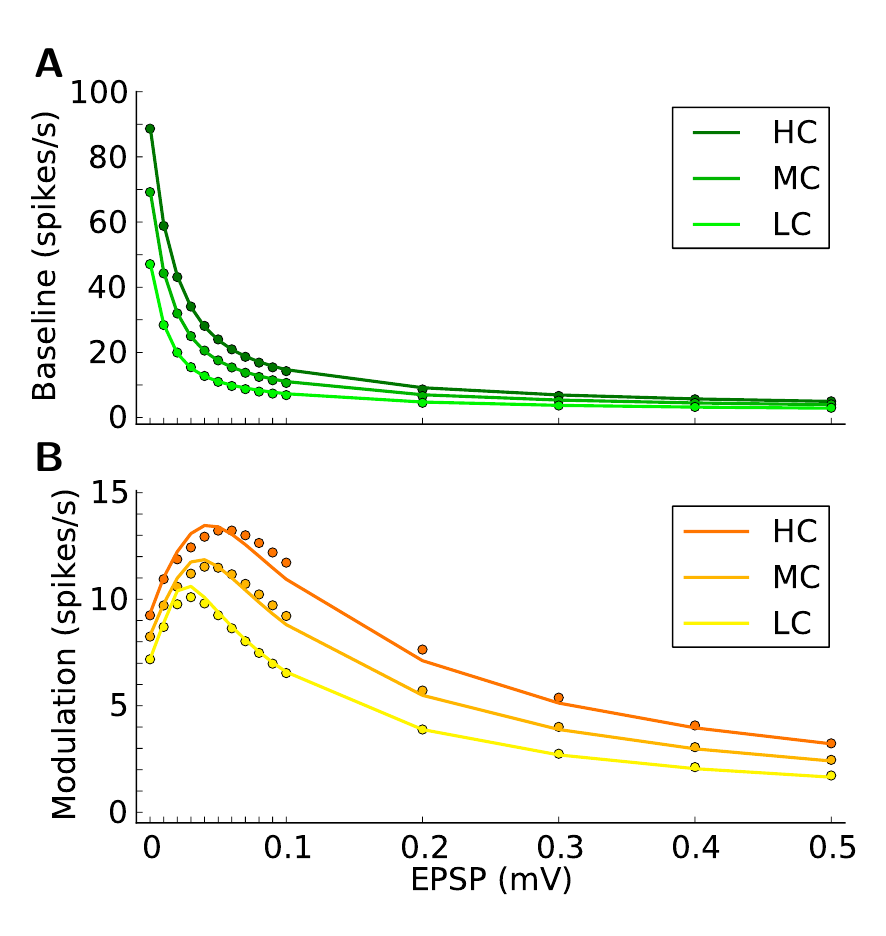} 
\caption{\textbf{Theoretical prediction of baseline and modulation gains.}    
    (\textbf{A,B})~The mean value of baseline (A) and modulation (B) components for each network at each contrast obtained by numerical simulation (circles), along with the values predicted by the theory decribed in the text (solid lines). } 
    \label{Fig_BasModPred} 

\end{figure}

To this end, we first compute the mean firing rate of output tuning curves, $r_B$. 
This is obtained by assuming no modulation in the input, i.e.\ as if all neurons are receiving the untuned component of the input tuning curve. 
This is justified by the fact that, in absence of strong non-linearities, the two pathways discussed above do not interfere with each other. 
We therefore employ mean field theory (see Sect.~\ref{Sec_scfr}) and compute the self-consistent baseline firing rates as done previously \cite{Brunel2000}. 
The predicted result, for networks with different couplings and for stimuli with different contrasts, is shown in Fig.~\ref{Fig_BasModPred}A, which matches the simulated results quite well.

Next, we compute the mean modulation component of output tuning curves, $r_M$. 
Here we make an approximation: 
We neglect the modulation of other neurons in the network and consider only the modulation of input to one neuron (see Sect.~\ref{Sec_MFRMMP}). 
This is equivalent to assuming `perfect balance' in terms of modulation, where all the modulation components of recurrent inputs from the network cancel each other perfectly ($\beta_M = 0$), such that only feedforward modulation remains. 

Based on this simplification, the mean modulation of the response, $r_M$, is already well predicted (Fig.~\ref{Fig_BasModPred}B). 
The residual small discrepancy as compared to numerical simulations, which is most pronounced for intermediate recurrence and high contrast, should be accounted for by including network interactions (Sect.~\ref{Sec_MxPFRRN}) that amplify the modulation, and spike correlations that are ignored in the simplified treatment presented here.  

%%%%%%%%%%%%%%%%%%%%%%%%%%%%%%%%%%%%%%%%%%%%%%%%%%%%%%%%%%%%%%%%%%%%%%%%%%%%%%%% 
%%%%%%%%%% Fig8 

%\begin{center} 
%\fbox{Fig.~8 approximately here}
%\end{center} 
%%%%%%%%%%%%%%%%%%%%%%%%%%%%%%%%%%%%%%%%%%%%%%%%%%%%%%%%%%%%%%%%%%%%%%%%%%%%%%%% 

\begin{figure}[h!] 
\centering\includegraphics[width=6.0in]{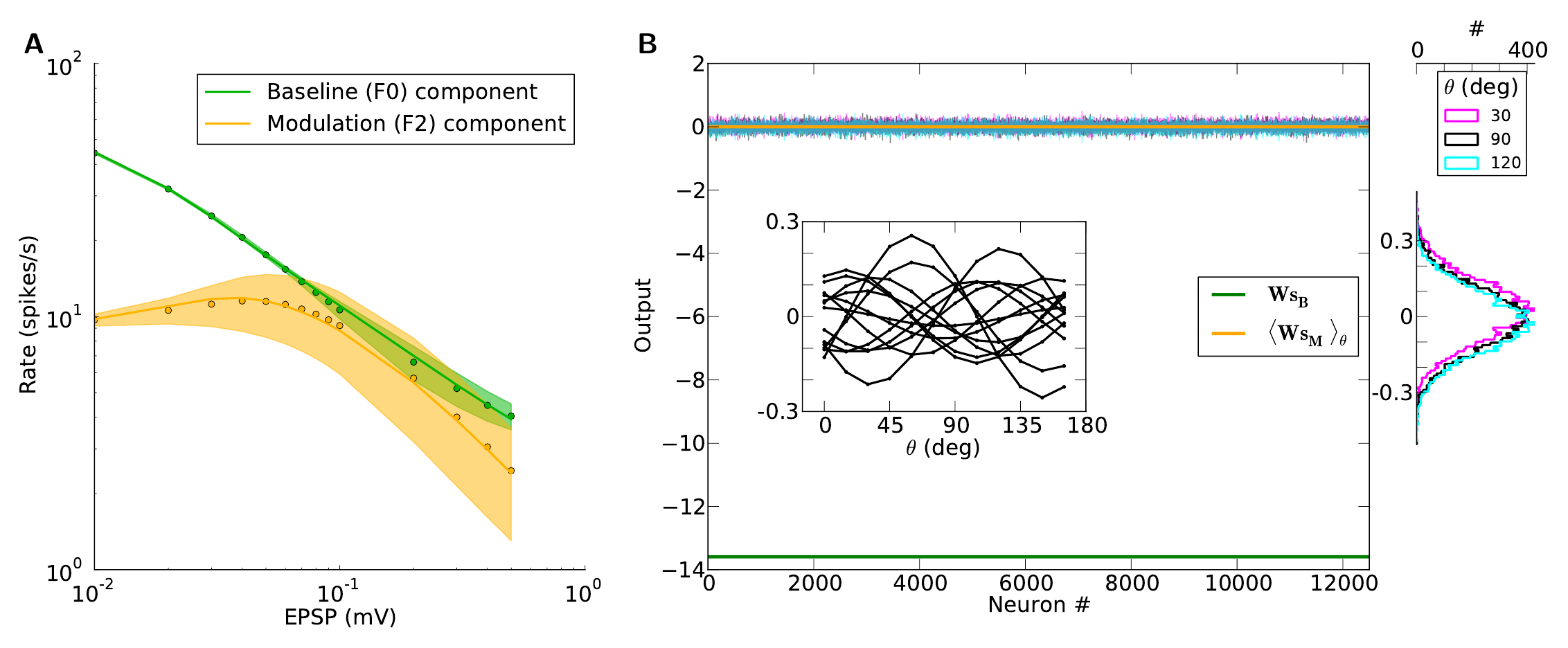} 
\caption{\textbf{Non-interference of baseline and modulation.}    
    (\textbf{A})~Mean and standard deviation of baseline (green) and modulation (orange) components for the medium contrast are plotted in logarithmic scales for comparison.     
    As in Fig.~\ref{Fig_BasModPred}, dots and solid lines indicate simulated and predicted values, respectively. 
    Shadings indicate $\mathrm{mean} \pm \mathrm{std}$. 
    As in Fig.~\ref{Fig_BasModGains}, mean and standard deviation are evaluated over all neurons in the network. 
    (\textbf{B})~The product of the weight matrix $\vec{W}$ with a baseline, $\vec{s}_B$, and a modulation input vector, $\vec{s}_M$. 
    The entries of the baseline vector are all normalized to one, i.e.\ the input to each neuron is $1$. 
    The operation of the weight matrix on the baseline vector, $\vec{W} \vec{s}_B$, is plotted in green. 
    The operation of the weight matrix on the modulation vector, $\vec{W} \vec{s}_M$, for three different orientations ($\theta = 30\deg, 90\deg, 120\deg$) are plotted in magenta, black and cyan, respectively. 
    The corresponding distributions of individual responses are plotted in the magnified histograms on the right. 
    Note that the assumption of `perfect balance' implies a very narrow distribution around zero.
    The average (over $12$ orientations) of the responses, $\langle \vec{W} \vec{s}_M \rangle_\theta$, are plotted in orange. 
    For $12$ sample neurons from the network the response vs.\ orientation of the stimulus are plotted in the inset (center). 
     } 
    \label{Fig_OrthVer} 
    
\end{figure}

From this result, we can also verify the non-interference property with regard to baseline and modulation. 
The separation of pathways proposed in Fig.~\ref{Fig_SepPath}B suggests that the modulated component of the input vector, $s_M$, does not affect the baseline component of output tuning curves, $r_B$. 
This we can verify directly by plotting the mean and standard deviation of baseline and modulation components of output tuning curves (Fig.~\ref{Fig_OrthVer}A). 
Although the variance of the modulation component increases by increasing the recurrence in the network, the variance of baseline firing rate is very small and does not change with recurrence. 
It only begins to increase when the mean value of baseline and modulation components become comparable in size (about $\mathrm{EPSP} = 0.1\,\mathrm{mV}$). 
This is the point at which tuning curves experience partial rectification: 
For the tuning curves with modulation component larger than the baseline component, some rectification is implied. 
Rectification, in turn, distorts the baseline component of tuning curves. 

The non-interference property of baseline and modulation can also be directly demonstrated from the weight matrix. 
The fact that baseline input does not have any component along the modulation vectors became clear from the eigenvector of $\vec{W}$ that corresponds to the exceptional eigenvalue. 
To show the opposite, namely that an input modulation vector induces none, or only negligible, baseline in the output response, explicit numerical simulations of the result of $\vec{W} \vec{s}_M$ are performed. 
The result is shown in Fig.~\ref{Fig_OrthVer}B, which demonstrates that the expected value of $\vec{W} \vec{s}_M$ (over orientation) is exactly zero, therefore not introducing any baseline component, as we discussed above. 

%%%%%%%%%%%%%%%%%%%%%%%%%%%%%%%%%%%%%%%%%%%%%%%%%%%%%%%%%%%%%%%%%%%%%%%%%%%%%%%% 
%%%%%%%%%% Fig9 

%\begin{center} 
%\fbox{Fig.~9 approximately here} 
%\end{center} 
%%%%%%%%%%%%%%%%%%%%%%%%%%%%%%%%%%%%%%%%%%%%%%%%%%%%%%%%%%%%%%%%%%%%%%%%%%%%%%%% 

Therefore, baseline and modulation are processed separately and independently, with no cross-talk involved, provided the network acts linearly on its inputs. In contrast, we currently cannot mathematically justify the assumption of perfect balance. 
The reason is that the modulation vectors, unlike the baseline vector, are not eigenvectors of the weight matrix, $\vec{W}$. 
As a result, it is not justified to replace $\vec{W}$ in the product $\vec{W} \vec{r}_M$ with a scalar value $\beta_M$. 
Treating the problem more rigorously could involve expanding the modulation vector in terms of the eigenvectors corresponding to the bulk eigenvalues of $\vec{W}$ (Fig.~\ref{Fig_SepPath}C), and obtaining the gain accordingly. 
This gain would not, in general, be a single scalar value, nor would it be exactly zero, as we have assumed here. 
We come back to a more precise treatment of the problem in Sect.~\ref{sect_LTRN}. 

%%%%%%%%%%%%%%%%%%%%%%%%%%%%%%%%%%%%%%%%%%%%%%%%%%%%%%%%%%%%%%%%%%%%%%%%%%%%%%%% 
%%%%%%%%%%%%%%%%%%%%%%%%%%%%%%%%%%%%%%%%%%%%%%%%%%%%%%%%%%%%%%%%%%%%%%%%%%%%%%%% 
\subsection{Tuning of recurrent inputs}
\label{Sec_TRI}

The assumption of perfect balance of modulation is the first-order approximation we make to obtain average gains of the network. 
Here we numerically check how far this assumption is from the result of our simulations. 
To answer this question, we investigate tuning of different components of the input to neurons in a network. 
We reconstruct the input from excitatory and inhibitory presynaptic sources by replacing each spike with the synaptic kernel (alpha function) and adding all the contributions to obtain a shot-noise signal. 
We then compute the mean value of this signal as the mean presynaptic excitation and inhibition, respectively.

%%%%%%%%%%%%%%%%%%%%%%%%%%%%%%%%%%%%%%%%%%%%%%%%%%%%%%%%%%%%%%%%%%%%%%%%%%%%%%%% 
%%%%%%%%%% Fig10 

\begin{figure}[h!] 
\centering\includegraphics[width=6.0in]{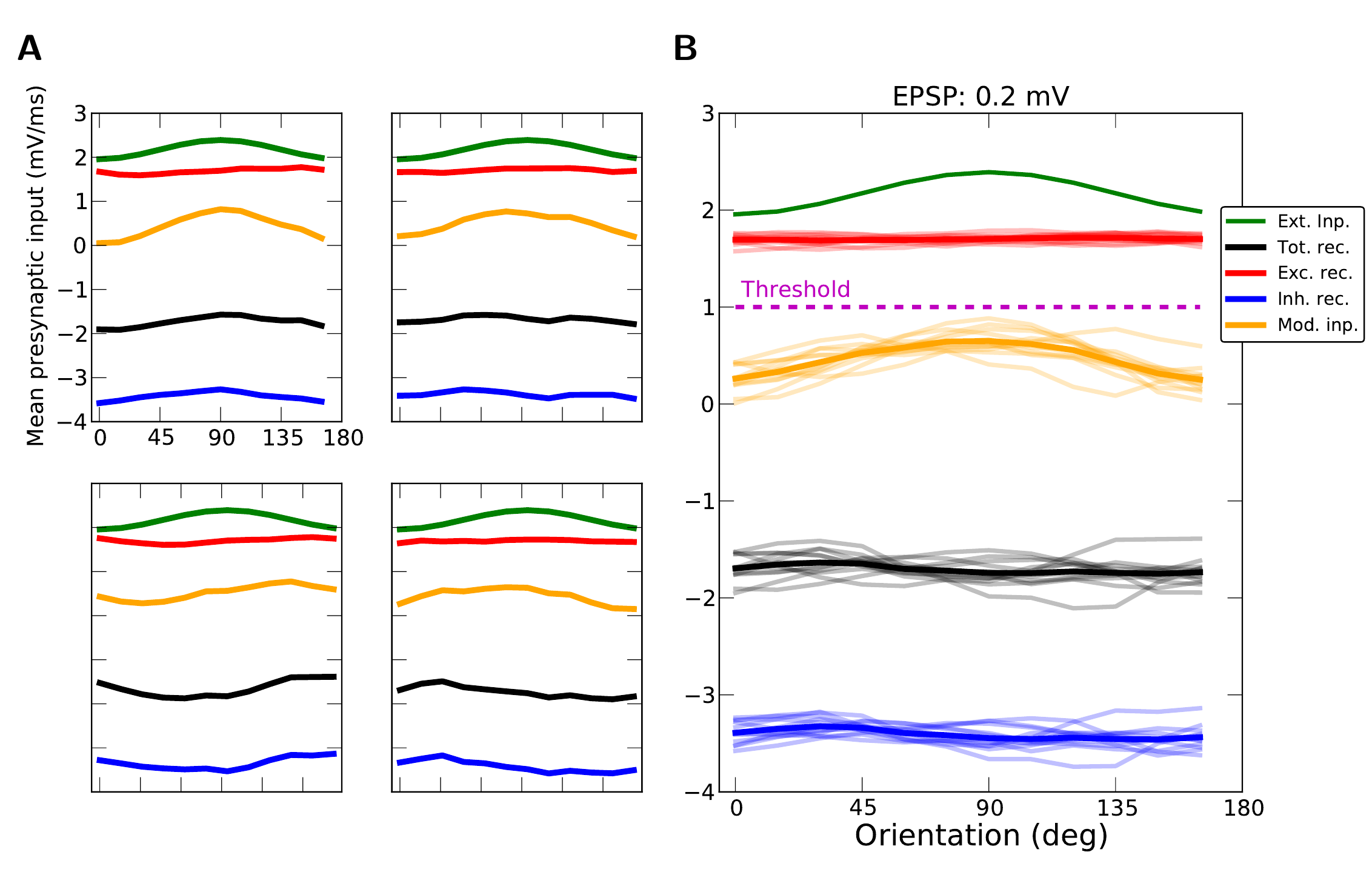} 
\caption{\textbf{Tuning of neuronal inputs.}    
    (\textbf{A})~Tuning of the mean input from excitatory (red) and inhibitory (blue) neuron populations, and of the external input (green). Shown are four sample neurons, for stimulations at the medium contrast. 
    $\mathrm{EPSP} = 0.2\,\mathrm{mV}$. 
    The total recurrent input (excitatory + inhibitory), and the net modulation of the input (external + total recurrent) are also plotted (in black and orange, respectively). 
    The input is computed by replacing each presynaptic spike by an alpha kernel and computing the mean amplitude of the shot-noise signal (in $\mathrm{mV}/\mathrm{ms}$). 
    $1\,\mathrm{mV}/\mathrm{ms}$ corresponds to a mean membrane potential at the spike threshold of the neurons in our simulations.    
    (\textbf{B})~Same for twelve sample neurons, along with their mean values (thicker lines). } 
    \label{Fig_InpTun} 
    
\end{figure} 

%\begin{center} 
%\fbox{Fig.~10 approximately here} 
%\end{center} 
%%%%%%%%%%%%%%%%%%%%%%%%%%%%%%%%%%%%%%%%%%%%%%%%%%%%%%%%%%%%%%%%%%%%%%%%%%%%%%%% 

Fig.~\ref{Fig_InpTun}A shows these inputs for four sample neurons from the network.
The tuning of excitatory input is very weak for all samples; the tuning of inhibition, however, is on average stronger.
The output tuning of a cell results from a combination of contributions from feedforward and from recurrent inputs. 
The recurrent input can therefore change the feedforward tuning:
It can either retain or change the preferred orientation of the feedforward input (upper and lower panels, respectively), and it can either amplify or attenuate the tuning strength (upper right and lower right, respectively).
On average, however, the initial shape of the tuning curve is maintained (Fig.~\ref{Fig_InpTun}B, traces and averages for $12$ cells), although the tuning of recurrent input leads to a deviation from the feedforward tuning, which creates a distribution of selectivity.

%%%%%%%%%%%%%%%%%%%%%%%%%%%%%%%%%%%%%%%%%%%%%%%%%%%%%%%%%%%%%%%%%%%%%%%%%%%%%%%% 
%%%%%%%%%%%%%%%%%%%%%%%%%%%%%%%%%%%%%%%%%%%%%%%%%%%%%%%%%%%%%%%%%%%%%%%%%%%%%%%% 
\subsection{Linear tuning in recurrent networks} 
\label{sect_LTRN} 

The simplified mean-field analysis discussed above accounts for the average tuning curve and the mean selectivity in the network. 
It does not, however, account for the distribution of orientation selectivity across neurons. 
In this section, we resort to a linear analysis of modulation processing, in order to provide an approximative analytic treatment of this distribution. 

For this linear analysis we need to make two additional assumptions. 
First, we assume that modulations in the network can be treated as small perturbations about the baseline, and that the dynamics can be linearized about this operating point. 
Note that this assumption implies that the contribution of nonlinearities like rectification is negligible.
Second, we assume that the mixture of tunings is linear. 
This assumption is justified as we have used cosine tuning (i.e.\ linear tuning) in the inputs (see Methods for details). 
This allows us to represent each tuning curve as a 2D feature vector.
Since the operation of network on feature vectors is linear,  the mixture of tunings is now reduced to the vectorial summation of corresponding tuning vectors (see below, and Methods). 

\begin{figure}[h!] 
\centering\includegraphics[width=4.5in]{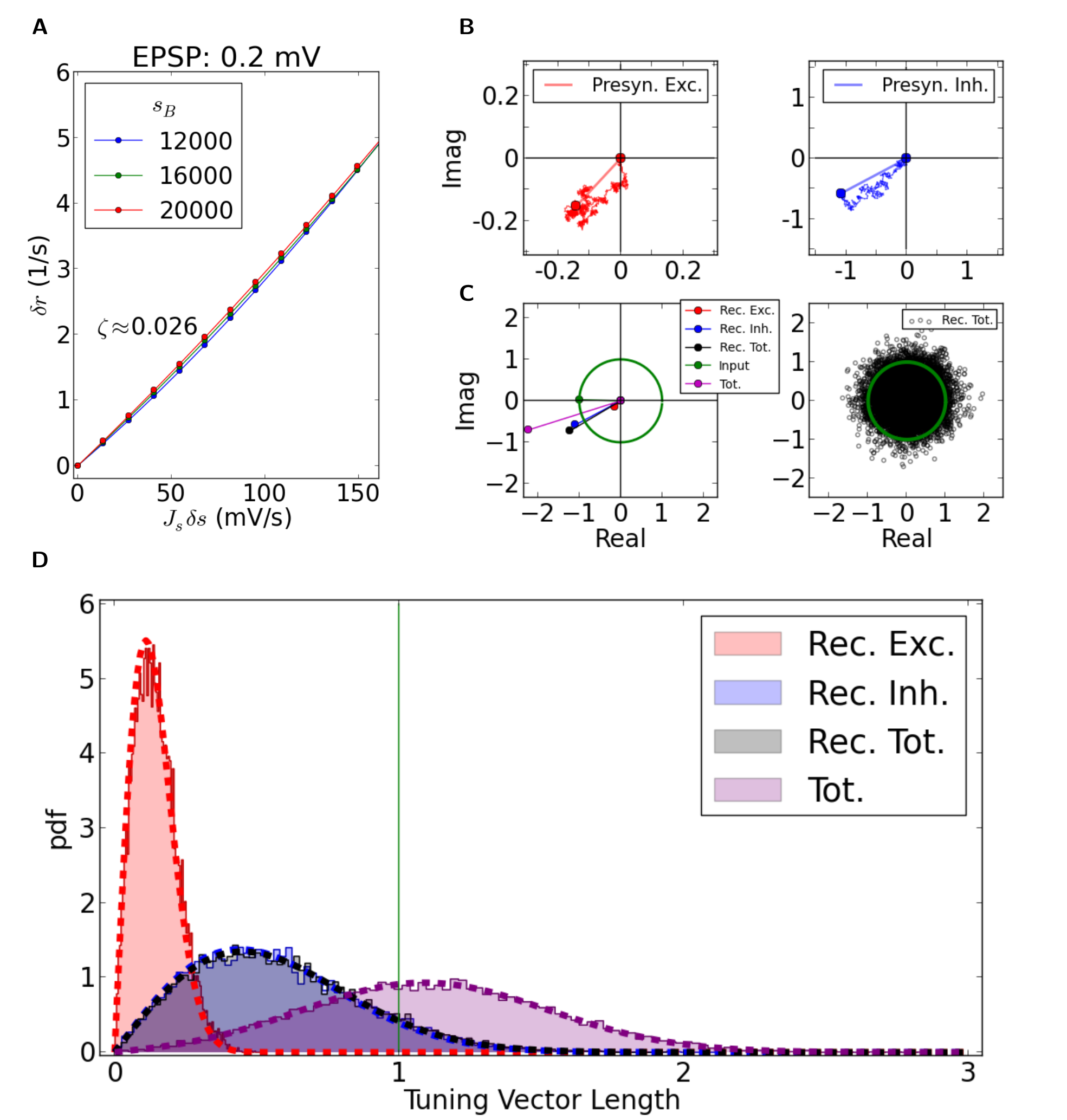} 
\caption{\textbf{Linear tuning in recurrent networks.}    
    (\textbf{A})~Linearized gains for single neurons embedded in the network. 
    The extra firing rate, $\delta r$, of a neuron produced in response to a small perturbation, $J_s \delta s$, in the input intensity, plotted for different baseline inputs corresponding to different contrasts. 
    The response is computed by numerically perturbing the mean-field equations (see Methods). 
    The linear gain, $\zeta = \delta r/(J_s \delta s)$, is then computed by linear regression of data points. 
    For this example with $\mathrm{EPSP} = 0.2\,\mathrm{mV}$, the value $\zeta = 0.026$ is obtained, which is also used for the next panels.    
    (\textbf{B})~For the sample neuron in Fig.~\ref{Fig_TunCurves}A, all presynaptic Tuning Vectors are extracted ($J \exp(j2\theta^*_i)$, weighted by the linear gain ($\zeta$), and vectorially added together, reflecting linear integration in neurons. 
    Although each presynaptic vector makes only a small contribution, the resulting random sum can lead to a large resultant Tuning Vector. 
    These are generally larger for presynaptic inhibition (Presyn.~Inh.) compared to presynaptic excitation (Presyn.~Exc.). 
    Note different scales of axes.    
    (\textbf{C})~Left panel: The resultant vectors for recurrent excitation (Rec.~Exc.), recurrent inhibition (Rec.~Inh.), total recurrent (Rec.~Tot. = Rec.~Exc. + Rec.~Inh.), feedforward input (Input), and the total input (Tot. = Input + Rec.~Tot.) are plotted. 
    All normalized input Tuning Vectors have the same length of one, denoted by the green circle. 
    Right panel: Total recurrent Tuning Vectors (Rec.~Tot.) for all neurons in the network are compared with the normalized length of their input Tuning Vectors (green circle).    
    \textbf{D})~Distribution of the length of all Tuning Vectors for all the neurons in the network. 
    Dashed lines show the predicted distributions of the linear analysis in each case (see text for details). 
    } 
    \label{Fig_LinTun}    

\end{figure}

We therefore start first by linearizing the dynamics about the operating point, i.e.\ the baseline. 
After computing the baseline firing rate, $r_B$, as described before, we compute the mean and standard deviation of the input to each neuron in the baseline state ($\mu_B$ and $\sigma_B$, respectively; Eq.~\eqref{Eq_inputMS} in Methods) as a function of baseline input $s_B$ and baseline firing rate $r_B$ 
\begin{align} 
  \mu_B &= \tau [J_s s_B + J r_B N \epsilon (f - g(1-f)]), \nonumber \\ 
  \sigma_B^2 &= \tau [J_s^2 s_B + J^2 r_B N \epsilon (f + g^2(1-f))]. \nonumber 
\end{align} 
We then compute the linear gains by perturbing the input with a small $\delta s$. 
We dismiss all contributions of order two or higher in the perturbation parameter and write 
\[ 
\delta r = \zeta J_s \delta s, 
\] 
where $\delta r$ is the change in the output firing rate resulting from a perturbation of the input rate $\delta s$. 
Here, we compute these gains numerically by solving the perturbed mean field equations about the basline (see Methods, and Eq.~\ref{Eq_zeta_num}). 
Fig.~\ref{Fig_LinTun}A shows the input-output relationship for the network with $\mathrm{EPSP} = 0.2\,\mathrm{mV}$ at the highest contrast. 

%%%%%%%%%%%%%%%%%%%%%%%%%%%%%%%%%%%%%%%%%%%%%%%%%%%%%%%%%%%%%%%%%%%%%%%%%%%%%%%% 
%%%%%%%%%% Fig11 

%\begin{center} 
%\fbox{Fig.~11 approximately here} 
%\end{center} 
%%%%%%%%%%%%%%%%%%%%%%%%%%%%%%%%%%%%%%%%%%%%%%%%%%%%%%%%%%%%%%%%%%%%%%%%%%%%%%%% 

After computing the linear gains, we can rewrite the linear rate equations of the network for the modulation pathway. 
The modulation in the firing rate of each neuron is caused by a linear mixture of external input and the input it receives from the recurrent network 
\[ 
\vec{r}_M = \zeta (\vec{W} \vec{r}_M + J_s \vec{s}_M). 
\] 
Both of these terms are now weighted with the extra factor, $\zeta$, from linearization, which is the effective gain of modulation at this operating point. 
If we rewrite the equation as 
\[ 
\vec{r}_M = (\one - \zeta \vec{W})^{-1} \zeta J_s \vec{s}_M = \zeta J_s \sum_{k=0}^{\infty} (\zeta \vec{W})^k \vec{s}_M, 
\label{Eq_pow_ser}
\] 
it can further be approximated by 
\[ 
\vec{r}_M \approx \zeta J_s (\vec{s}_M + \zeta \vec{W} \vec{s}_M).
\label{lin_mix} 
\] 
%assuming $\rho(\zeta \vec{W}) \ll 1$. 
Here we are neglecting the contribution of higher-order recurrent inputs ($(\zeta \vec{W})^2 \vec{s}_M$, $(\zeta \vec{W})^3 \vec{s}_M$, ...) in the processing. 

Next step is to interprete the above equation for tuning curves. 
Since we have assumed cosine tuning for the input, we can represent each input tuning curve by a vector, $S_i$.
Likewise, we represent output tuning curves by vectors $R_i$.
We refer to these vectors as the Tuning Vectors.
For a 2D feature like orientation selectivity we obtain 2D Tuning Vectors.
The angle of this vector represents the input preferred orientation $\theta_i^*$, and the length of it is a measure of orientation selectivity (it is indeed equal to OSI before normalization).
For notational convenience, we represent these 2D vectors by complex numbers.
We identify real parts with x-directions, and imaginary parts with y-directions.
Any 2D vector then corresponds to a complex number in a one to one fashion.
 
The mixture of tuning curves in cosine tuning is now simplified to the summation of the corresponding Tuning Vectors in the complex plane (see Methods for further details). 
We can therefore rewrite the linear rate equation Eq.~\eqref{lin_mix} for Tuning Vectors 
\[ 
\vec{R} = \zeta J_s (\vec{S} + \zeta \vec{W} \vec{S}). 
\label{eq_linear_mixing} 
\] 
Here $\vec{R}$ and $\vec{S}$ are vectors with complex elements, representing output and input tuning of all neurons in the network, respectively. 
This is now an equation which expresses the output Tuning Vectors in terms of a linear mixture of input Tuning Vectors.
Similar to Eq.~\eqref{lin_mix}, we are neglecting higher-order terms ($(\zeta \vec{W})^2 \vec{S}$, $(\zeta \vec{W})^3 \vec{S}$, ...) here.

An individual output Tuning Vector, $R_i$, is then given by $R_i = \zeta J_s (S_i + \zeta \sum_j W_{ij} S_j)$: 
The output tuning of each neuron is a mixture of its input tuning and weighted vectors of all presynaptic sources. 
For the specific example of the neuron in Fig.~\ref{Fig_TunCurves}A, all the contributions of presynaptic sources are shown in Fig.~\ref{Fig_LinTun}B, for excitatory and inhibitory populations separately. 
Each small jump in the space represents the contribution of a presynaptic Tuning Vector, the size of which is $\zeta J$ or $- \zeta g J$ for excitatory and inhibitory presynaptic sources, respectively. 
$S_i$ is normalized to $1$. 

Although each presynaptic contribution is much smaller than the input (of order $\zeta J = \O(10^{-3})$ in this case), the resultant vector (dashed lines) can be large. 
In particular, the resultant vector of inhibition is comparable to the length of the input vector for this specific example (Fig.~\ref{Fig_LinTun}C, left). 
Since the angle of the vector is close to the input angle, this leads to an amplification of the resultant tuning, although it typically also changes the preferred orientation (Fig.~\ref{Fig_LinTun}C, left). 
This explains why the OSI of this neuron ($0.65$) is larger than the mean OSI of the network ($0.42$, see Fig.~\ref{Fig_TunCurves} and \ref{Fig_OsiPop}). 

Not all the vectors resulting from recurrent contributions, however, are large, nor do all have a similar preferred orientation as the input tuning. 
Indeed, as the connectivity is random and the initial preferred orientations are assigned randomly to each neuron, the preferred orientation of the resultant vectors are also random. 
This is shown in Fig.~\ref{Fig_LinTun}C, right panel, where all the recurrent Tuning Vectors are explicitly computed (by reading the input Tuning Vectors and the connectivity), and plotted in the complex plane. 
The distribution of the vectors in this plane is a 2D Gaussian distribution, according to the Central Limit Theorem. 
As a result, most of the Tuning Vectors have small magnitude, and only a few of greater magnitude contribute to the tail of the distribution. 

The distribution of this length is plotted in Fig.~\ref{Fig_LinTun}D, for different subpopulations of neurons. 
The peak of the distribution for excitatory neurons is at smaller values than for inhibitory neurons, and the overall length of recurrent tuning is mainly determined by inhibition in the network. 
Knowing the standard deviation of distributions of Tuning Vectors, one obtains the distribution of vector lengths according to $\frac{x}{\sigma^2} e^{-x^2/(2\sigma^2)}$ (see Methods, Eq.~\eqref{Weibull}). 
In this example, the standard deviations are $\sigma_\mathrm{exc} = 0.11$, $\sigma_\mathrm{inh} = 0.44$ and $\sigma_\mathrm{tot} = 0.45$, for excitation, for inhibition, and for all recurrent neurons, respectively. 
The length of tuning vectors predicted by this result is plotted in Fig.~\ref{Fig_LinTun}D (dashed lines). 
The length of overall tuning, i.e.\ recurrent Tuning Vectors vectorially combined with the input Tuning Vector (green), can also be computed (see Methods, Eq.~\eqref{Tot_tuning}). 
Normalizing the input Tuning Vectors to length $1$, the distribution amounts to $\frac{x}{\sigma^2} e^{-(x^2 + 1)/(2\sigma^2)} I_0 (\frac{x}{\sigma^2})$, plotted in the same figure (dashed purple line). 

From Fig.~\ref{Fig_LinTun}D one can compare the strength of the input tuning (green line) with the tuning generated within the random network (black distribution). 
The mean length of the recurrent tuning is smaller than the feedforward, and only few neurons show comparable tuning strength. 
The distribution of the combined tuning strength (purple line), which is a mixture of feedforward and recurrent components, has now a broad distribution, where many neurons show less tuning than their input (less than $1$, attenuated), and many more have enhanced selectivity (greater than $1$, amplified). 
In general, amplification happens when the randomly generated Tuning Vector within the network is roughly aligned with the initial Tuning Vector, and attenuation happens for recurrent Tuning Vectors in the opposite directions.\footnote{Note that, as a $\pi$-periodic parameter $\theta$ is now mapped to a $2\pi$-periodic parameter $2\theta$, an opposite direction here implies an orthogonal orientation in the original parameter space.}
A random recurrent network, thus, in itself is capable of attenuating and amplifying orientation selectivity. This is a mechanism in addition to the selective gains of baseline and modulation described before. 
This mechanism, however, comes at the expense of shifting the tuning curves of neurons from their initial, feedforward preferred orientations. 

%%%%%%%%%%%%%%%%%%%%%%%%%%%%%%%%%%%%%%%%%%%%%%%%%%%%%%%%%%%%%%%%%%%%%%%%%%%%%%%% 
%%%%%%%%%%%%%%%%%%%%%%%%%%%%%%%%%%%%%%%%%%%%%%%%%%%%%%%%%%%%%%%%%%%%%%%%%%%%%%%% 
\subsection{Regimes of orientation selectivity} 
\label{Sec_RegOS} 

\begin{figure}[h!] 
\centering\includegraphics[width=6.in]{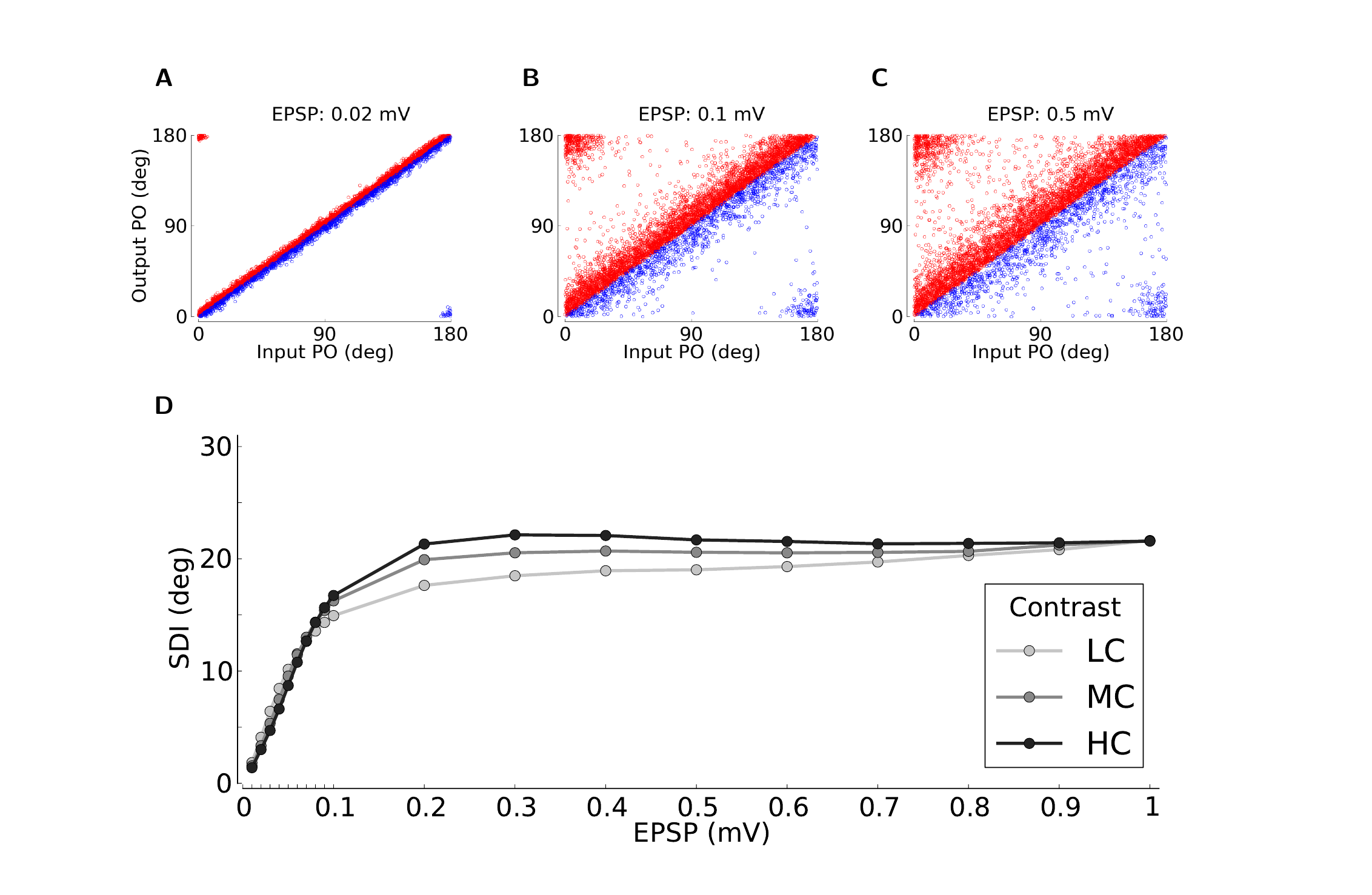} 
\caption{\textbf{Input vs.\ output preferred orientation of neurons in the network.}    
    (\textbf{A})~Output PO vs.\ Input PO for three values of recurrence, at the medium contrast. 
    Increasing the recurrence leads to more scatter about the diagonal. 
    For illustration purposes, the excitatory (red) and inhibitory (blue) neurons have been plotted only above or below the diagonal, respectively.     
    (\textbf{B})~To quantify the amount of PO change when going from input to output, the Scatter Degree Index (SDI) is plotted as the angular deviation of $\Delta \mathrm{PO} =  \mathrm{Output PO} - \mathrm{Input PO}$ (see Methods). 
    The maximum value of this index is $\approx 40.5\deg$, which corresponds to a uniform distribution of $\Delta \mathrm{PO}$.  
    Darker colors show higher contrasts, respectively. 
    } 
    \label{Fig_PO_Scat} 
    
\end{figure}

\begin{figure}[h!] 
\centering\includegraphics[width=6.0in]{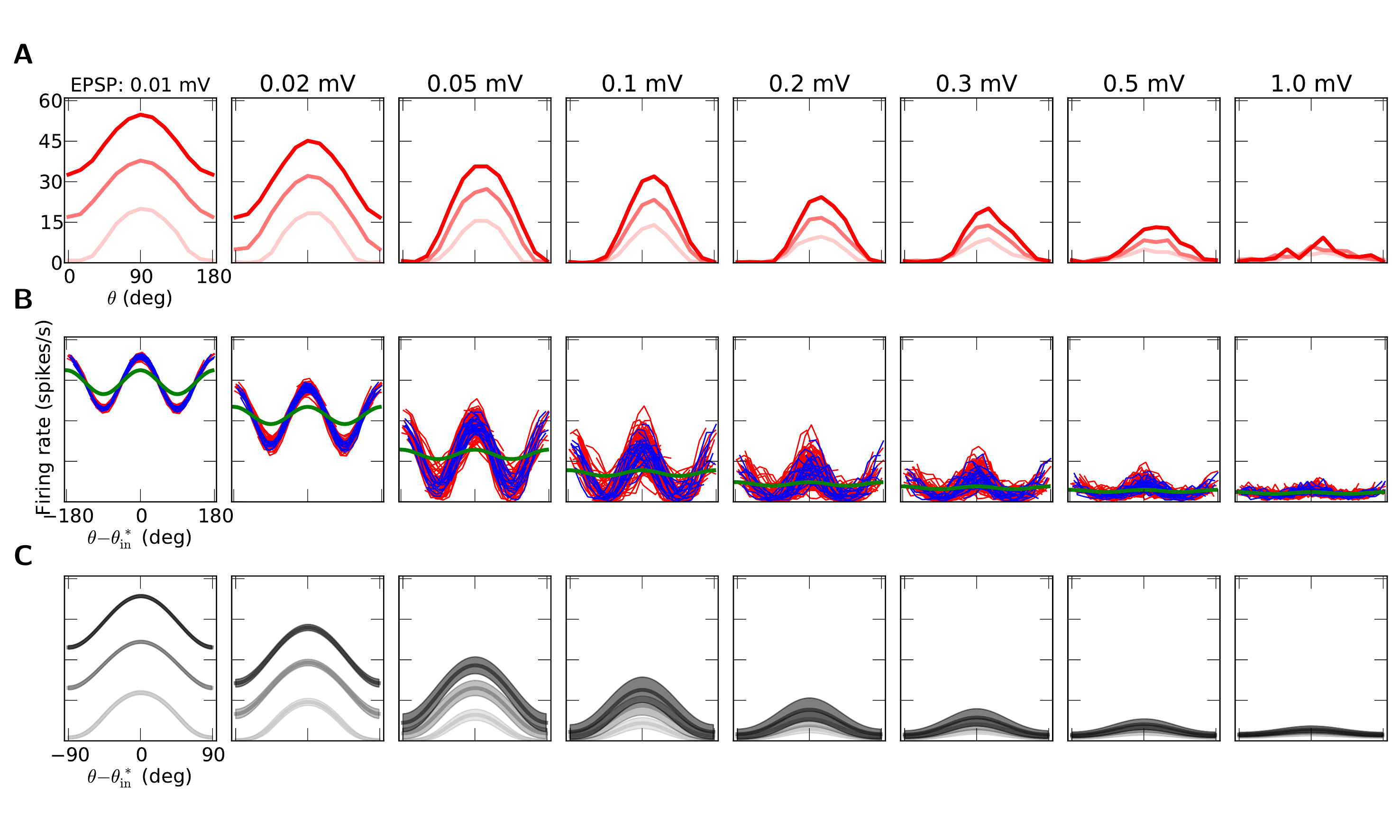} 
\caption{\textbf{Neuronal tuning curves in weakly (left) and strongly (right) recurrent networks.}    
    (\textbf{A})~Tuning curves of a sample neuron (same as in Fig.~\ref{Fig_TunCurves}A), for different degrees of recurrence in the network, as indicated by different EPSP amplitudes.    
    (\textbf{B})~More ($125$) sample tuning curves from the network, aligned to their Input PO. 
    Red and blue curves show excitatory and inhibitory output tuning curves for the medium contrast, respectively. 
    %The maximum value on y-axis in each case denotes the maximum output firing rate of all tuning curves in a given network, respectively. 
    Shown in green is the input tuning curve, normalized to the average (over the population) of the mean (over all orientations) of all tuning curves in the network. 
    (\textbf{C})~Mean and standard deviation (across neurons) of all aligned tuning curves, for networks with different degrees of recurrence. 
    Lighter shadings denote lower contrasts, respectively.} 
    \label{Fig_TC_Shapes} 
    
\end{figure}

\begin{figure}[h!] 
\centering\includegraphics[width=6.in]{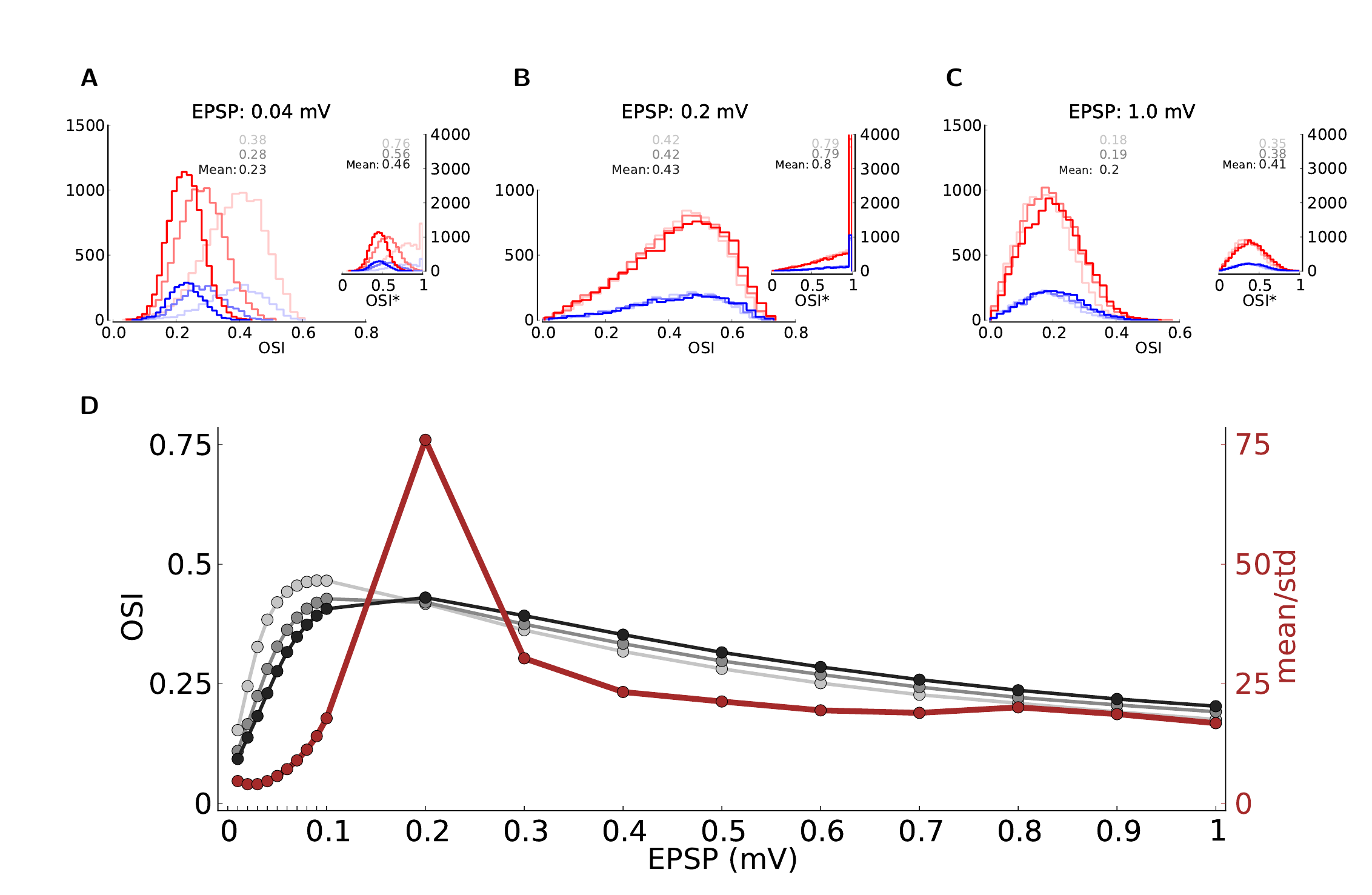} 
\caption{\textbf{Distribution and contrast invariance of selectivity.}            
    (\textbf{A})~Distribution of orientation selectivity in three different regimes of recurrence. 
    %Increasing the recurrence makes the distribution broader and more stable. 
    Lighter colors code for lower contrasts. 
    The mean OSI for each distribution is indicated in the corresponding brightness, respectively. 
    OSI* is computed from the cosine fit, as in the main panel of Fig.~\ref{Fig_OsiPop}B.    
    (\textbf{B})~The mean OSI of all neurons in the network is shown for different levels of recurrence, at three different contrasts. 
    Lower contrasts are plotted in lighter colors. 
    %The shaded region indicates the range of variation in the mean OSI. 
    Increasing the recurrence makes the OSI less susceptible to changes in contrast. 
    For high recurrences, this invariance comes at the expense of a decreased selectivity, as the mean OSI in the network decreases. 
    This trade-off between selectivity and invariance is quantified (in brown) by the average (over contrasts) of mean OSI (across neurons) divided by the standard deviation (over contrasts) of the average OSI (across neurons).} 
    \label{Fig_OSI_trdoff}    

\end{figure}

As Fig.~\ref{Fig_PO_Scat}A shows, in a weakly coupled network the PO of neurons are hardly changed with respect to their input PO. 
This is a regime where feedforward projections are dominant with respect to functional properties of neurons. 
Under these conditions, the recurrent network cannot compensate the increase in the baseline, and both baseline and modulation are scaled identically (Fig.~\ref{Fig_TC_Shapes}). 
As a result, neuronal tuning curves tend to simply reflect the tuning of the input (Fig.~\ref{Fig_TC_Shapes}), and tuning curves reduce their selectivity for high-contrast inputs, because the feedback compensation is weak. 
Although the average orientation selectivity index (OSI) of neurons in the network could be high for the lowest contrast in a weakly connected network, this selectivity is lost when the contrast is increasing (Fig.~\ref{Fig_OSI_trdoff}A). 

%%%%%%%%%%%%%%%%%%%%%%%%%%%%%%%%%%%%%%%%%%%%%%%%%%%%%%%%%%%%%%%%%%%%%%%%%%%%%%%% 
%%%%%%%%%% Fig12 

%\begin{center} 
%\fbox{Fig.~12 approximately here} 
%\end{center} 
%%%%%%%%%%%%%%%%%%%%%%%%%%%%%%%%%%%%%%%%%%%%%%%%%%%%%%%%%%%%%%%%%%%%%%%%%%%%%%%% 

As the strength of recurrent couplings increases, the contribution of the network becomes more effective to attenuate the baseline and selectively enhance the modulation (Fig.~\ref{Fig_TC_Shapes} and \ref{Fig_BasModGains}). 
This leads to more stable OSI distributions across different contrasts (Fig.~\ref{Fig_OSI_trdoff}B and \ref{Fig_OSI_trdoff}C), and hence makes the selectivity more robust. 
Moreover, as a consequence of stronger recurrence in the network, output POs deviate more from their initial PO (Fig.~\ref{Fig_PO_Scat}B and \ref{Fig_PO_Scat}C), since the strength of recurrent contributions (recurrent Tuning Vectors) has now increased. 
This is summarized for all networks in Fig.~\ref{Fig_PO_Scat}D by a scatter degree index (SDI), which quantifies the degree of PO deviation in each network. 

%%%%%%%%%%%%%%%%%%%%%%%%%%%%%%%%%%%%%%%%%%%%%%%%%%%%%%%%%%%%%%%%%%%%%%%%%%%%%%%% 
%%%%%%%%%% Fig13 

%\begin{center} 
%\fbox{Fig.~13 approximately here} 
%\end{center} 
%%%%%%%%%%%%%%%%%%%%%%%%%%%%%%%%%%%%%%%%%%%%%%%%%%%%%%%%%%%%%%%%%%%%%%%%%%%%%%%% 

Notably, SDI does not linearly increase with recurrent connection strength. Rather, it saturates for rather weak connections, and then reaches an asymptotic value. 
The reason for this behavior is that the contribution of recurrent Tuning Vectors in the final tuning depends on $J$ and the linear gain, $\zeta$, as we described in the previous section (see Eq.~\eqref{eq_linear_mixing}). 
Although $J$ is monotonically increasing in Fig.~\ref{Fig_PO_Scat}D, the effective strength of recurrent Tuning Vectors depends on the product $J \zeta$. 
It appears, therefore, that the linear gains are not increasing as $J$ increases, or they are even decreasing. 

This trend is further supported by shape and size of the tuning curves (Fig.~\ref{Fig_TC_Shapes}). 
For networks with strong recurrent coupling, the maximum firing rate of tuning curves decreases and the modulation component becomes smaller. 
Since the linear gains determine the embedded gain of neurons in the network in response to modulations, this trend also suggests that these gains are decreasing in the highly recurrent regime. 
This was indeed visible in the behavior of gains and firing rates for modulations, shown in Fig.~\ref{Fig_BasModGains}A and \ref{Fig_BasModPred}A, respectively. 

%%%%%%%%%%%%%%%%%%%%%%%%%%%%%%%%%%%%%%%%%%%%%%%%%%%%%%%%%%%%%%%%%%%%%%%%%%%%%%%% 
%%%%%%%%%% Fig14 

%\begin{center} 
%\fbox{Fig.~14 approximately here} 
%\end{center} 
%%%%%%%%%%%%%%%%%%%%%%%%%%%%%%%%%%%%%%%%%%%%%%%%%%%%%%%%%%%%%%%%%%%%%%%%%%%%%%%% 

Although increasing the recurrence stabilizes the OSI, it makes the neurons of the network less feature selective, if the recurrence is too large (Fig.~\ref{Fig_OSI_trdoff}, A-D). 
There is, therefore, a trade-off between orientation selectivity and contrast invariance in the networks. 
Increasing the recurrence makes the negative feedback in the baseline pathway stronger, making the divisive suppression of the baseline -- and hence the contrast invariance -- more effective. 
This comes, however, at the expense of a decrease in the gain in the modulation pathway, which makes the responses weaker and less selective. 
We have quantified this trade-off by dividing the mean OSI of individual tuning curves in a network by its standard deviation across different contrasts. 
The intermediate recurrent regime shows optimal behavior with large and stable OSI (Fig.~\ref{Fig_OSI_trdoff}D), and it more or less coincides with the region of optimal tuning amplification (Fig.~\ref{Fig_BasModGains}). 

%%%%%%%%%%%%%%%%%%%%%%%%%%%%%%%%%%%%%%%%%%%%%%%%%%%%%%%%%%%%%%%%%%%%%%%%%%%%%%%% 
%%%%%%%%%%%%%%%%%%%%%%%%%%%%%%%%%%%%%%%%%%%%%%%%%%%%%%%%%%%%%%%%%%%%%%%%%%%%%%%% 
\subsection{Tuning and invariance of membrane potentials} 
\label{Sec_TIMP} 

\begin{figure}[h!] 
\centering\includegraphics[width=6.0in]{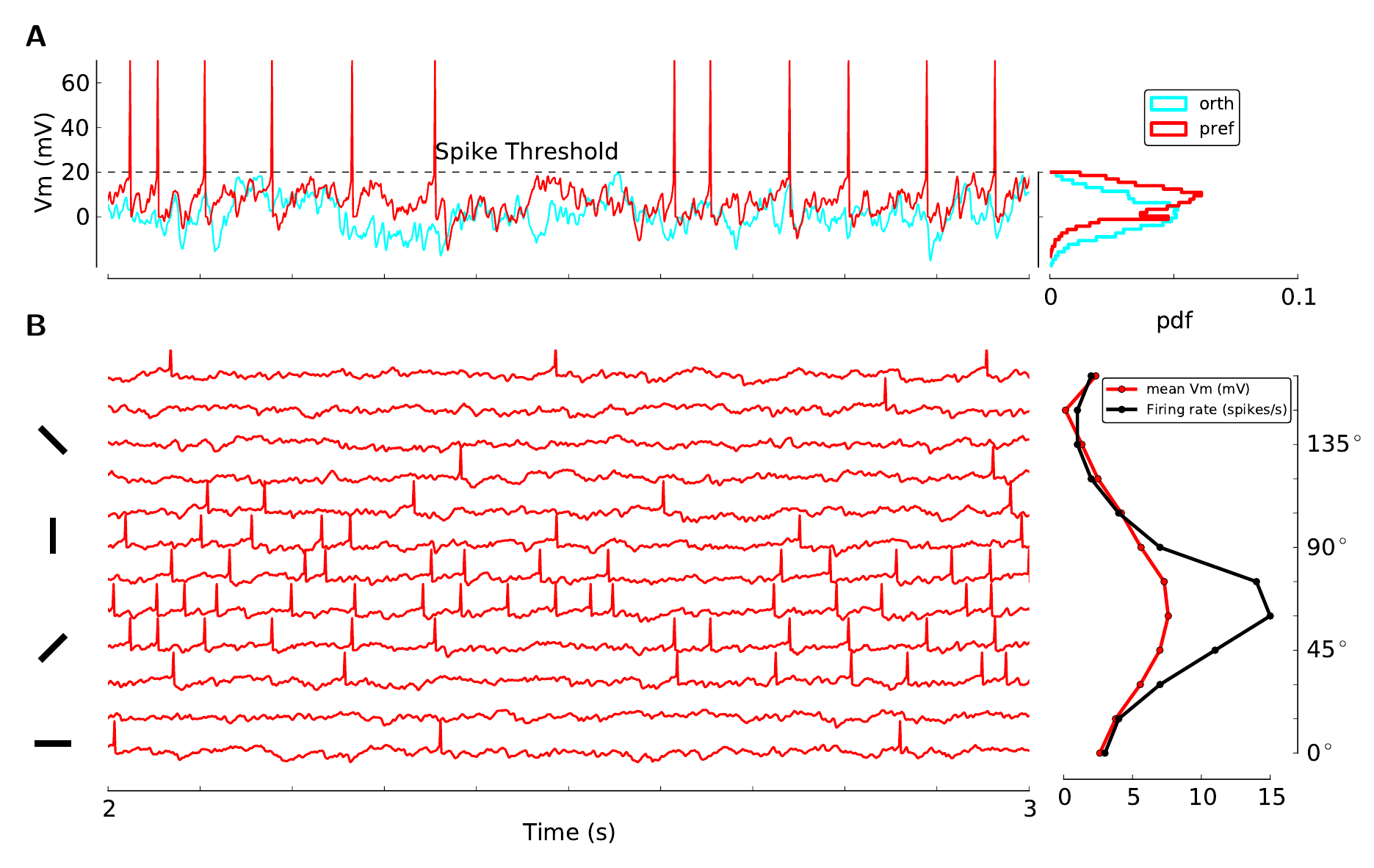} 
\caption{\textbf{Tuning of the membrane potential.}    
    (\textbf{A})~Membrane potential of a sample excitatory neuron from the same network with $\mathrm{EPSP} = 0.2\,\mathrm{mV}$, at the medium contrast. 
    Traces of the membrane potential are plotted for $1\,\mathrm{s}$ of response to the preferred (red) and its orthogonal (cyan) stimulus orientation. 
    The histograms of the membrane potential for $6\,\mathrm{s}$ of stimulation are shown on the right.     
    (\textbf{B})~Traces of the membrane potential along with the elicited spikes for $12$ orientations, for $1\,\mathrm{s}$ of recording. 
    The tuning curves of the mean membrane potential (red) and the corresponding firing rate (black) is computed from $6\,\mathrm{s}$ of stimulation in the right panel.} 
    \label{Fig_VmTrace} 

\end{figure}

It is also informative to look at the membrane potential dynamics of neurons in the network. 
Fig.~\ref{Fig_VmTrace}A shows the membrane potential of a sample neuron in response to a stimulus of its preferred orientation, and the orthogonal one. 
The mean membrane potential remains, on average, below threshold, as it is reflected in the distribution of membrane potential (Fig.~\ref{Fig_VmTrace}A, right panel), and only fluctuations in the input leads to spiking activity. 
For the orthogonal orientation, this activity is very sparse; for the preferred orientation it leads to a reasonable firing rate. 
If we plot the mean membrane potential for each orientation, it shows the same tuning as of the spiking activity (Fig.~\ref{Fig_VmTrace}B). 

%%%%%%%%%%%%%%%%%%%%%%%%%%%%%%%%%%%%%%%%%%%%%%%%%%%%%%%%%%%%%%%%%%%%%%%%%%%%%%%% 
%%%%%%%%%% Fig15 

%\begin{center} 
%\fbox{Fig.~15 approximately here} 
%\end{center} 
%%%%%%%%%%%%%%%%%%%%%%%%%%%%%%%%%%%%%%%%%%%%%%%%%%%%%%%%%%%%%%%%%%%%%%%%%%%%%%%% 

\begin{figure}[h!] 
\centering\includegraphics[width=5.0in]{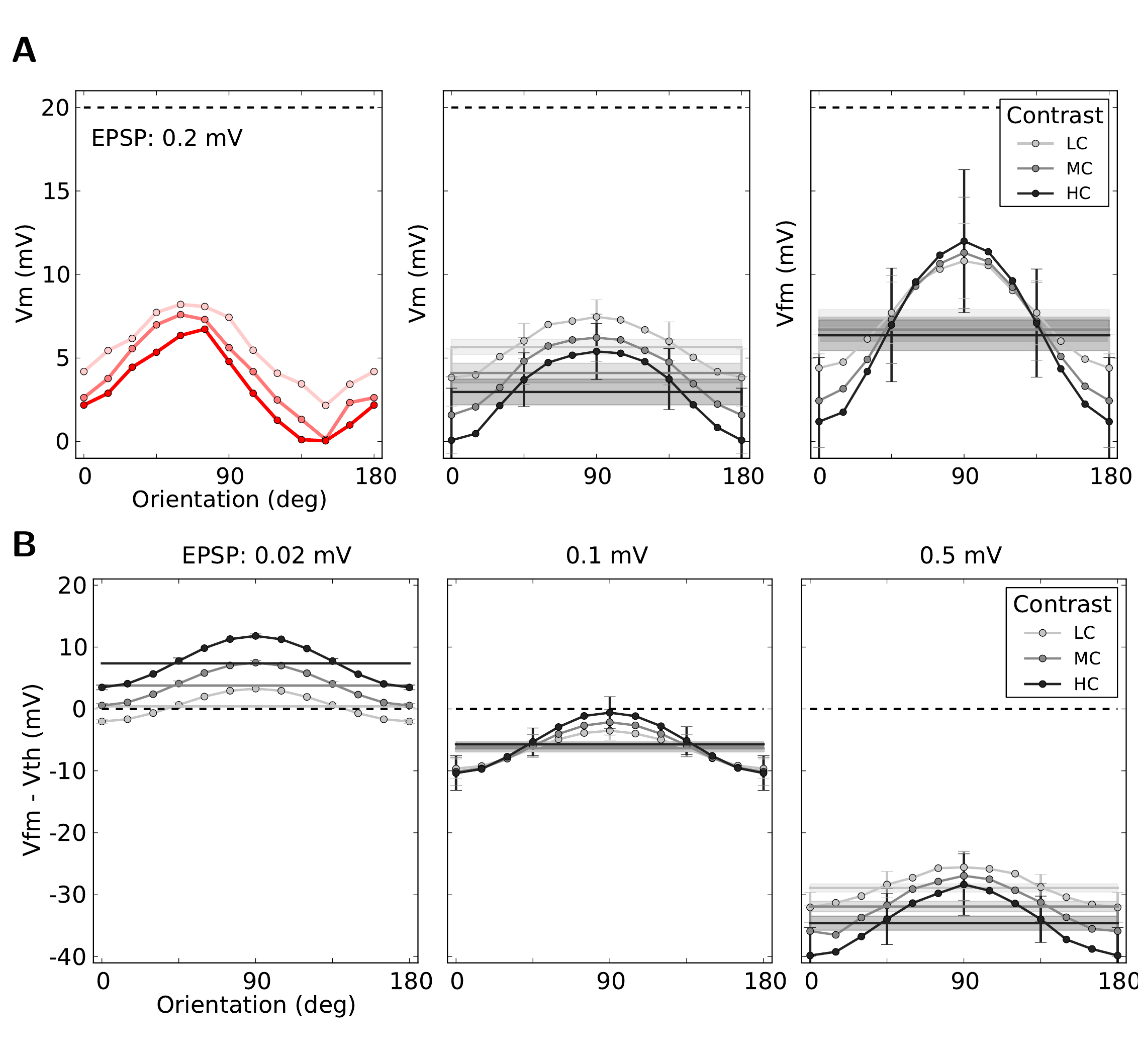} 
\caption{\textbf{Membrane potential at different contrasts.}    
    (\textbf{A})~Left: Tuning of the membrane potential of the same neuron as in Fig.~\ref{Fig_VmTrace} for different contrasts. 
    Center: Average tuning curve of $24$ ($12$ excitatory, $12$ inhibitory) randomly sampled neurons, ranging over all POs. 
    The tuning curves are aligned at a PO of $90\deg$.     
    Error bars indicate $\mathrm{mean} \pm \mathrm{std}$ over sampled neurons. 
    To improve the display, they are plotted only for every third data point. 
    The mean membrane potential (over all neurons and all orientations) is indicated by solid, horizontal lines in each case. 
    The shading represents the standard deviation (over neurons) of the mean (over orientations) of the average membrane potential ($v_B$). 
    Right: Same as the panel in the center, for the free membrane potential, $V_\mathrm{fm} = V + \tau r V_\mathrm{reset}$.     
    (\textbf{B})~The same $\mathrm{mean} \pm \mathrm{std}$ (over sampled neurons) of tuning curves for the distance to threshold of free membrane potentials ($V_\mathrm{fm} - V_\mathrm{th}$), for different regimes of recurrence. 
    For the lowest contrast, most of the error bars are smaller than the size of markers and hence not visible. 
    The mean membrane potential and its standard deviation over the (sampled) population is shown, as before, by horizontal lines and shadings, respectively.} 
    \label{Fig_VmTC} 

\end{figure}

Plotting the free membrane potential for this neuron (Fig.~\ref{Fig_VmTC}A, left), and more samples from the same network (Fig.~\ref{Fig_VmTC}A, center), verifies this observation. 
The free membrane potential (Fig.~\ref{Fig_VmTC}A, right) remains below threshold and has a similar tuning. Indeed, the tuning is slightly enhanced, since the negative contribution of the reset voltage is now corrected for.
This behavior is consistent across different contrasts. 
Increasing the contrast shifts the mean tuning curve as a whole down to more negative values, as a result of more negative feedback recruitment in the network. 
This, in turn, compensates for the increase in baseline, and suppresses the response to non-preferred orientations. 
The tuned part, however, is still capable of generating spikes, which leads to the tuning of spiking activity. 

%%%%%%%%%%%%%%%%%%%%%%%%%%%%%%%%%%%%%%%%%%%%%%%%%%%%%%%%%%%%%%%%%%%%%%%%%%%%%%%% 
%%%%%%%%%% Fig16 

%\begin{center} 
%\fbox{Fig.~16 approximately here} 
%\end{center} 
%%%%%%%%%%%%%%%%%%%%%%%%%%%%%%%%%%%%%%%%%%%%%%%%%%%%%%%%%%%%%%%%%%%%%%%%%%%%%%%% 

If the recurrent compensation was not effective, a different behavior would emerge. 
In fact, if the recurrent coupling is very weak (Fig.~\ref{Fig_VmTC}B, left), the free membrane potential is above threshold for almost all orientations. 
In such a case, the response to all orientations is in the mean-driven regime, which yields significant firing rates for the preferred as well as orthogonal orientations. 
As a result, the so-called iceberg effect broadens the tuning curves significantly upon increasing the contrast. 
Increasing the recurrent coupling shifts the mean membrane potential down and the network operates in the fluctuation driven regime; this makes the tuning of membrane potential and the resulting spiking activity more robust and contrast invariant (Fig.~\ref{Fig_VmTC}B, center and right panels). 
Indeed, for the intermediate recurrent regime (Fig.~\ref{Fig_VmTC}B, center) the tuning is perfectly contrast invariant. 

%%%%%%%%%%%%%%%%%%%%%%%%%%%%%%%%%%%%%%%%%%%%%%%%%%%%%%%%%%%%%%%%%%%%%%%%%%%%%%%% 
%%%%%%%%%%%%%%%%%%%%%%%%%%%%%%%%%%%%%%%%%%%%%%%%%%%%%%%%%%%%%%%%%%%%%%%%%%%%%%%% 
\subsection{Spiking activity in inhibition-dominated networks} 
\label{Sec_SAIDN} 

The recurrent excitation in inhibition-dominated networks of the sort we are studying here is over-compensated by the surplus recurrent inhibition. 
Some residual inhibition remains as the net effect of recurrent interactions. 
If the recurrent coupling is strong enough, this net inhibition determines the effective threshold of neurons in the network. 
Therefore, it is not the threshold mechanism of neurons which cuts off the responses at non-preferred orientations. 
Balance of excitation and inhibition within the network, in contrast, governs the generation of spikes and, hence, attenuation and amplification of the baseline and modulation components, respectively. 

%%%%%%%%%%%%%%%%%%%%%%%%%%%%%%%%%%%%%%%%%%%%%%%%%%%%%%%%%%%%%%%%%%%%%%%%%%%%%%%% 
%%%%%%%%%% Fig17 

\begin{figure}[h!] 
 \centering\includegraphics[width=4.5in]{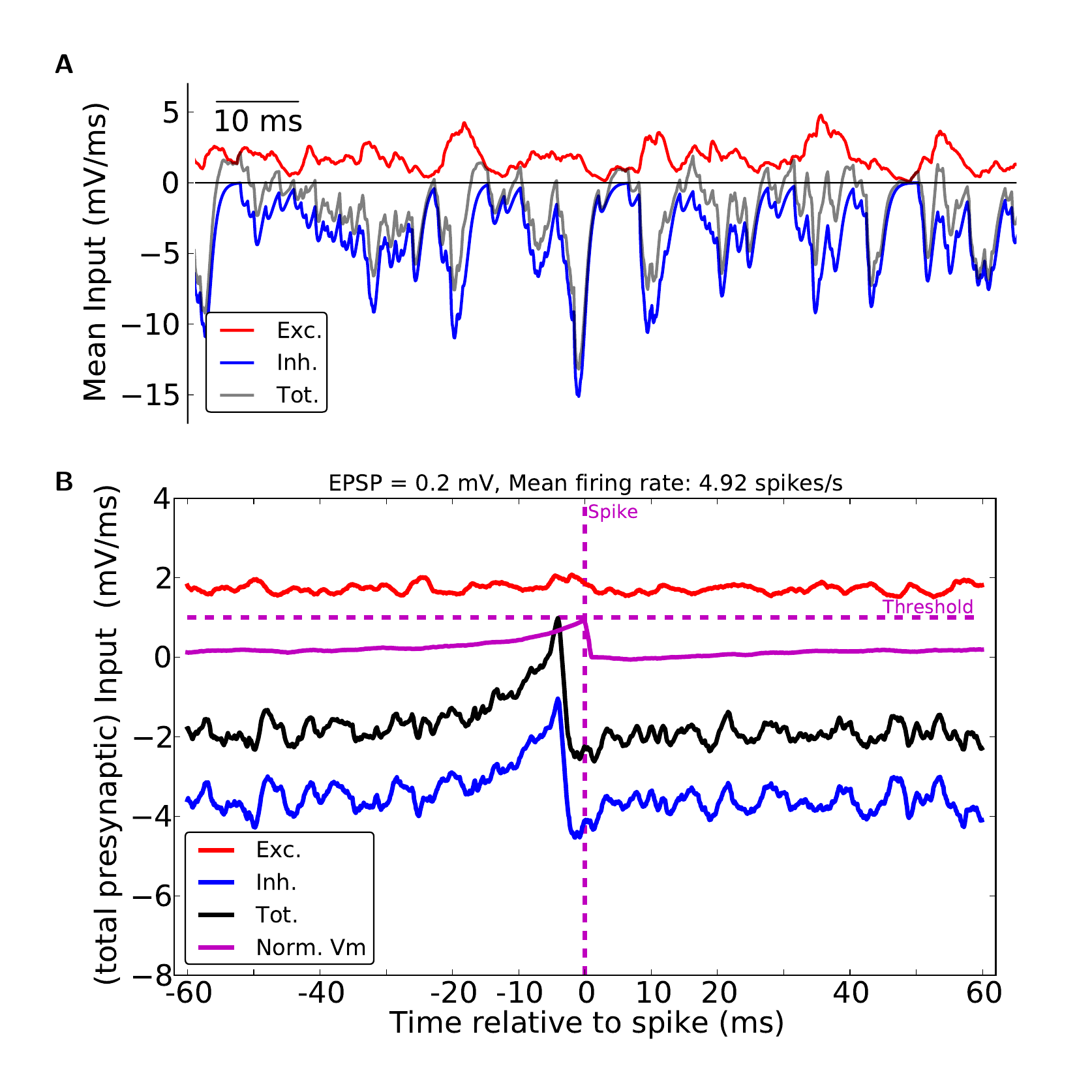} 
\caption{\textbf{Spiking activity in the inhibition dominated regime.}    
    (\textbf{A})~Statistical balance of overall excitatory (red) and inhibitory (blue) input from the recurrent network. The dominant inhibition keeps the net recurrent input (gray) negative on average, and only occasional transients lead for a moment to a net excitatory drive from the network. 
    $\mathrm{EPSP} = 0.2\,\mathrm{mV}$.    
    (\textbf{B})~Spike triggered averages (STA) of excitatory (Exc.) and inhibitory (Inh.) recurrent inputs are plotted from spikes of $12$ randomly sampled neurons in response to one stimulation ($6\,\mathrm{s}$). 
    The total recurrent input (Tot.) is plotted in black. 
    The membrane potential is normalized by $V_\mathrm{th}$ (Norm.~Vm.). 
    } 
    \label{Fig_STA} 

\end{figure} 

%\begin{center} 
%\fbox{Fig.~17 approximately here} 
%\end{center} 
%%%%%%%%%%%%%%%%%%%%%%%%%%%%%%%%%%%%%%%%%%%%%%%%%%%%%%%%%%%%%%%%%%%%%%%%%%%%%%%% 

A sample of this temporal balance is shown in Fig.~\ref{Fig_STA}A for the net excitatory and inhibitory inputs from the network to a neuron. 
Although the net excitation is on average above the firing threshold, the net inhibition is twice as large on average, as a result of the parameter configuration used. 
Moreover, occasional deflections in excitation and inhibition follow each other on a fine time scale. 
The net recurrent input to the neuron is on average negative. Occasional disbalance, however, provides a net excitatory drive for brief periods of time. 

Considering spike-triggered averages of the net excitatory and inhibitory input (Fig.~\ref{Fig_STA}B) reveals that spike emission in the network is mainly governed by recurrent inhibition, rather than recurrent excitation. 
Therefore, in the inhibition-dominated regime, fluctuations of inhibition is the main determining factor for spiking activity, in agreement with the results of experimental studies \cite{Rudolph2007}. 

%%%%%%%%%%%%%%%%%%%%%%%%%%%%%%%%%%%%%%%%%%%%%%%%%%%%%%%%%%%%%%%%%%%%%%%%%%%%%%%% 
%%%%%%%%%%%%%%%%%%%%%%%%%%%%%%%%%%%%%%%%%%%%%%%%%%%%%%%%%%%%%%%%%%%%%%%%%%%%%%%% 
\subsection{Modulation gain depends on the operating point of the network} 
\label{Sec_MGOPN} 

\begin{figure}[h!] 
\centering\includegraphics[width=6.0in]{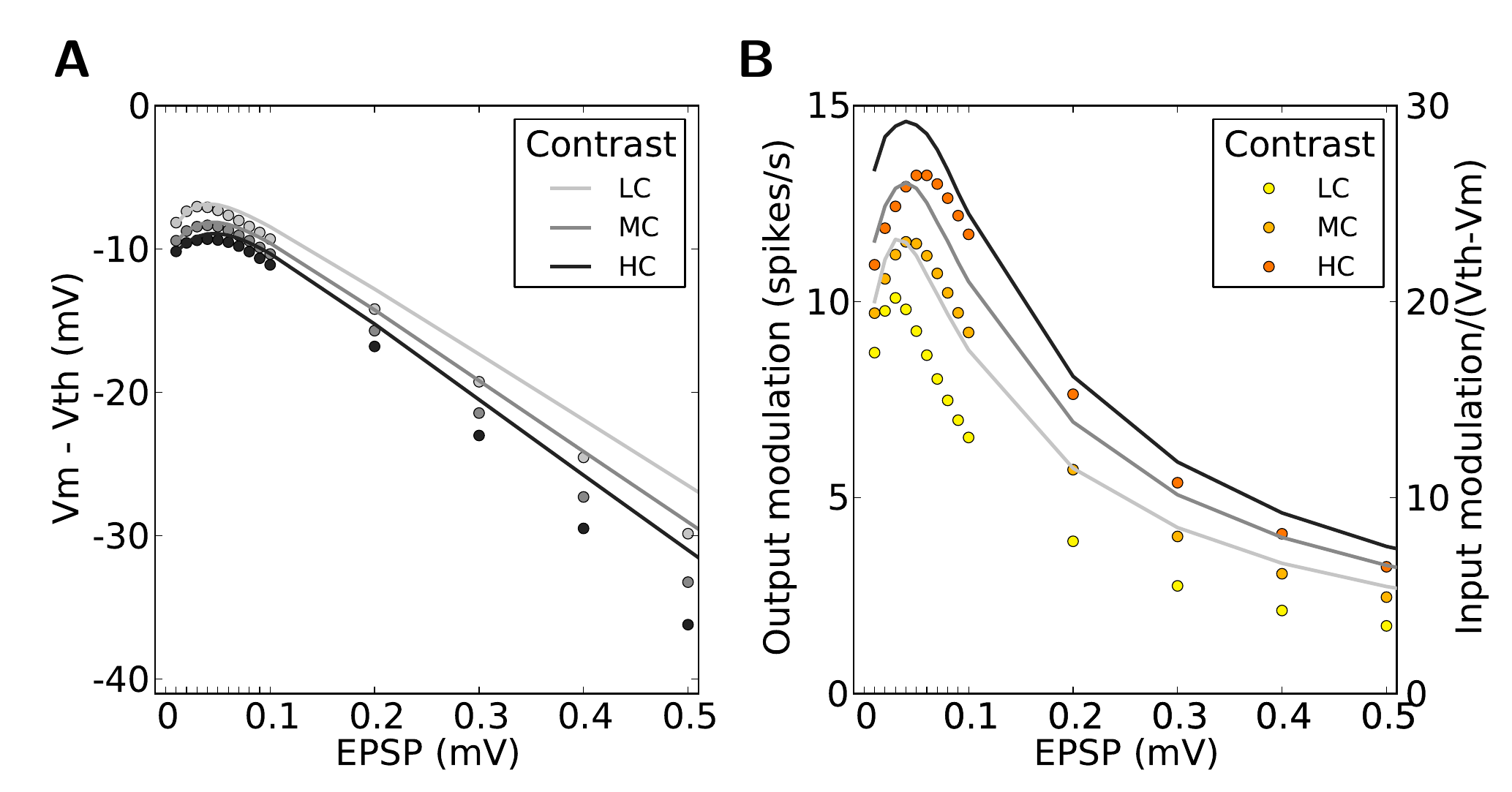} 
\caption{\textbf{Mean distance to threshold sets the operating point of the network.}    
    (\textbf{A})~The mean membrane potential for networks with different $\mathrm{EPSP}$ sizes (circles), at different contrasts, along with the predicted values (solid lines). 
    Mean membrane potential is computed as the average (over orientations) of mean tuning curves of $12$ sample neurons (the same as in Fig.~\ref{Fig_VmTC}A). 
    Lighter colors belong to lower contrasts, respectively. 
    %The discrepancy for higher recurrences is due to neglecting correlations.     
    (\textbf{B})~Input modulation normalized by the distance to threshold, $\mathrm{Vth} - \mathrm{Vm}$, (solid lines) compared to the output modulation (orange circles; same as Fig.~\ref{Fig_BasModPred}B). 
    Input modulation is given as the input modulation rate times its efficacy ($J_s m s_B$). } 
    \label{Fig_Vm_ModGain} 

\end{figure}

The strength of net inhibitory feedback also affects the selective gains. 
Whereas the mean inhibitory recurrent input sets the divisive gain in the baseline pathway (Fig.~\ref{Fig_SepPath}B), it affects the modulation gain by determining the mean distance of the membrane potential to threshold. 
As suggested by Fig.~\ref{Fig_VmTC}B, increasing the level of recurrence induces larger average distances of the mean membrane potential from threshold. 
This is shown for all networks in Fig.~\ref{Fig_Vm_ModGain}A. 

Using the simplified mean-field analysis provided in Section \ref{Sec_MFA}, we can predict the mean membrane potential of a network. 

Knowing the baseline firing rate of the network, $r_B$, the mean baseline membrane potential of the network, $v_B$, is obtained (see Eq.~\eqref{Eq_StatRate} in Methods) as 
\[ 
v_B = \tau \bigl[ -V_\mathrm{th} r_B + \mu_B \bigr], 
\label{Eq_MemPot} 
\] 
where $\mu_B = J N \epsilon (f - g(1-f)) r_B + J_s s_B$ (see Methods, Eq.~\eqref{Eq_inputMS}). 
Note that, as the input is shunted during the refractory period, the shunted feedforward and recurrent input should be subtracted from the total input in this expression. 
The shunted input can be computed as $r_B t_\mathrm{ref} \mu_B$ \cite{Murthy1994, Kuhn2003}. 
The corrected membrane potential, after considering the effect of refractory period, is then given as 
\[ 
v_B = \tau \bigl[ -V_\mathrm{th} r_B + (1-r_B t_\mathrm{ref}) \mu_B \bigr]. 
\label{Eq_MemPot_RefPer} 
\] 
The result of this prediction for different contrasts is plotted in solid lines in Fig.~\ref{Fig_Vm_ModGain}A.

%%%%%%%%%%%%%%%%%%%%%%%%%%%%%%%%%%%%%%%%%%%%%%%%%%%%%%%%%%%%%%%%%%%%%%%%%%%%%%%% 
%%%%%%%%%% Fig18 

%\begin{center} 
%\fbox{Fig.~18 approximately here} 
%\end{center} 
%%%%%%%%%%%%%%%%%%%%%%%%%%%%%%%%%%%%%%%%%%%%%%%%%%%%%%%%%%%%%%%%%%%%%%%%%%%%%%%% 

Note that this prediction is obtained under the assumption of a Gaussian distribution of input to all neurons. 
The prediction, therefore, fails to be exact if this assumption is violated. 
The distribution is, in fact, skewed as a result of correlations in the network \cite{Kuhn2003}.
The deviation from a Gaussian distribution increases for higher recurrences, explaining the discrepancy of our predictions for highly recurrent regimes. 

The mean membrane potential is crucial in determining the overall gain of the linearized dynamics. 
A very depolarized membrane potential reduces the chance of an input perturbation to reach the firing threshold and to elicit a spike. 
Therefore, it affects the gain of modulation, as shown in Fig.~\ref{Fig_Vm_ModGain}B: 
The mean output modulation of tuning curves, $\gamma_M$, is indeed inversely related to the mean distance to threshold. 
This suggests that the embedded gain of neurons in the network in response to modulation is also inversely proportional to the distance to threshold. 
Although the mean modulation in the input is below the spike threshold of a single neuron, fluctuations are nevertheless capable of eliciting reasonable firing rates. 
The resultant linearization of the $f$-$I$ curve, as shown in Fig.~\ref{Fig_LinTun}A, is a result of the noise, $\sigma_B$, generated within the network due to the balance of excitation and inhibition. 

%%%%%%%%%%%%%%%%%%%%%%%%%%%%%%%%%%%%%%%%%%%%%%%%%%%%%%%%%%%%%%%%%%%%%%%%%%%%%%%% 
%%%%%%%%%% Fig19 

\begin{figure}[h!] 
   \centering\includegraphics[width=4.5in]{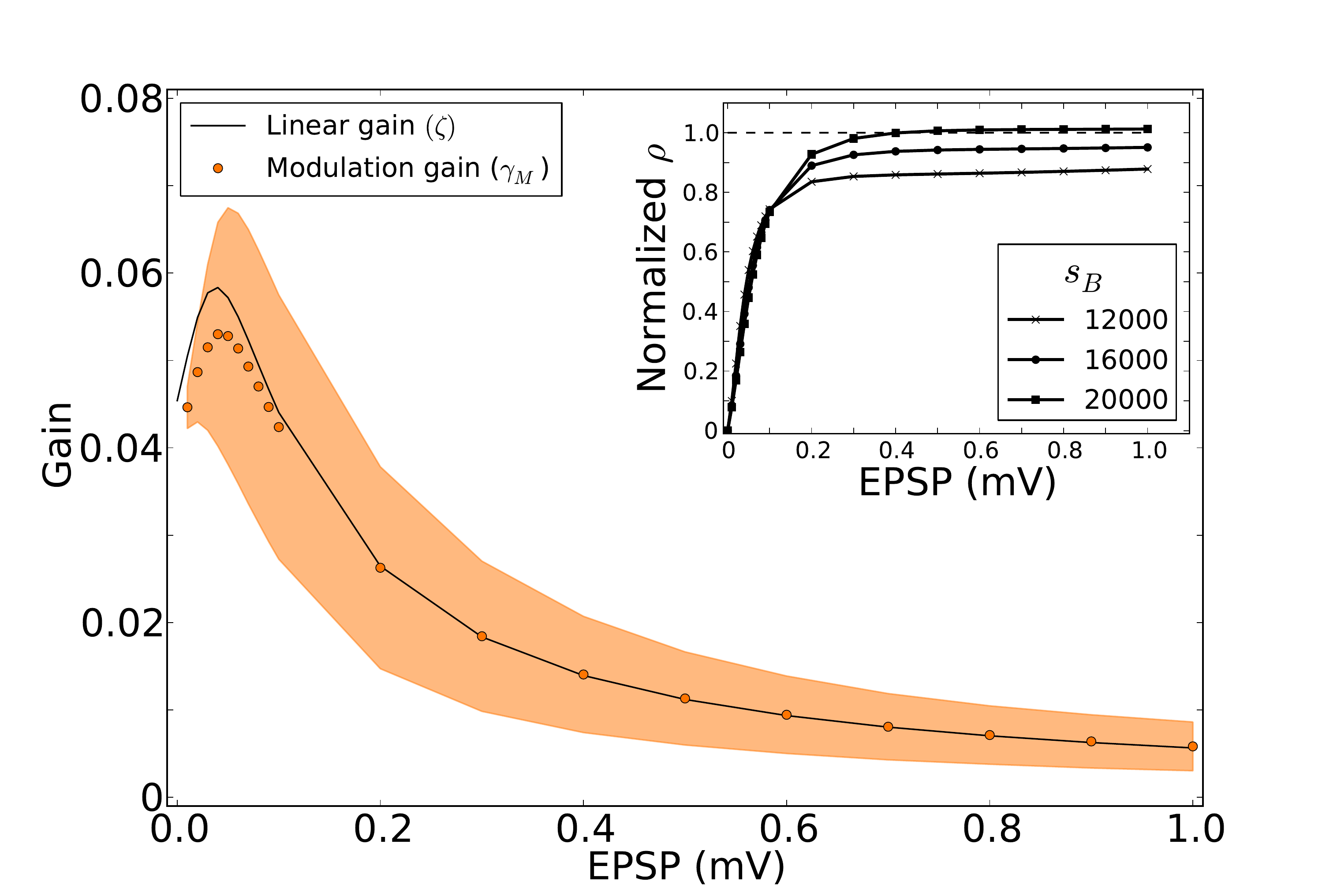} 
\caption{\textbf{Linear gains determine the gain and stability of networks in response to modulation.} 
    For the medium contrast, linear gain ($\zeta$, black line) is computed at each $\mathrm{EPSP}$ and is compared with the corresponding modulation gain, $\gamma_M = \text{Output~modulation}/\text{Input~modulation}$ (orange). 
    $\zeta$ is computed as $\zeta = \delta r/(J_s \delta s)$, for a small perturbation of the input, $\delta s = 100\,\mathrm{spikes}/\mathrm{s}$. 
    Inset: The radius, $\rho$, of bulk spectrum of $W$, normalized by the linear gain ($\zeta$) at each $\mathrm{EPSP}$. 
    Instead of dividing $J$ by $V_\mathrm{th}$ (as in Fig.~\ref{Fig_SepPath}C), $J$ is now multiplied by $\zeta$. 
    As a result, the normalized radius is now obtained as $\rho^\mathrm{norm} = \zeta J \sqrt{N \epsilon (1-\epsilon) \bigl[ f + g^2 (1-f) \bigr]}$.} 
    \label{Fig_LinGain_StabSpec} 
    
\end{figure} 

%\begin{center} 
%\fbox{Fig.~19 approximately here} 
%\end{center} 
%%%%%%%%%%%%%%%%%%%%%%%%%%%%%%%%%%%%%%%%%%%%%%%%%%%%%%%%%%%%%%%%%%%%%%%%%%%%%%%%

If we now compute these gains for networks with different degrees of recurrence, the linearized gains match well with the mean modulation gain in the network (Fig.~\ref{Fig_LinGain_StabSpec}, for the middle contrast). 
This suggests that the mean amplification of modulation in the network is fully accounted for by the linear gain, which in turn depends on the operating point of the network 
as defined by the mean, $\mu_B$, and the standard deviation, $\sigma_B$, of the input to neurons from the network in the baseline state. 
The linear gains are obtained by perturbing $r = f(\mu, \sigma)$ at this operating point 
\[ 
\delta r = \frac{\partial f(\mu, \sigma)}{\partial \mu} \delta \mu + 
\frac{\partial f(\mu, \sigma)}{\partial \sigma} \delta \sigma, 
\] 
and using $J_s \zeta = \delta r / \delta s$. 
Here, we determine this quantity numerically as 
\[ 
\delta s J_s \zeta = f(\mu(s+\delta s), \sigma(s+\delta s)) - f(\mu(s), \sigma(s)) 
\label{Eq_zeta_num}
\] 
for $\delta s = 100\,\mathrm{spikes}/\mathrm{s}$ (Fig.~\ref{Fig_LinGain_StabSpec}). 

In summary, the inhibitory feedback in a recurrent network contributes to orientation selectivity in crucial ways. 
First, it provides a negative feedback which offsets the baseline component of the input tuning curves and leads to divisive attenuation of the common mode (selective attenuation in the baseline pathway). 
Second, it sets the operating point of the network and determines the linearized, embedded gain, which in turn determines the modulation gain (selective amplification in the modulation pathway). 
Moreover, the feature selectivity generated by the recurrent network as a result of summing many inputs of random selectivity leads to either amplification, or attenuation, of the feedforward tuning (random summation of recurrent tuning). 
Since the contribution of each presynaptic modulation vector must be weighted according to the linearized gain, the bulk spectrum of the connectivity matrix $W$ must also to be weighted accordingly. 
The spectrum of the network with $\mathrm{EPSP} = 0.2\,\mathrm{mV}$ shown in Fig.~\ref{Fig_InstabDyn}A, for instance, was obtained by normalizing $J$ by $V_\mathrm{th}$. 
The linear gain suggests now that $J$ should be multiplied by $\zeta = 0.026$, which is a factor $2$ smaller than $1/V_\mathrm{th} = 0.05$. 
This leads to a decrease in the radius of the bulk from $\rho = J/V_\mathrm{th} \sqrt{N \epsilon (1-\epsilon) \bigl[ f + g^2 (1-f) \bigr]} \approx 1.68$ to $\rho = \zeta J \sqrt{N \epsilon (1-\epsilon) \bigl[ f + g^2 (1-f) \bigr]} \approx 0.87$. 
This implies that none of the modulation modes corresponding to the bulk are actually unstable, at this operating point of the network. 
Indeed, if we plot the normalized radius, $\rho^\mathrm{norm}$, for all the networks at different contrasts, the $\rho^\mathrm{norm}$ never exceeds one (Fig.~\ref{Fig_LinGain_StabSpec}, inset). 
This means that, although the coupling strength is monotonically increasing, the network dynamics stabilizes the spectrum in inhibition-dominated networks (see \cite{Pernice2012} for related observations).

%%%%%%%%%%%%%%%%%%%%%%%%%%%%%%%%%%%%%%%%%%%%%%%%%%%%%%%%%%%%%%%%%%%%%%%%%%%%%%%% 
%%%%%%%%%%%%%%%%%%%%%%%%%%%%%% Discussion 
%%%%%%%%%%%%%%%%%%%%%%%%%%%%%%%%%%%%%%%%%%%%%%%%%%%%%%%%%%%%%%%%%%%%%%%%%%%%%%%% 

\section{Discussion}

Using large-scale simulations and associated mean-field analysis of networks of spiking neurons, we have demonstrated how highly selective neuronal responses can be obtained in random networks without any spatial or feature specific structure.
Our mathematical analysis pinpoints the mechanisms responsible for selective attenuation of the common mode and selective amplification of modulation, and predicts some essential properties of these networks.

\subsection{A generic model of local circuitry}

Here we discussed the specific case of orientation selectivity in the early visual system, as we were able to link our findings to an ample body of experimental literature. However, our model could potentially be of a much broader scope.
It proposes a general mechanism for the emergence of strong feature selectivity, which could actually be at work in other sensory modalities as well.
Our network model can thus be conceived as a generic model for the local cortical circuitry, which enhances feature selectivity and ensures contrast invariance of processing, without resorting to feature specific structure or experience-dependent fine tuning.

Our analysis suggests that a randomly connected network with dominant inhibition is already capable of selectively removing the uninformative common mode of a stimulus that is represented by the network in a distributed fashion, while preserving the informative modulations in the response pattern induced by stimulation.
This way, the tuned part gains salience, and the signal-to-noise ratio improves.
The network also amplifies the tuned component (signal) by two mechanisms:
First, by modulating the embedded gain through adjusting the operating point of the network, and second, by recurrently mixing presynaptic selectivities and thereby amplifying weakly tuned inputs in some neurons.

\subsection{Regimes of orientation selectivity}

The same mechanisms could also lead to attenuation of the signal in the network.
First, increasing the recurrent coupling in the network increases the mean distance of the membrane potential to the firing threshold, which in turn decreases the modulation gain in the network.
Second, the recurrent mixing of weak tunings in the network generates a distribution of selectivity, with attenuation in many neurons.
As a result, this mechanism cannot increase the selectivity of a certain fraction of the neuronal population beyond the input selectivity, unless this is compromised by a decrease in the selectivity of another fraction of the population.

In addition, there is a trade-off between selectivity and contrast invariance of the neuronal responses.
Increasing the degree of recurrence in the network makes the selectivity more invariant and the distribution of it more robust with regard to variations of contrast, but it decreases the overall gain of modulation.
As a result, there is a region of intermediate recurrence in our networks, where tuning amplification is most pronounced, and orientation selectivity has the largest value with the lowest sensitivity to contrast.

Our computational study, therefore, suggests an intermediate regime of recurrence as the optimal regime of feature selectivity for early sensory processing.
This is the state of the network which exhibits the stimulus driven properties of neurons observed in experiments \cite{Hofer2011}, while preserving other important features like strong and contrast invariant orientation tuning.
In this regime, the feature selectivity of neurons would exhibit the least deviation from their input selectivities, essentially reflecting the tuning of the feedforward input.
The role of the recurrent network at this stage would then be to enhance this selectivity, by performing operations like increasing the signal-to-noise ratio  and contrast invariant gain control.

This scenario might best explain the state of the input layer L4, in which orientation selectivity first emerges in the cortex. The same is not necessarily true for more recurrent layers like L2/3 or L5, which are involved in later stages of sensory processing like learning, association and motor control.
It is, therefore, plausible that different regimes of recurrence exist in different layers, which may be suited to perform different types of processing.
One measure for the degree of recurrence in a network is the tuning of the total recurrent input.
As the recurrent coupling increases in a network, the tuning of the recurrent input generated within the network increases as well, and the assumption of untuned total input becomes a questionable approximation.

There are indeed contradictory results reported in experimental studies on the tuning of input in rodents:
Both untuned inhibition \cite{Li2012a, Li2012b, Liu2011} and co-tuning of excitation and inhibition \cite{Tan2011} have been reported by different laboratories.
Our results show that even in absence of significant tuning of the total input received from the recurrent network, another mechanism of selective attenuation and amplification can lead to strong selectivity and contrast invariance.
This, however, does not exclude a random summation of selectivity within the recurrent network as a contributing mechanism.
Indeed, in the first example network we investigated here, both mechanisms were at work.
It is, therefore, plausible that both tuned and untuned components exist to some degree in such networks, but the exact mixing depends on the operating point and, specifically, on the degree of recurrence.
This would therefore suggest that in more recurrent layers like L2/3 more neurons with strong input tuning should be recorded, while in the input layer L4 untuned inputs preponderate.\footnote{One should also consider the possible effect of feature-specific connectivity \cite{ko2011, ko2013} on this behavior, which predicts a preponderance of connections between neurons with similar selectivity and hence a more tuned input to neurons.}

It should be noted, however, that our discussion here applies to recurrent excitation and inhibition.
Tuned excitation and inhibition, when measured in terms of the total excitatory and inhibitory conductances in intracellular recordings, are the total excitatory and inhibitory input that a neuron observes.
It is therefore possible that the tuning of the feedforward input is dominant in the tuning.
Likewise, feedforward inhibition, mediated by disynaptic inhibition, can have the same tuning as the feedforward excitatory input, as the former is mediating it.
The mechanisms discussed here, however, apply to recurrent excitation and inhibition, since they are a consequence of the dynamics of a network of synaptically connected neurons and, in particular, recruitment of feedback inhibition within the network.

\subsection{Recurrent vs.\ feedforward inhibition}

Recurrent inhibition in our networks selectively feeds the mean signal back and subtracts it from the tuning curves. The overall effect of this subtraction results in a divisive attenuation of the baseline.
This untuned suppression has been experimentally demonstrated to play a crucial role in increasing the selectivity \cite{Shapley2003, Xing2011}.
More specifically, it has been recently shown that in L4 of mouse visual cortex, it underlies sharpening of orientation selectivity \cite{Li2012a, Li2012b}.
This is also consistent with the results of a recent study on the role of somatostatin expressing, SOM$^+$, GABAergic neurons in orientation selectivity \cite{Wilson2012} (but see \cite{Lee2012}).
The subtraction of the baseline, attributed to this specific subtype of inhibition by \cite{Wilson2012}, effectively leads to the selective attenuation/division of the baseline, as described in the present article.
Reduction of this inhibition would therefore lead to a constant increase in the baseline activity of the tuning curves, which has indeed been recently reported in Dlx1(-/-) mice with selective reduction of activity in dendrite targeting inhibitory interneurons \cite{Mao2012}.

This mechanism is in contrast to the role of feedforward inhibition of fast spiking interneurons, parvalbumin expressing, PV$^+$, GABAergic neurons.
As opposed to SOM$^+$ neurons, which are more recurrent, these neurons are mainly driven by feedforward input \cite{Adesnik2012}.
Unlike SOM$^+$ neurons, which are involved in dendritic computation and controlling the input, PV$^+$ neurons are better suited for controlling the output, as they innervate the peri-somatic regions \cite{DiCristo2004, Ma2010}.
Consistently, they are also more effective during the transient responses, as reflected in their activation pattern \cite{Tan2008}.
In contrast, SOM$^+$ neurons are better suited for sustained activity.

These properties might then suggest that feedforward inhibition is primarily involved in gain control \cite{Atallah2012}, which uniformly rescales all components of the tuning curves.
The attenuation of the baseline, therefore, comes at the expense of attenuating the modulation.
It cannot be selective to the baseline, in contrast to the recurrent mechanism we have modeled here.
Note that, for simplicity, we have not considered feedforward inhibition in our model.
However, feedforward inhibition could easily be added to the model by mediating the same excitatory input to each neuron by an inhibitory neuron.
Doing so would effectively lead to a change in the overall feedforward gain, provided that inhibition is not strong enough to result in rectification (compare with \cite{Lee2012}).

If PV$^+$ neurons are strongly driven by feedforward input, and if the feedforward input is only slightly tuned, as we assumed here, the responses of PV$^+$ neurons should be only weakly modulated.
In contrast, as the results of our simulation showed here, inhibitory neurons involved in recurrent computations can be highly selective, although they receive weakly modulated inputs.
In agreement with this, SOM$^+$ inhibitory neurons involved in recurrent inhibition have been reported to be as selective as excitatory neurons, in contrast to PV$^+$ neurons with broader selectivity \cite{Ma2010}.
However, it should be possible to make PV$^+$ interneurons more selective, by providing more recurrent inhibition to them.
In fact, it has recently been reported that activating SOM$^+$ inhibitory neurons can unmask and enhance the selectivity of PV$^+$ cells by suppressing untuned input \cite{Cottam2013}.

Note that, as contrast invariance depends on the selective attenuation of the baseline, it should be the result of a recurrent mechanism.
We therefore suggest that the constant increase of untuned inhibition that neurons receive upon increasing the contrast \cite{Li2012b} should be a result of the recurrent, and not of the feedforward, inhibition.
This may explain why dark reared mice in this experiment, which lacked a broadening of PV$^+$ responses, still show an aggregate untuned input from the network and hence highly selective responses \cite{Li2012b}.
Although individual inhibitory neurons were on average highly selective in our networks, the emergent result of the interaction of excitation and inhibition lead to an effective untuned inhibition, which increases proportionally with contrast.
This is again consistent with the results of \cite{Li2012b}, who demonstrated that ``blocking the broadening of output responses of individual inhibitory neurons does not block the broadening of the aggregate inhibitory input to excitatory neurons''.
It is also consistent with the results of a previous report, demonstrating that ``broad inhibitory tuning'' of fast spiking cells is ``not required for developmental sharpening of excitatory tuning'' \cite{Kuhlman2011}.
Based on these results, we therefore hypothesize that untuned inhibition might be an emergent property of an inhibition dominated network, and not a feedforward consequence of broadly tuned fast spiking neurons.

\subsection{Comparison with other models}

Most existing recurrent theories of orientation selectivity consider the case of species like carnivores and primates, with a clustered organization of selectivity in orientation maps \cite{Ben-Yishai1995, Somers1995}.
Consistent with the proximity of neurons with similar selectivity, these theories assume a feature specific connectivity of neurons.
The Mexican hat profile of connectivity which they assume leads to a more broadly tuned inhibition, which suppresses the mean, and to a sharper tuning of excitatory input, which amplifies the modulation.
Therefore, these models cannot be applied to the case of a salt-and-pepper structure, as found in rodents, with no apparent spatial or feature specific connectivity.

Even in species with orientation maps, there seem to be some issues with these models.
First, they rely on -- and predict -- a sharpening mechanism of selectivity due to tuned recurrent excitation.
However, the late (presumably recurrent) sharpening of selectivity, which these theories predict, has not been observed in experiments \cite{Sharon2002, Gillespie2001}.
Also, the orientation selectivity of neurons seem to be the same as their feedforward input, since the preferred orientation of neurons does not change with recurrent interactions \cite{Gillespie2001}.
Rather than sharpening of tuning curves, a more plausible function of the recurrent network is increasing the modulation ratio, by suppressing the baseline \cite{Sharon2002}.

This was indeed the main mechanism of orientation selectivity we described in our networks here.
As it is based on essentially linear processing, our model predicts no sharpening of the tuning as a result of recurrent interactions, but only an increase in modulation depth, not affecting the tuning width \cite{Sharon2002}.\footnote{Tuning width is unchanged if the baseline activity is removed, and half-width at half-height (HWHH) is computed from the Gaussian fit to the tuned part, as done in \cite{Sharon2002}. Tuning amplification would effectively decrease the tuning width, however, if the baseline is also taken into consideratio (as in \cite{Wilson2012}).}
Sharpening of tuning curves would only be a consequence of the feedforward nonlinearity, reflected in a nonlinear transfer function of neurons.
As our results do not depend on the power-law transfer function of single neurons, our model would also work if the operating regime of the cortex suggested a smaller exponent of the power-law \cite{Xing2011}.\footnote{Note that we do not exclude the presence of nonlinear mechanisms in the biological cortex and their contribution to orientation selectivity. The nonlinearity of the neuronal transfer function, as well as other nonlinear mechanisms like nonlinear dendritic amplification \cite{Jia2010, Lee2012a, Lavzin2012} or synchronization of thalamic inputs \cite{Stanley2012, Bruno2006}, may contribute in addition to obtain higher selectivity.}
Moreover, as we demonstrated above, this mechanism does not have to be accompanied, on average, by a large shift between the input and output preferred orientations.

Another consequence of the sharpening theories is the emergence of strong pairwise correlations in the network \cite{Series2004}.
This seems not to be consistent with the very low correlations reported in the neocortex \cite{Ecker2010}.
More specifically, it has recently been shown that highly selective neurons in the input layer of monkey V1 exhibit very low noise correlations \cite{Hansen2012}.
This imposes an important constraint on recurrent models which need sharper tuning of excitatory input to the neurons as compared to inhibition, as this sharper tuning leads to higher noise correlations in the local network (see Figure 5 in \cite{Hansen2012}).
Hence, it might be difficult for these models to simultaneously account for both sharp orientation selectivity and low pairwise correlations in the input layers.

The mechanism we discussed here, in contrast, does not need -- and not predict -- strong pairwise correlations in the network.
In fact, our mean field analysis was even based on the assumption of no correlations in the network.
As the network operates in the AI state, the amount of linear read-out of information from our networks would therefore be several times higher than in sharpening theories, comparable to feedforward models \cite{Series2004}.

In comparison to feedforward models, however, our analysis suggests that contrast invariant tuning of both membrane potentials and spiking activity \cite{Anderson2000} can robustly and reliably emerge through the action of a recurrent network.
Contrast invariance is a critical property of feature selectivity, which ensures reliable and consistent feature detection for a wide range of different stimuli.
Without a network mechanism of the sort described here, neurons would need a specific fine-tuning for each contrast, in order to be selective for the same feature.
The network mechanism proposed here provides a generic mechanism to dynamically achieve contrast invariance, without the need for feature specific wiring, special correlation structure, power-law transfer function, contrast-dependent variability, shunting inhibition, synaptic depression, adaptation or learning.
However, it remains to be experimentally verified whether intra-cortical recurrent connectivity is indeed necessary for contrast invariance, or whether feedforward mechanisms are enough to account for this phenomenon \cite{Finn2007, Sadagopan2012, Priebe2012}.
A crucial experiment would be to test whether the tuning of the membrane potential is still contrast invariant if lateral interactions in the cortex are deactivated \cite{Kara2002, Chung1998, Ferster1996}.

\subsection{Future studies}

There are, however, several issues which need to be further examined in future works.

\subsubsection{Extending the scope of linear analysis}
First, our analysis is mainly provided to compute the mean values of baseline and modulation gains in the network.
It is therefore necessary to extend the analysis such that it accounts for the distribution of these gains.
Also, the model assumes cosine tuning of inputs (linear tuning), and linear network operation (e.g.\ no rectification in the tuning curves).
It would therefore be revealing to see if, and to which degree, the results of the linear analysis hold in the presence of nonlinearities reported to exist in the biological cortex \cite{Anderson2000}.

\subsubsection{Neuron model}
We used current based LIF neurons with unconstrained membrane potentials in our simulations, since this gave us the opportunity to perform a theoretical analysis of network dynamics.
It is however necessary to test to which degree the results of our study change by recruiting a different neuron model.
For instance, using an alternative neuron model like concuctance-based LIF neurons might not allow the distance to threshold of the membrane potential to increase unboundedly.
This was not the case here, as we did not impose any minimum bounday condition on our current-based LIF neurons.

Such a difference may change the effective gain of neurons and, as a result, a different eigenvalue spectrum might be obtained.
This, in turn, may change network dynamics and lead to a qualitatively different behavior of the network.
It might also have consequences for the structure of correlations in the network, and may affect the AI state.
Our preliminary results indicate that imposing a lower boundary condition on current based LIF neurons can amplify correlations and synchrony in the network, to the extent that the network does not operate anymore in the AI state.
Feature selectivity is still obtained and even enhanced in the network.
Tuning curves show a higher average OSI and maximum firing rates, more rectification and reduced tuning widths follow from this, while contrast invariance is preserved (not shown).
Such a scenario should be analyzed in more detail in future studies.

\subsubsection{Network connectivity}
It is also important to study the effect of different connectivity patterns in the network.
Here, we have modeled the dominance of inhibition by increasing the relative strength of IPSPs.
An alternative implementation is to increase the density of inhibitory axonal projections, which increases the density of connectivity.
Dense pattern of connectivity has been reported for inhibition in different cortices \cite{Fino2011, Packer2011, Hofer2011}, and seems to be a common motif.
Such a change in network connectivity might have consequences for sensory processing.
For instance, a decrease in temporal fluctuations of the local inhibitory population and, likewise, in the quenched noise of preferred orientations is implied, which can, in turn, affect the tuning of inputs and amplification or attenuation of orientation selectivity.

Also, the model and the analysis provided here should be extended to account for networks with spatial structure.
It should be analytically studied how distance dependent connectivity, and in particular different connectivity profiles for excitation and inhibition, affect the results obtained here.
It has been shown, for instance, that balanced networks can show topologically invariant statistics \cite{Yger2011}.
It would therefore be interesting to see if the same analysis also applies to functional properties of these networks.
Of special interest would be to study how the spectrum of the network changes \cite{Voges2011}, and to which degree this predicts the operation of the network,
in particular, how the spatial extent of excitation and inhibition affects this behavior.
Experimentally, it has been reported that inhibition is more local than excitation in terms of anatomical projections. Many theoretical models, however, assume broader inhibition for convenience.
Studying the functional properties of networks with realistic patterns of connectivity might therefore shed light on this aspect.

\subsubsection{Orientation selectivity and orientation maps}
As we were primarily interested in the emergence of orientation selectivity in species without orientation maps, we studied here random networks with salt-and-pepper structure.
However, the model could be extended to networks with spatial or functional maps, which imply feature specific connections.
As opposed to the Mexican hat profile assumed in the ring model, if one now assumes a more realistic pattern of local inhibition and longer range excitation, different dynamic properties might follow \cite{McLaughlin2000, Hansen2012, Pernice2011}.
The analysis of the new regime of orientation selectivity therefore calls for a further study.
The results of this modeling would in turn help identifying the basic mechanisms that are responsible for the emergence of orientation selectivity in different species with different structures, and to eventually provide an answer to the question whether common design principles exist, or whether different strategies have been recruited by different species.

%%%%%%%%%%%%%%%%%%%%%%%%%%%%%%%%%%%%%%%%%%%%%%%%%%%%%%%%%%%%%%%%%%%%%%%%%%%%%%%% 
%%%%%%%%%%%%%%%%%%%%%%%%%%%%%% Methods 
%%%%%%%%%%%%%%%%%%%%%%%%%%%%%%%%%%%%%%%%%%%%%%%%%%%%%%%%%%%%%%%%%%%%%%%%%%%%%%%% 

% You may title this section "Methods" or "Models". 
% "Models" is not a valid title for PLoS ONE authors. However, PLoS ONE
% authors may use "Analysis" 
\section{Methods}

\label{Sec_Methods} 

%%%%%%%%%%%%%%%%%%%%%%%%%%%%%%%%%%%%%%%%%%%%%%%%%%%%%%%%%%%%%%%%%%%%%%%%%%%%%%%% 
%%%%%%%%%%%%%%%%%%%%%%%%%%%%%%%%%%%%%%%%%%%%%%%%%%%%%%%%%%%%%%%%%%%%%%%%%%%%%%%% 
\subsection{Simulation and analysis of network dynamics} 
\label{Sec_SAND} 

%%%%%%%%%%%%%%%%%%%%%%%%%%%%%%%%%%%%%%%%%%%%%%%%%%%%%%%%%%%%%%%%%%%%%%%%%%%%%%%% 
\subsubsection{Neuron model and surrogate spike trains} 
\label{NMSST} 

We studied networks of leaky integrate-and-fire (LIF) neurons.  
For this spiking neuron model, the sub-threshold dynamics of the membrane potential $V_i(t)$ of neuron $i$ is described by the leaky-integrator equation 
\[\tau \dot{V}_i(t) + V_i(t) = RI_i(t).	 
  \label{Eq_NeuronModel} 
\] 
The current $I_i(t)$ represents the total input to the neuron, the integration of which is governed by the leak resistance $R$, and the membrane time constant $\tau = 20\,\mathrm{ms}$.  
When the voltage reaches the threshold at $V_\mathrm{th} = 20\,\mathrm{mV}$, a spike is generated and transmitted to all postsynaptic neurons, and the membrane potential is reset to the resting potential at $V_0 = 0\,\mathrm{mV}$.  
It remains at this level for short absolute refractory period, $t_\mathrm{ref} = 2\,\mathrm{ms}$, during which all synaptic currents are shunted. 

To simulate spiking inputs to neurons from outside the network (e.g.~from the lateral geniculate nucleus, LGN), we resorted to the conceptually simpler model of a Poisson process. 
The associated surrogate spike trains have the property that spikes are generated randomly and independently with a prescribed firing rate at each point in time. 
The linear superposition of an arbitrary number of Poisson processes (as in the case of multiple afferents) is again a Poisson process. 
The rate of the superposition process is exactly the linear sum of the rates of its components, and it can be effectively simulated as a single process with high rate. 

%%%%%%%%%%%%%%%%%%%%%%%%%%%%%%%%%%%%%%%%%%%%%%%%%%%%%%%%%%%%%%%%%%%%%%%%%%%%%%%% 
\subsubsection{Network connectivity and activity dynamics} 
\label{Sec_NCAD} 

The networks considered in this study comprised $N = 12\,500$ neurons, $f = 80\%$ of which were excitatory and $1-f = 20\%$ inhibitory. 
Synaptic connections were drawn randomly and independently, such that each neuron received exactly $1\,000$ inputs from the excitatory and $250$ from the inhibitory neuron population, respectively.  
This amounted to an overall connectivity of $\epsilon = 10\%$, as suggested by statistical neuroanatomy of local cortical networks \cite{Braitenberg1998}.  
The wiring imposed in our model was in accordance with Dale's principle, i.e.~each neuron formed the same type of synapse with all its postsynaptic partners, either excitatory or inhibitory \cite{Kriener2008}. 
Self-connections were excluded. 

If $t_i^k$ is the time of a spike elicited by neuron $i$, we use a Dirac delta-function $\delta(t-t_i^k)$ to represent it as a time dependent signal.  
The sum $Y_i(t) = \sum_k \delta(t-t_i^k)$ then stands for a spike train.  
The input $I_i(t)$ to each neuron is the sum of all excitatory and inhibitory postsynaptic currents (PSCs) induced by presynaptic spikes that arrive at its dendrite, and the hyperpolarizing currents responsible for the reset after each output spike.  
Assuming that all currents are pulse-like, the dynamic equation for the network is obtained from 
\[ R I_i(t) = \tau \Bigl[ -V_\mathrm{reset} Y_i(t) + \sum_j W_{ij} Y_j(t-D) +  J_s X_i(t) \Bigr]. 
  \label{Eq_NetworkModel} 
\] 
After each spike, the membrane potential was reset to the resting potential at $0\,\mathrm{mV}$, therefore the size of the voltage jump is $V_\mathrm{reset} = V_\mathrm{th} - V_0 = V_\mathrm{th}$. 
The cross-neuron coupling $W_{ij}$ encodes the amplitude of the postsynaptic potential (PSP), corresponding to a synapse from neuron $j$ (source) to neuron $i$ (target).  
A uniform transmission delay of $D = 1.5\,\mathrm{ms}$ was assumed for all recurrent synapses in the network.  
The spike train $X_i(t)$ stands for the accumulated external input to neuron $i$, and the corresponding synapses have connection strength of amplitude $J_s$. 

In all the simulations described in our paper, in fact, we used stereotyped synaptic transients of finite width, instead of normalized impulses, as postsynaptic currents.  
All synaptic kernels had the shape of an alpha-function $J_\alpha \frac{e}{\tau_\mathrm{syn}} t e^{-t/\tau_\mathrm{syn}}$, with a fixed time constant $\tau_\mathrm{syn} = 0.5\,\mathrm{ms}$, replacing the delta-functions in the spike trains. 
The peak amplitude of the kernel is $J_\alpha$, to which we refer as $\mathrm{EPSP}$ to denote the strength of post-synaptic potentials. 
The parameter $W_{ij}$ corresponds to the integral of the PSP, which is $J = e \tau_\mathrm{syn} \mathrm{EPSP}$. 

In this model, keeping all time constants at fixed values, the efficacy of a synaptic connection is determined by the peak amplitude of the PSP. 
For any specific network, we assumed that all recurrent excitatory synapses induce excitatory postsynaptic potentials of the same peak amplitude, $\mathrm{EPSP}$. 
The peak amplitudes of inhibitory postsynaptic potentials were taken to be a fixed multiple, $g$, of the excitatory ones, such that $\mathrm{IPSP} = - g \, \mathrm{EPSP}$.  
For all our results presented in the main text, individual inhibitory couplings were assumed to be much more effective than excitatory ones: 
The excitation-inhibition ratio was fixed at $g=8$.  
As a consequence, recurrent connectivity in our networks was characterized by a net surplus of inhibition, since the small number of inhibitory neurons was over-compensated by the strength of individual inhibitory couplings.  
Fixing the balance between recurrent excitation and inhibition in this way is an important concept in models of cortical dynamics, although measurements in real brains are difficult (see e.g.\ \cite{Okun2008}). 

In different simulations, we used excitatory synapses with an EPSP amplitude in the range between $0\,\mathrm{mV}$ and $1.0\,\mathrm{mV}$.  
Accordingly, inhibitory synapses had IPSP amplitudes between $0\,\mathrm{mV}$ and $-8.0\, \mathrm{mV}$.  
All external inputs in our simulations were excitatory, and the amplitude of their synapses, $\mathrm{EPSP}_\mathrm{ffw}$, was fixed at $0.1\,\mathrm{mV}$ throughout all simulations. 

This configuration of parameters, combined with a stationary driving input to each neuron in the network, was previously shown to induce relatively low rates in all neurons, while spike trains are irregular, and pairwise correlations remain weak \cite{Brunel2000}. 
These properties are known to be a result of complex recurrent network dynamics, and not a consequence of random inputs (e.g.~Poisson spike trains) that drive the network \cite{VanVreeswijk1996, Kriener2008}. 
Inhibitory feedback actively decorrelates the network activity \cite{Renart2010, Pernice2011, Tetzlaff2012}.  
The resulting states of network dynamics are dubbed asynchronous-irregular, AI, and they are thought to closely resemble the dynamic states of neocortical networks recorded with extracellular electrodes \cite{Ecker2010}. 

In this parameter setting, the degree of recurrence in any specific network is essentially determined by the amplitude of excitatory postsynaptic potentials, $\mathrm{EPSP}$, of the recurrent connections. 
Recurrence can be effectively quantified by the spectral radius of the connectivity matrix $\vec{W}$, which scales linearly with the EPSP amplitude.  
This fact is explained in more detail below. 

%%%%%%%%%%%%%%%%%%%%%%%%%%%%%%%%%%%%%%%%%%%%%%%%%%%%%%%%%%%%%%%%%%%%%%%%%%%%%%%% 
\subsubsection{Neuronal tuning} 
 \label{Sec_NT} 

To explore the tuning curves of neurons in a network, we simulated their responses to stimuli with different orientations. 
Beyond excitatory and inhibitory input from the recurrent network, each neuron received extra excitatory input, the firing rate of which exhibited a slight dependence on stimulus orientation. 
This external input can be conceived as the overall effect of stimulation, and it includes inputs from LGN, and possibly afferents from other, non-local cortical sources. 
In our simulations, the input was implemented as a homogeneous Poisson process, with an average firing rate, $s$, depending on the stimulus orientation $\theta$ according to 
\[s(\theta) = s_B \bigl[ 1 + m \cos(2(\theta - \theta^*)) \bigr]. 
\label{Eq_S} 
\] 
The baseline $s_B$ is the mean level of input across all orientations.  
We used a logarithmic relation between input contrast $C$ and input baseline, $s_B \propto \log_{10}(1+100 C)$, as a practical way to specify the input intensity, inspired by biology. 

In all our simulations, the relative amplitude, $m$ of the stimulus dependent modulation was fixed to a fraction of $10\%$ of the baseline level, corresponding to setting $m = 0.1$.  
The parameter $\theta^*$ signified the stimulus orientation at which the neuron received its maximal input, $s_\mathrm{max} = (1+m) s_B$.  
It represented the initial preferred orientation, $\text{Input PO}$, of the neuron, a parameter that was randomly and independently assigned to each neuron in the population. 

To measure the output tuning curves in numerical simulations, we stimulated the networks for $12$ different stimulus orientations, covering the full range between $0\deg$ and $180\deg$ in steps of $15\deg$.  
Each simulation was run for $6.3\,\mathrm{s}$, using a simulation time step of $0.1\,\mathrm{ms}$.  
In order to include only steady state activity into our analysis, and to avoid onset transients, the first $300\,\mathrm{ms}$ in each simulation were discarded.  
The output tuning curve of any neuron in the network was obtained in terms of its average firing rate $r$ in response to each stimulus orientation $\theta$, and normally plotted as a curve $r(\theta)$.  
An output tuning curve would be termed contrast invariant, if its overall shape does not depend on the contrast, $C$, of the stimulus.  

To explore the interaction between feedforward and recurrent connectivity on orientation selectivity, we systematically changed two parameters in our networks: 
The mean input firing rate, $s_B$, and the EPSP amplitude as a measure for the strength of recurrent coupling. 
We changed these two parameters in a network, while fixing all other parameters, including the network topology given by a specific realization of the random synaptic connectivity, $\vec{W}$, the inhibition-excitation ratio, $g$, and the input modulation ratio, $m$. 
We used three different values for the baseline intensity $s_B = 12\,000$, $16\,000$, and $20\,000\,\mathrm{spikes}/\mathrm{s}$. 
This is corresponding to low, medium, and high contrast, $C \approx 9\,\%$, $39\,\%$, and $99\,\%$, respectively. 

%%%%%%%%%%%%%%%%%%%%%%%%%%%%%%%%%%%%%%%%%%%%%%%%%%%%%%%%%%%%%%%%%%%%%%%%%%%%%%%% 
\subsubsection{Free membrane potential} 
\label{Sec_FMP} 

Contrast invariance of the membrane potential tuning is, in the case of a tuned spike response, compromised by the reset mechanism in our neuron model: 
After each spike, the membrane potential is reset to its resting value, which exerts a negative contribution to the membrane potential, which effectively imposes the opposite tuning as compared to the output spiking. 
As a result, when the neuron fires more (higher contrasts), it inevitably attains a more negative membrane potential. 
To correct for this phenomenon, we add this negative contribution back to the membrane potential ($V_\mathrm{fm} = V + \tau r V_\mathrm{reset}$) or, equivalently, keep the neuron from spiking by raising its threshold to very high levels. 
In membrane potential recordings from real neurons, tt is also common to correct for this phenomenon by cutting out the spikes including their aftereffects. 
If used with care, this procedure is essentially equivalent to the correction we applied here. 

%%%%%%%%%%%%%%%%%%%%%%%%%%%%%%%%%%%%%%%%%%%%%%%%%%%%%%%%%%%%%%%%%%%%%%%%%%%%%%%% 
\subsubsection{Numerical methods and simulation software} 
\label{Sec_NMSS} 

The implementation of the LIF model employed in the present study is based on a numerical method known as ``exact integration'' \cite{Rotter1999, Diesmann2001}. 
Numerical simulations of all networks were performed using the neuronal simulation environment NEST \cite{Gewaltig2007}. 
This tool has been developed to support the reliable, precise and performant numerical simulations of networks of spiking neurons. 

%%%%%%%%%%%%%%%%%%%%%%%%%%%%%%%%%%%%%%%%%%%%%%%%%%%%%%%%%%%%%%%%%%%%%%%%%%%%%%%% 
%%%%%%%%%%%%%%%%%%%%%%%%%%%%%%%%%%%%%%%%%%%%%%%%%%%%%%%%%%%%%%%%%%%%%%%%%%%%%%%% 
\subsection{Data analysis} 
\label{Sec_DA} 

%%%%%%%%%%%%%%%%%%%%%%%%%%%%%%%%%%%%%%%%%%%%%%%%%%%%%%%%%%%%%%%%%%%%%%%%%%%%%%%% 
\subsubsection{Orientation selectivity} 
\label{Sec_OS} 

To quantify orientation selectivity, we computed the Preferred Orientation, PO, and the Orientation Selectivity Index, OSI, of each neuron from its respective tuning curve, $r(\theta)$, obtained in numerical simulations.  
To this end, we first computed the circular mean \cite{Batschelet1981} of the firing rate measured at each orientation, which we call the Orientation Selectivity Vector, OSV, 
\[\vec{\mathrm{OSV}} = \frac{\sum_\theta r(\theta)\exp(2\pi i\theta/180\deg)}{\sum_\theta r(\theta)}. 
\label{OSV} 
\] 
The PO, $\theta^*$, was then extracted as the angle, $\arg(\vec{\mathrm{OSV}})$, of the OSV. 
Its length, $|\vec{\mathrm{OSV}}|$, yielded a measure for the degree of orientation selectivity, OSI \cite{Ringach2002}. 
For a highly selective neuron, which is only active for one orientation, and remains silent for all other orientations, the OSI would be $1$; 
for a completely unselective neuron responding with the same firing rate for all orientations, this measure returns $0$. 

For better comparison with the experimental literature (see e.g.\ \cite{Niell2008}), an alternative measure of orientation selectivity has also been computed for the tuning curves obtained in our simulations. 
It is given by 
\[\mathrm{OSI}^* = \frac{r_\mathrm{pref} - r_\mathrm{orth}} {r_\mathrm{pref} + r_\mathrm{orth}}, 
\] 
where $r_\mathrm{pref} = r(\theta^*)$ is the firing rate at the preferred stimulus orientation, $\theta^*$, and the $r_\mathrm{orth} = r(\theta^*+90\deg)$ is the firing rate for the orientation orthogonal to it. 

Since the output preferred (and hence the orthogonal) orientations of a neuron are computed from Eq.~\eqref{OSV}, we need to interpolate between the data points of a tuning curve to obtain $r_\mathrm{pref}$ and $r_\mathrm{orth}$. 
To do this, we fit a cosine function to the tuning curve sampled at $12$ equidistant orientations, employing a nonlinear least squares method. 
Then, the cosine fit of the tuning curve is evaluated at $\mathrm{PO}$ and $\mathrm{PO}+90\deg$ to obtain $r_\mathrm{pref}$ and $r_\mathrm{orth}$, respectively. 
Negative numbers, whenever occurring, were replaced by $0$. 
This is similar to what experimentalists typically do (see e.g.\ \cite{Niell2008}, although by fitting different functions, like a Gaussian or a von Mises probability density; also see \cite{Hansel2012}), therefore allows us to compare our distributions to their reported results.\footnote{We verified however, if the cosine fit imposes a general constraint on the tuning curves and changes the OSI*. To check this, we use an alternative way to obtain $r_\mathrm{pref}$ and $r_\mathrm{orth}$ from the smoothened tuning curves, namely by linear interpolation between the data points. The result of this is compared with the result of the cosine fit in Fig.~\ref{Fig_OsiPop}B, and does not change the conclusions qualitatively.} 

For a perfect cosine tuning curve according to Eq.~(\ref{Eq_S}), we obtain 
\[\mathrm{OSI} = \frac{m}{2} 
\qquad\text{and}\qquad 
\mathrm{OSI}^\ast = m 
\] 
irrespective of the baseline firing rate $s_B$, and of the preferred orientation $\theta^*$.  
Thus, in the case of our input tuning curves with $m = 0.1$, we have $\mathrm{OSI} = 0.05$ and $\mathrm{OSI}^\ast = 0.1$, respectively. 

%%%%%%%%%%%%%%%%%%%%%%%%%%%%%%%%%%%%%%%%%%%%%%%%%%%%%%%%%%%%%%%%%%%%%%%%%%%%%%%% 
\subsubsection{Baseline and modulation gain} 
\label{Sec_BMG} 

To quantify the processing of baseline and modulation in a specific network, we compute the mean baseline and mean modulation gains over all neurons. 
To obtain this, we first compute baseline and modulation gain for individual tuning curves, $r_n(\theta)$, as follows. 
The baseline is obtained by averaging the tuning curve over all orientations 
\[r_n^B = \frac{1}{K}{\sum_{k=1}^K r_n(\theta_k)}, 
\label{def_baseline} 
\] 
where $r_n(\theta_k)$ is the firing rate of $n$-th neuron in the network in response to a stimulus with the $k$-th orientation, $\theta_k$, and $K = 12$ is the number of different orientations 
considered in the simulation. 
We refer to this as the F0 component of tuning curves. 
The modulation is conveniently obtained from the absolute value of the second Fourier component of the tuning curve 
\[r_n^M = | \frac{2}{K} \sum_{k=1}^K r_n(\theta_k) \exp(-2\pi i\, \theta/180\deg) |. 
\label{def_mod} 
\] 
We refer to this as the F2 component of a tuning curve. 

The baseline gain and the modulation gain are then defined as the output value divided by the input value, respectively 
\[\gamma_n^B = r_n^B/s_n^B 
\qquad\text{and}\qquad 
\gamma_n^M = r_n^M/s_n^M. 
\] 
The input baseline and modulation, $s_n^B$ and $s_n^B$, are obtained from input tuning curves in the same fashion as $r_n^B$ and $r_n^B$ were obtained from the output tuning curves, respectively (Eq.~\eqref{def_baseline} and \eqref{def_mod}, respectively). 
The corresponding mean values, $\gamma_B$ and $\gamma_M$, for any network are the average over all individual gains. 
Finally, each gain is normalized by its value for a network with no recurrence, $\gamma^0_B$ and $\gamma^0_M$, which are obtained from the simulation of a network with no recurrence, $\mathrm{EPSP} = 0\,\mathrm{mV}$ 
\[\gamma^\mathrm{norm}_B = \frac{\gamma_B}{\gamma^0_B} 
\qquad\text{and}\qquad 
\gamma^\mathrm{norm}_M = \frac{\gamma_M}{\gamma^0_M}. 
\] 

%%%%%%%%%%%%%%%%%%%%%%%%%%%%%%%%%%%%%%%%%%%%%%%%%%%%%%%%%%%%%%%%%%%%%%%%%%%%%%%% 
\subsubsection{PO scatter} 
\label{Sec_POS} 

The transformation of preferred orientation induced by a recurrent network is visualized by a scatter plot showing Output PO vs.\ Input PO for all neurons (Fig.~\ref{Fig_PO_Scat}). 
Weakly recurrent networks essentially preserve the preferred orientations of the input to each neuron, leading to scatter plots centered about the diagonal. 
For networks with increased recurrence, the output PO deviates from the input PO, and off-diagonal elements occur more frequently.  
To quantify the deviation, we first compute the difference, $\Delta \mathrm{PO}_n = \mathrm{OutputPO}_n - \mathrm{InputPO}_n$, for each neuron. 
Observe that orientation should be taken modulo $180\deg$ and $\Delta \mathrm{PO}_n = 90\deg$ represents the largest possible difference between input and output PO. 

As a numerical measure for the total degree of PO scatter in a network, we computed the Scatter Degree Index, SDI (Fig.~\ref{Fig_PO_Scat}B).  
Its definition is based on the circular mean 
\[R_{\Delta} = \frac{1}{N}\sum_{n=1}^N \exp(2\pi i\, \Delta\mathrm{PO}_n/180\deg), 
\] 
where $N$ is the number of neurons in the network. 
The SDI is then given by the angular deviation \cite{Batschelet1981}, which can be computed from the length $|R_{\Delta}|$ according to 
\[\mathrm{SDI} = \frac{90\deg}{\pi} \sqrt{2 (1 - |R_{\Delta}|)}. 
\] 
Note that as $\Delta \mathrm{PO}$ spans the half-cricle, i.e. the range $[0\deg, 180\deg]$, we have taken half the resultant angle as the SDI. 
If all Output POs are exactly the same as Input POs, SDI returns zero; the maximum scatter from the Input POs corresponds to a uniform distribution of $\Delta \mathrm{PO}$, for which SDI returns $\approx 40.5\deg$.

%\clearpage 

%%%%%%%%%%%%%%%%%%%%%%%%%%%%%%%%%%%%%%%%%%%%%%%%%%%%%%%%%%%%%%%%%%%%%%%%%%%%%%%% 
%%%%%%%%%%%%%%%%%%%%%%%%%%%%%% Theoretical analysis 
%%%%%%%%%%%%%%%%%%%%%%%%%%%%%%%%%%%%%%%%%%%%%%%%%%%%%%%%%%%%%%%%%%%%%%%%%%%%%%%% 
\section{Theoretical analysis} 
\label{Sec_THA} 

%%%%%%%%%%%%%%%%%%%%%%%%%%%%%%%%%%%%%%%%%%%%%%%%%%%%%%%%%%%%%%%%%%%%%%%%%%%%%%%%  
%%%%%%%%%%%%%%%%%%%%%%%%%%%%%%%%%%%%%%%%%%%%%%%%%%%%%%%%%%%%%%%%%%%%%%%%%%%%%%%% 
\subsection{Firing rates and membrane potential statistics} 
\label{Sec_FRMPS} 

%%%%%%%%%%%%%%%%%%%%%%%%%%%%%%%%%%%%%%%%%%%%%%%%%%%%%%%%%%%%%%%%%%%%%%%%%%%%%%%% 
\subsubsection{Mean firing rates in recurrent networks} 
\label{Sec_tempmean} 

The tuning curves considered in this work reflect time-averaged firing rates of neurons in a recurrent network.  
From our numerical simulations, it became clear that the time averaged membrane potential is indicative of the operating point of the network with regard to the tuning properties of its neurons (see Sect.~\ref{Sec_Results}). 
Therefore, we begin our analysis by considering time averaged equations. 

Assuming stationarity, we form temporal averages $\langle\,.\,\rangle$ of all dynamic variables that occur in Eq.~(\ref{Eq_NeuronModel}) and (\ref{Eq_NetworkModel}). 
Since there can be no drift of the time averaged membrane potential in this case, we have $\langle \dot{V}_i \rangle = 0$. 
We write $v_i = \langle V_i \rangle$ for the time averaged membrane potential, $r_i = \langle Y_i \rangle$ for the mean firing rate of neuron $i$ in the network, and $s_i = \langle X_i \rangle$ for the mean firing rate of its external input, respectively. 
We obtain an equation that relates the stationary firing rates of all neurons in the network with their mean membrane potentials \cite{Hansel2002, Kriener2008} 
\[  v_i = \tau \Bigl[ - V_\mathrm{th} r_i + \sum_j W_{ij} r_j + J_s s_i  \Bigr]. 
    \label{Eq_StatRate} 
\] 
Observe that transmission delays do not matter for temporal averages, and that the above equation holds for networks of LIF neurons with arbitrary connectivity. 

From now on, we rescale Eq.~(\ref{Eq_StatRate}) such that all firing rates are expressed in units of $1/\tau$, and all voltages are given in units of $V_\mathrm{th}$.  
For a network of $N$ neurons, the recurrent synaptic connectivity is encoded by a fixed $N \times N$ coupling matrix $\vec{W} = (W_{ij})$. 
The external inputs, the firing rates and the membrane potentials of all neurons are represented by the $N$-dimensional vectors $\vec{s} = (s_i)$, $\vec{r} = (r_i)$ and $\vec{v} = (v_i)$, respectively. 
The time averaged equation above then reads, in matrix-vector notation 
\[ \vec{v} = -\vec{r} + \vec{W} \vec{r} + J_s \vec{s}. 
  \label{Eq_StatRate_a} 
\] 
Solving for the vector $\vec{r}$ of recurrent firing rates, we obtain 
\[ \vec{r} = \vec{A} (J_s \vec{s} - \vec{v}),  
\qquad\text{where}\quad  \vec{A} = (\vec{\one} - \vec{W})^{-1}. 
  \label{Eq_StatRate_b} 
\] 

We assume that the matrix $\vec{\one} - \vec{W}$ is always invertible, with inverse $\vec{A}$.  
If two out of the three variables $\vec{s}$, $\vec{r}$ and $\vec{v}$ are known, the third one can then be computed in a straightforward fashion.  

%%%%%%%%%%%%%%%%%%%%%%%%%%%%%%%%%%%%%%%%%%%%%%%%%%%%%%%%%%%%%%%%%%%%%%%%%%%%%%%%  
\subsubsection{Eigenvalue spectrum of homogeneous random networks} 
\label{Sec_evrn} 

We now specifically consider a recurrent network of excitatory and inhibitory neurons, as discussed above. 
It is assumed that $N_E = f N$ neurons are excitatory, forming synapses of uniform strength $J$ with their postsynaptic targets. 
The remaining neurons $N_I = N - N_E = (1-f) N$ are inhibitory, forming synapses of uniform strength $-g J$.  
The factor $g > 0$ describes the relative strength of inhibitory synapses. 
We refer to the network as being ``inhibition dominated'', if the lower number of inhibitory neurons is compensated for by stronger inhibitory weights. 
In the case considered here, this amounts to the condition $g > N_E/N_I = f/(1-f)$ \cite{Brunel2000}. 

The connectivity of the network is set to $\epsilon$, such that each neuron receives input from exactly $N_E \epsilon$ excitatory neurons and $N_I \epsilon$ inhibitory neurons. 
The presynaptic sources are randomly selected from the available pool, multiple synaptic contacts are excluded. 
The graph underlying such a network is a specific type of random graph \cite{Erdos1959, Bollobas2001}. 
The connectivity matrix $\vec{W}$ is a random matrix with two types of entries, organized in homogeneous columns.  
The entries in positive columns of this matrix, corresponding to excitatory neurons, have a mean of $\eta_E = J \epsilon$ and a variance of $\sigma^2_E = J^2 \epsilon(1-\epsilon)$, whereas the entries in negative columns, corresponding to inhibitory neurons, have a mean of $\eta_I = -g J \epsilon$ and a variance of $\sigma^2_I = g^2 J^2 \epsilon(1-\epsilon)$, respectively. 
The matrix $\vec{W}$ has an eigenvalue spectrum with two components that are, for large and not too sparse networks, described as follows \cite{Rajan2006}: 
There is one exceptional eigenvalue, proportional to the mean recurrent input to each neuron 
\[ \lambda_0 = N_E \eta_E + N_I \eta_I  = J N \epsilon \bigl[ f - g (1-f) \bigr]. 
\] 
It belongs to uniform eigenvectors with all components being equal. 
They represent a $1$-dimensional subspace, spanned by the uniform vector $\vec{u} = (1,\ldots,1)^T$.  
In a balanced random network, we have $\lambda_0 < 0$.  The bulk spectrum $\Lambda = \{\lambda_1,\ldots,\lambda_{N-1}\}$ covers a circular region in the complex plane, centered at the origin. 
Its radius $\rho$ satisfies 
\begin{align} 
\rho^2 &= N_E \sigma^2_E + N_I \sigma^2_I  %\nonumber \\ 
       &= J^2 N \epsilon (1-\epsilon) \bigl[ f + g^2 (1-f) \bigr]. 
\label{Eq_rho}
\end{align} 
The density of eigenvalues within the circle is in general non-uniform, and it can be approximated by a density derived in \cite{Rajan2006}. 

Eq.~(\ref{Eq_StatRate_a}), which relates input, output and membrane potentials under stationary conditions, has an effective coefficient matrix $\vec{W}-\vec{\one}$.  
Its eigenvalue spectrum consists of numbers $\lambda-1$, where $\lambda$ is from the spectrum of $\vec{W}$. 
Likewise, the eigenvalues of $(\one-W)^{-1}$ are $(1-\lambda)^{-1}$. 
This can either be derived directly, or it can be implied by the spectral mapping theorem \cite{Higham2008}. 
The associated eigenvectors are the same in each case. 

%%%%%%%%%%%%%%%%%%%%%%%%%%%%%%%%%%%%%%%%%%%%%%%%%%%%%%%%%%%%%%%%%%%%%%%%%%%%%%%% 
\subsubsection{Self-consistent firing rates in homogeneous random networks} 
\label{Sec_scfr} 

Under the same conditions on homogeneity as made above, explicit solutions for the response rates, and for the mean membrane potentials, can be obtained by resorting to additional constraining assumptions. 
Specifically, one can analytically describe the response statistics of a leaky integrate-and-fire neuron, which is driven by randomly fluctuating input. 
If inputs are uncorrelated, and synaptic couplings are weak, the lumped synaptic input current may be approximated by a Gaussian white noise with appropriate parameters $\mu$ and $\sigma^2$ 
\[RI(t) = \mu + \sigma \sqrt{\tau} \eta(t) 
\] 
where $\eta(t)$ is a stationary Gaussian white noise with zero mean and unit power spectral density. 
Assuming stationarity and a fixed voltage threshold, the associated first-passage time problem can in fact be solved: 
The membrane potential dynamics of the neuron can be conceived as a diffusion process, and the time evolution of the membrane potential distribution is given by a Fokker-Planck equation with specific boundary conditions.  
Its solution yields explicit expressions for the moments of the inter-spike interval distribution \cite{Siegert1951, Ricciardi1977}. 
In particular, the mean response rate of the neuron, $r$, in terms of its input statistics 
\begin{align} 
 r &= f(\mu, \sigma) \nonumber \\ 
   &= \left[ t_\mathrm{ref} +\tau \sqrt{\pi} \int_{\tilde{V}_0}^{\tilde{V}_\mathrm{th}} e^{u^2} (1+\erf(u)) \,du \right]^{-1}  
 \label{Eq_mT} 
\end{align} 
with $\tilde{V}_\mathrm{th} = (V_\mathrm{th}-\mu) / \sigma$ and $\tilde{V}_0 = (V_0 - \mu) / \sigma$. 

Employing a mean field ansatz, the above theory can be applied to networks of identical pulse-coupled LIF neurons, randomly connected with homogeneous in-degrees, and driven by external excitatory input of the same strength. 
Under these circumstances, all neurons exhibit the same mean firing rate, which can be determined by a straight-forward self-consistency argument \cite{Amit1997, Brunel2000}: 
The firing rate $r$ is a function of the first two moments of the input fluctuations, $\mu$ and $\sigma^2$, as described by Eq.~(\ref{Eq_mT}). 
Both parameters are, in turn, functions of the firing rate $r$. 
This leads to a fixed point equation, the root of which can be found numerically. 
Here we employed Newton's method, verifying the convergence of the iteration by appropriate means. 

For networks of the type described here, we have specifically 
\begin{align} 
  \mu &= \tau [J_s s + J r N \epsilon (f - g(1-f))], \nonumber 
  \\ \sigma^2 &= \tau [J_s^2 s + J^2 r N \epsilon (f + g^2(1-f))], 
  \label{Eq_inputMS} 
\end{align} 
where $s$ is the input (stimulus) firing rate, and $r$ is the mean response rate of all neurons in the network, respectively. 
Here, $J_s$ denotes the EPSP amplitude of external inputs, and $J$ denotes the amplitude of recurrent EPSPs. 
The inhibition-excitation ratio $g$ has been introduced above. 
The remaining structural parameters are the number of neurons in the network, $N$, the connection probability, $\epsilon$, and the fraction $f$ of neurons in the network that are excitatory, implying that a fraction $1-f$ is inhibitory. 

%%%%%%%%%%%%%%%%%%%%%%%%%%%%%%%%%%%%%%%%%%%%%%%%%%%%%%%%%%%%%%%%%%%%%%%%%%%%%%%% 
\subsubsection{Correction for $\alpha$-synapses}
\label{Sec_AlphaCorr} 

The treatment described above is only approximating the networks considered in numerical simulations, since we chose biologically more realistic LIF neurons with alpha-synapses.  
In order to make use of the same analytical framework as just described, we made the simplifying assumption that all the presynaptic current is delivered immediately, and that the input current to each neuron is still white. 
We therefore need to obtain the effective values for mean and variance. 

To obtain the effective value of the mean, we match the area under the PSC kernel of $\alpha$-shape with a corresponding $\delta$-synapse 
\[\int_0^{\infty} \frac{t}{\tau_\mathrm{syn}^2} e^{-t/\tau_\mathrm{syn}}\,dt = 1. 
\] 
The actual $\alpha$-synapse with a peak amplitude $J_\alpha$ would then be matched to a $\delta$-PSC as follows 
\begin{align} 
\int_0^{\infty} J_\alpha \frac{e}{\tau_\mathrm{syn}} t e^{-t/\tau_\mathrm{syn}}\,dt = \nonumber 
\\ (J_\alpha e \tau_\mathrm{syn}) \int_0^{\infty} \frac{t}{\tau_\mathrm{syn}^2} e^{-t/\tau_\mathrm{syn}}\,dt. 
\end{align} 
Therefore, we choose the value $J = J_\alpha e \tau_\mathrm{syn}$ as the effective value for the mean input. 
This is equivalent to the integral under the PSC, i.e.\ the total amount of current that is delivered by an alpha synapse with peak $J_\alpha$. 

The effective value of the variance can be obtained in the same fashion by matching the integral of the squared PSC, of the $\alpha$-PSC with the $\delta$-PSC 
\[\int_0^{\infty} \frac{4}{\tau_\mathrm{syn}^3} \bigl[ t e^{-t/\tau_\mathrm{syn}} \bigr]^2 \, dt = 1, 
\] 
and 
\begin{align} 
\int_0^{\infty} \bigl[ J_\alpha \frac{e}{\tau_\mathrm{syn}} t e^{-t/\tau_\mathrm{syn}} \bigr]^2 \, dt = \nonumber 
\\ (\frac{J_\alpha^2 e^2 \tau_\mathrm{syn}}{4}) \int_0^{\infty} \frac{4}{\tau_\mathrm{syn}^3} \bigl[ t e^{-t/\tau_\mathrm{syn}} \bigr]^2 \, dt. 
\end{align} 
This suggests $J^2 = \frac{1}{4} J_\alpha^2 e^2 \tau_\mathrm{syn}$. 

%%%%%%%%%%%%%%%%%%%%%%%%%%%%%%%%%%%%%%%%%%%%%%%%%%%%%%%%%%%%%%%%%%%%%%%%%%%%%%%% 
%%%%%%%%%%%%%%%%%%%%%%%%%%%%%%%%%%%%%%%%%%%%%%%%%%%%%%%%%%%%%%%%%%%%%%%%%%%%%%%% 
\subsection{Transformation of tuning by a recurrent network} 
\label{Sec_TTRN} 

%%%%%%%%%%%%%%%%%%%%%%%%%%%%%%%%%%%%%%%%%%%%%%%%%%%%%%%%%%%%%%%%%%%%%%%%%%%%%%%% 
\subsubsection{Baseline and modulation} 
\label{Sec_BM} 

We now consider the case of tuned input to a recurrent network. 
The input $s_i$ to each neuron in the network, as well as its firing rate response $r_i$ and its membrane potential $v_i$, may depend on a given feature $\vec{\phi} \in \F$ of a sensory stimulus.  
The functions $s_i(\vec{\phi})$, $r_i(\vec{\phi})$ and $v_i(\vec{\phi})$ will be called the input, output and membrane potential tuning curves of neuron $i$, respectively. 
Let now the input to each neuron in a recurrent network be tuned, i.e.\ $\vec{s} = \vec{s}(\vec{\phi})$. 
After relaxation to equilibrium, the output of the recurrent network is given by Eq.~(\ref{Eq_StatRate_b}), and the tuning curves are obtained by 
\begin{align} 
 \vec{r}(\vec{\phi}) &= (\vec{\one} - \vec{W})^{-1} \bigl[ J_s \vec{s}(\vec{\phi}) - \vec{v}(\vec{\phi}) \bigr] \nonumber \\ 
  &= \vec{A} \bigl[ J_s \vec{s}(\vec{\phi}) - \vec{v}(\vec{\phi}) \bigr]. 
  \label{TunMix} 
\end{align}  

Assume now that a stimulus ensemble has been fixed for an experiment, think of a uniform distribution for the orientation of a stimulus offered, for example. 
Mathematically, this is described by a suitable probability distribution on the set of possible values for the feature $\vec{\phi}$, the stimulus ensemble. 
Thereby, any stimulus dependent quantity $\vec{x}$ (like~$\vec{s}$, $\vec{r}$ and $\vec{v}$) turns into a real-valued random variable. 
Its expected value $\E[\vec{x}]$ then corresponds to the component of the tuning curve $\vec{x}(\vec{\phi})$ that is common to all parameter values, the baseline of the tuning curve. 
Note that this concept depends on the stimulus ensemble, and it suggests the following further terminology: 
\begin{align} 
  \text{\textbf{baseline}}:\qquad & \vec{x}_B = \E[\vec{x}] 
  \\ \text{\textbf{modulation}}:\qquad & \vec{x}_M = \vec{x} - \E[\vec{x}] 
\end{align} 
Evidently, the decomposition $\vec{x} = \vec{x}_B + \vec{x}_M$ is fully specified by these settings. 
Moreover, by linearity of $\E[\,.\,]$ Eq.~(\ref{Eq_StatRate_b}) implies 
\[  \vec{r}_B = \vec{A} (J_s \vec{s}_B - \vec{v}_B), 
  \label{Eq_StatRate_B} 
\] 
and, therefore, 
\begin{align} 
 \vec{r}_M = \vec{r} - \vec{r}_B &= \vec{A} \bigl[J_s (\vec{s}-\vec{s}_B) - (\vec{v}-\vec{v}_B)\bigr] \nonumber \\ 
 &= \vec{A} (J_s \vec{s}_M - \vec{v}_M). 
  \label{Eq_StatRate_M} 
\end{align} 
This means that the recurrent network defined by the coupling matrix $\vec{W}$, and the matrix $\vec{A}$ derived from it, processes baseline and modulation components separately and independently, with no cross-talk involved. 
In other words, pure modulation input will not attain any baseline through network processing, and \textit{vice versa}. 
This is exactly the meaning of Figure~\ref{Fig_SepPath}B. 

Note, however, that, as the mean membrane potential $\vec{v}$ actually depends on the input $\vec{s}$ in a highly nonlinear fashion, the above equations determine the network response only implicitly. 
Moreover, for the processing to be independent, it is necessary that $\vec{v}_B$ and $\vec{v}_M$ depends only on $s_B$ and $s_M$, respectively, with no cross-talk. 
For the baseline firing rates, this implies that $\vec{v}_B$ is not affected by the modulation in the input. 
As baseline firing rates are the same in our homogeneous networks, the mean $\vec{v}_B$ (over neurons in the network) should therefore be more or less constant in one experiment with fixed $s_B$. 
We have checked this numerically in our simulations by plotting the standard deviation (over neurons) of the mean membrane potential (over time and over orientation) for $24$ sampled excitatory and inhibitory neurons (Fig.~\ref{Fig_VmTC}). 
The variance is indeed much smaller than $v_M$, the modulation of the membrane potential due to modulation in the input, $s_M$, and this is consistent for all recurrent regimes. 

%%%%%%%%%%%%%%%%%%%%%%%%%%%%%%%%%%%%%%%%%%%%%%%%%%%%%%%%%%%%%%%%%%%%%%%%%%%%%%%% 
\subsubsection{Baseline attenuation by inhibition dominated random networks} 
\label{Sec_BAIDRN} 

We will now consider the case of homogeneous properties of all inputs to the network. 
Specifically, we assume that the input baseline is a uniform vector 
\[\E[\vec{s}] = \vec{s}_B \sim \vec{u}. 
\] 
%notwithstanding the fact that individual inputs have different preferred orientations. 
We now consider an inhibition dominated random network as described in Sect.~\ref{Sec_evrn}, where all neurons have identical parameters, and the connectivity matrix is statistically homogeneous. 
As we have already discussed, under the assumption of orthogonality, the firing rate responses of all neurons to a homogeneous stimulus, and the corresponding mean membrane potentials, are all identical 
\[\vec{r}_B \sim \vec{u} \qquad\text{and}\qquad \vec{v}_B \sim \vec{u}. 
\] 
In other words, homogeneous vectors, like baseline input, are eigenvectors of the coupling matrix $\vec{W}$.  
They are also eigenvectors of the matrix $\vec{A} = (\vec{\one} - \vec{W})^{-1}$ which determines the stationary firing rates. 
The corresponding eigenvalue $\lambda_0$ of the matrix $\vec{W}$ is in fact negative, and the corresponding eigenvalue of $\vec{A}$ is $1/(1-\lambda_0)$ is positive, but much smaller than $1$, since $|\lambda_0|$ is typically large (of order $N J \epsilon$).  
The overall amplification or attenuation of the baseline is given by 
\[ \vec{r}_B = \vec{A} (J_s \vec{s}_B - \vec{v}_B) = \frac{1}{1-\lambda_0}(J_s \vec{s}_B -  \vec{v}_B). 
 \] 

In the case of uniform input $\vec{s}_B$ an explicit solution of the mean firing rate $\vec{r}_B$ can be obtained by the mean field approximation, as described in Sect.~\ref{Sec_scfr}. 
The mean membrane potential $\vec{v}_B$ is then determined by Eq.~(\ref{Eq_StatRate_a}). 
We will now discuss how the modulation part of tuned inputs can be approximated. 

%%%%%%%%%%%%%%%%%%%%%%%%%%%%%%%%%%%%%%%%%%%%%%%%%%%%%%%%%%%%%%%%%%%%%%%%%%%%%%%% 
\subsubsection{Modulation of the firing rate and of the membrane potential} 
\label{Sec_MFRMMP} 

We used the theory described in the previous sections to approximately determine the response of our networks, when they are processing non-uniform inputs, tuned to some stimulus feature. 
Specifically, the mean (over the network) modulation component (F2 component) of the output tuning curves in response to a tuned input can be obtained approximately.
Under the assumption of stability (see Sect.~\ref{Sec_OpRN}), and for the tuned inputs considered here, it seems justified to start with the approximation of ``perfect balance'' of the recurrent modulations: 
\[ 
\vec{W} \vec{r}_M \approx 0. 
\] 
Although this approximation is not strictly true, it holds on average:
As the result of numerical simulation in Fig.~\ref{Fig_OrthVer}B demonstrated, $\vec{W} \vec{s}_M$ had a narrow distribution around zero. 
Similar distribution is expected for $\vec{W} \vec{r}_M$, as $\vec{r}_M$ can be expanded in terms of powers of $\vec{s}_M$ (Eq.~\eqref{Eq_pow_ser}) under the assumption of stability.

In terms of diagrammatic illustration of Fig.~\ref{Fig_SepPath}B, this is equivalent to $\beta_M \approx 0$. 
This in turn implies that the net input from the recurrent network is on average untuned. 
For $\mathrm{EPSP} = 0.2\,\mathrm{mV}$, this input tuning is shown in Fig.~\ref{Fig_InpTun}. 
Although the tuning of recurrent input is (compared to feedforward tuning) not negligible for all neurons, it holds on average, such that average tuning curves have the same shape as the input tuning (Fig.~\ref{Fig_InpTun}B). 
We therefore use this approximation to compute the mean modulation gain of the network. 

Under this assumption, the computation of modulation rate vector, $\vec{r}_M$, is simplified to 
\[ \vec{r}_M \approx J_s \vec{s}_M - \vec{v}_M. 
\] 
The linear mixture of tuning curves described by Eq.~(\ref{TunMix}) is reduced to an amplification or attenuation of the respective input tuning curves. 

Since $\vec{v}_M$ depends nonlinearly on input parameters, we again need compute the self-consistent firing rates employing mean field theory. 
Mean $\mu$ and variance $\sigma^2$ of the input current analogous to Eq.~(\ref{Eq_inputMS}), however, are now computed by approximating the recurrent firing rate by the baseline firing rate, $r_B$, which is the same for all neurons in the network, as discussed above. 
The external input to this specific neuron, in contrast, experiences a feature specific modulation $s = s_B + s_M$.  
As a result, we let 
\begin{align} 
  \mu &= \tau [J_s (s_B + s_M) + J r_B N \epsilon (f - g(1-f)]), \nonumber \\ 
  \sigma^2 &= \tau [J_s^2 (s_B + s_M) + J^2 r_B N \epsilon (f + g^2(1-f))]. 
  \label{Eq_inputPB} 
\end{align} 
The parameters $J_s$, $J$, $g$, $N$, $\epsilon$ and $f$ are the same as above. 

The resultant firing rate of the neuron according to Eq.~(\ref{Eq_mT}) now differs from its baseline firing rate. 
The difference is, in fact, a good estimate for the modulation in the output firing rate of this particular neuron. 
Due to our general homogeneity assumptions, all neurons in the network will have the same output modulation, notwithstanding the fact that they all have different preferred orientations. 
Note that, to be consistent, the correction of $J$ due to the refractory period should be performed based on the modulated firing rate. 
This becomes specifically important for low recurrences, where the modulation rate is higher and, as a result, the effect of shunting of input due to refractory period ($J r t_\mathrm{ref}$) becomes more prominent. 
 
If the firing rates have been determined self-consistently, Eq.~(\ref{Eq_StatRate_a}) yields the corresponding membrane potentials directly. 
This is true for both the baseline and for the modulation, which are defined in the obvious way also for the membrane potential. 

%%%%%%%%%%%%%%%%%%%%%%%%%%%%%%%%%%%%%%%%%%%%%%%%%%%%%%%%%%%%%%%%%%%%%%%%%%%%%%%% 
\subsubsection{Operating regime of network} 
\label{Sec_OpRN} 

The modulation gains can also be computed by linearizing the dynamics around the baseline operating regime of the network. 

Our results revealed that the mean modulation gain, $\gamma_M$, in the network depends on the mean distance of the membrane potential from the threshold (Fig.~\ref{Fig_Vm_ModGain}). 
Subthreshold modulations (with regard to the mean-driven threshold) were capable of eliciting output firing activity, and the input-output relationship (the gain) was inversely proportional to the distance to threshold. 
One way to interpret these results is to summarize them in terms of the mean and standard deviation of the input that a neuron receives on average from the network, in the baseline state. 
To this, and alternatively to the mean and standard deviation of membrane potential, which is uniquely determined by the input, we refer as the operating point of the network. 

The effect of mean, $\mu_B$, and standard deviation of input, $\sigma_B$, on the modulation gain can be described, respectively, as shifting the mean membrane potential (and hence determining the mean distance to threshold), and smoothing (linearizing) the $f$-$I$ curve \cite{Miller2002, Anderson2000}. 
The linearized gains can then be obtained by perturbing the input around the baseline state, as it was described in Sect.~\ref{Sec_TIMP} and \ref{sect_LTRN}. 
This total embedded gain of the neuron in response to this perturbation determines the effective coupling strength and, as Fig.~\ref{Fig_LinGain_StabSpec} demonstrated, predicts the mean modulation gain in these networks quite well. 

The embedded gain modulates both feedforward and recurrent couplings. 
The fact that recurrent connections are now effectively weighted by these linear gains might suggest an explanation why the network exhibits stable activity even for highly recurrent regimes. 
As Fig.~\ref{Fig_InstabDyn}A showed, if the spectrum of $\vec{W}$ is computed from the weight matrix normalized by $V_\mathrm{th}$, the radius of the bulk of eigenvalues, $\rho$, would be larger than one already for an intermediate recurrent regime ($\mathrm{EPSP} = 0.2$, $J = 0.27\,\mathrm{mV}$, and $\rho = 1.68$ in this example). 
If one now computes the normalized radius by weighting the coupling strength according to linear gains (i.e.\ $W_{ij} = \zeta J$, instead of $J/V_\mathrm{th}$), the new normalized radius, $\rho^\mathrm{norm}$, is not unstable anymore ($\rho^\mathrm{norm} \approx 0.87 $ in this example). This coincides with our observations of the numerical simulations.

This enhanced stability has indeed been demonstrated to be the case for all networks we have studied here (Fig.~\ref{Fig_LinGain_StabSpec}, inset). 
If one now add to this that $\zeta$ is inversely proportional to distance to threshold, it follows that the network dynamically settles in a regime of operation which stabilizes the bulk of eigenvalues. 
This is due to the fact that in inhibition dominated networks, increasing the recurrent coupling also increases the negative feedback within the network, which results in more hyperpolarized average membrane potentials of the neurons.\footnote{If there is no minimum boundary condition, as it is the case for our neuron model, this hyperpolarization can grow beyond all bounds. Note that this would not be the case if one uses another neuron model, like conductance-base LIF neurons. The treatment of the problem in those cases might, therefore, be different.} 
This in turn leads to a smaller relative contribution of each spike from a presynaptic source to the firing activity of the postsynaptic neuron, since the distance to threshold has effectively increased. 
The overall increase or decrease in the effective gain, $\zeta J$, depends on how exactly the mean membrane potential, $v$, is affected by $J$ and how $\zeta$ is in turn depending on $v$. 

%%%%%%%%%%%%%%%%%%%%%%%%%%%%%%%%%%%%%%%%%%%%%%%%%%%%%%%%%%%%%%%%%%%%%%%%%%%%%%%% 
%%%%%%%%%%%%%%%%%%%%%%%%%%%%%%%%%%%%%%%%%%%%%%%%%%%%%%%%%%%%%%%%%%%%%%%%%%%%%%%% 
\subsection{Linear tuning in the strongly recurrent regime} 
\label{Sec_LTSRR} 
For networks with weak to intermediate recurrence, the assumption of ``perfect balance'' allowed a rather accurate prediction of the gain for the tuned part of network activation (``modulation''). 
This assumption, however, fails in the case of strongly recurrent networks. 
Under the constraints of ``linear tuning'' some aspects of the problem can be nevertheless treated. 

%%%%%%%%%%%%%%%%%%%%%%%%%%%%%%%%%%%%%%%%%%%%%%%%%%%%%%%%%%%%%%%%%%%%%%%%%%%%%%%% 
\subsubsection{Vectorial features and linear tuning} 
\label{Sec_VFLT} 

We now consider stimulus features that can be represented by vectors $\vec{\phi}$ in $\F \subset \R^n$ for some $n \geq 1$. 

\paragraph{Example 1.} 
The direction of a moving light dot stimulus in the visual field is represented by an angle in $[0,2\pi)$ or, alternatively, by a vector in $\F = S^1 \subset \R^2$, the $1$-dimensional sphere. 
The speed of the movement can be considered simultaneously with its direction, if encoded by the length of the vector. 
Any vector in $\R^2$ is then corresponding to a valid stimulus. 

\paragraph{Example 2.} 
The orientation of a moving grating in the visual field corresponds to vectors in $\F = S^1 \subset \R^2$ via the bijective mapping 
\[ [0,\pi) \to S^1,  \quad \theta \mapsto \bigl( \cos(2\theta), \sin(2\theta) \bigr). 
\] 
The factor $2$ in the argument of the cosine and the sine function makes sure that a rotation of the grating by $\pi$ is mapped to the initial orientation again.  

\paragraph{Example 3.} 
A stimulus for studying color vision is represented by the activation profiles of the different types of receptor cells in the retina, distinguished by their specific light absorption spectrum.  
For example, trichromacy in humans and closely related monkeys involves the differential activation of the three different types of cones $(S,M,L)$. 
This leads, in a natural way, to a representation of a color stimulus in terms of a vector in $\R^3$. 

A simple but relevant model of specific tuning curves is linear tuning 
\[ T_i(\vec{\phi}) = \psi_i^\ast + \langle \vec{\phi}_i^\ast, \vec{\phi} \rangle. 
\] 
The parameters $\psi_i^\ast \geq 0$ and $\vec{\phi}_i^\ast \in \F$ are fixed and specific for each neuron: $\psi_i^\ast$ is the baseline rate in absence of stimulation, the vector $\vec{\phi}_i^\ast$ is the preferred feature. 
In fact, stimulating with $\vec{\phi} = \vec{\phi}_i^\ast$ produces the highest, and stimulating with $\vec{\phi} = -\vec{\phi}_i^\ast$ the lowest firing rates. 
More generally, if $\xi$ denotes the angle between the vectors $\vec{\phi}_i^\ast$ and $\vec{\phi}$, linear tuning is equivalent to cosine tuning 
\[  \langle \vec{\phi}_i^\ast,\vec{\phi} \rangle  = \|\vec{\phi}_i^\ast\| \|\vec{\phi}\| \cos (\xi) . 
\] 
The length of the vector that represents the preferred feature $\|\vec{\phi}_i^\ast\|$ is the tuning strength, it satisfies 
\[\|\vec{\phi}_i^\ast\| = \max_{\vec{\phi}} \langle \vec{\phi}_i^\ast,\vec{\phi} \rangle / \| \vec{\phi} \| 
\] 
where $\| \vec{\phi} \|$ is the stimulus strength. 
To ensure that the firing rate $T_i(\vec{\phi})$ remains positive 
\[0 \leq \min_{\vec{\phi}} T_i(\vec{\phi}) = \psi_i^\ast - \|\vec{\phi}_i^\ast\| \, \|\vec{\phi}\| 
\] 
the strength of the admitted stimuli must be limited, so we admit only stimuli that are weak enough such that linearity of the tuning and positivity of firing rates remain compatible 
\[ 
\F = \bigl\{ \vec{\phi} \in \R^n : \|\vec{\phi}\| \leq \psi_i^\ast / \|\vec{\phi}_i^\ast\| \bigr\}. 
\] 

%%%%%%%%%%%%%%%%%%%%%%%%%%%%%%%%%%%%%%%%%%%%%%%%%%%%%%%%%%%%%%%%%%%%%%%%%%%%%%%% 
\subsubsection{Linear tuning in recurrent networks} 
\label{Sec_LTRN} 

Assume now that the individual inputs $s_i$ are tuned with respect to an $n$-dimensional feature $\vec{\phi}$.  
The same stimulus is ``seen'' by all neurons, but each neuron responds with its private tuning curve 
\[s_i(\vec{\phi}) = T_i(\vec{\phi}). 
\] 
As described by Eq.~(\ref{TunMix}), the responses of neurons in a recurrent network have tuning curves that are, in general, linear sums of the tuning curves of the input channels.  
If the recurrent interactions are strong, many input channels contribute indirectly to the tuning of every output channel, recruiting multi-synaptic pathways.  
Assume now that all inputs $s_i$ are linearly tuned to the stimulus $\vec{\phi}$ according to 
\[s_i(\vec{\phi}) = \psi_i^\ast + \langle \vec{\phi}_i^\ast , \vec{\phi}\rangle 
\] 
for parameters $\psi_i^\ast$ and $\vec{\phi}_i^\ast$. 
Then, exploiting the two-fold linearity, we obtain 
\begin{align} 
  \vec{r}(\vec{\phi}) 
     &= \vec{A} \bigl[ J_s \vec{s}(\vec{\phi}) - \vec{v}(\vec{\phi}) \bigr] \nonumber \\ 
    &= \vec{A} \bigl[ J_s (\vec{\psi}^\ast + \vec{\Phi}^\ast \vec{\phi}) 
                                     - \vec{v}(\vec{\phi}) \bigr]\nonumber\\ 
    &= \vec{A} J_s \vec{\psi}^\ast + \vec{A} J_s \vec{\Phi}^\ast \vec{\phi} - \vec{A} \vec{v}(\vec{\phi}) 
  \label{RLT} 
\end{align} 
where $\vec{\psi}^\ast$ is the vector of baseline activities and $\vec{\Phi}^\ast$ is a matrix the rows of which are given by the transposed preferred features $(\vec{\phi}_i^\ast)^T$.  
Therefore, apart from nonlinear distortions induced by nonzero mean membrane potentials, all neurons in the recurrent network are again linearly tuned, with baselines given by the components of the vector $\vec{A} J_s \vec{\psi}^\ast$, and preferred features encoded by the rows of the matrix $\vec{A} J_s \vec{\Phi}^\ast$. 

%%%%%%%%%%%%%%%%%%%%%%%%%%%%%%%%%%%%%%%%%%%%%%%%%%%%%%%%%%%%%%%%%%%%%%%%%%%%%%%% 
\subsubsection{Mixing of preferred features in random recurrent networks} 
\label{Sec_MxPFRRN} 

Because linear tuning curves are linearly transformed according to Eq.~(\ref{RLT}), we can actually compute the recurrent preferred features that result from this transformation.  
Note, however, that the actual tuning curves will, in general, be contaminated by nonlinear distortions by $\vec{A} \vec{v}(\vec{\phi})$ that are not reflected by the linear mix $\vec{A} J_s \vec{\Phi}^\ast$ of preferred features of the inputs to the network. 
To keep the present discussion simple, we ignore this complication here. 

A network with zero recurrent interaction would be described by the matrix $J_s \vec{\one}$.  
Therefore, the perturbation of each neuron's private preferred input feature resulting from recurrent network action is given by $\vec{A} J_s \vec{\Phi}^\ast - J_s \vec{\Phi}^\ast = \vec{R} J_s \vec{\Phi}^\ast$, where 
\[\vec{R} = J_s \vec{A} - J_s \vec{\one} = J_s \vec{W} (\vec{\one} - \vec{W})^{-1}. 
\] 
If we can assume that the preferred features $\vec{\Phi}^\ast$ of inputs are all chosen independently from some common distribution, the sums of preferred vectors that result from the action of the matrix $\vec{R}$ will, according to the Central Limit Theorem, be normally distributed vectors in $\R^n$.  
In that case, it suffices to compute the covariance matrix $\vec{C}_\mathrm{ptb}$ of the perturbations of the preferred features of all output tuning curves, and to compare it with the covariance matrix $\vec{C}_\mathrm{in}$ of the inputs 
\begin{align} 
  \vec{C}_\mathrm{in} &= \frac{1}{N-1} (\vec{\Phi}^\ast)^T \vec{\Phi}^\ast \nonumber\\ 
  \vec{C}_\mathrm{ptb} &= \frac{1}{N-1} (\vec{R} J_s \vec{\Phi}^\ast)^T (\vec{R} J_s 
  \vec{\Phi}^\ast) \nonumber \\&= \frac{J_s^2}{N-1} (\vec{\Phi}^\ast)^T (\vec{R}^T \vec{R}) \vec{\Phi}^\ast. 
\end{align} 
Specifically, if the distribution of input features is isotropic with covariance $\vec{C}_\mathrm{in} = \sigma_\mathrm{in}^2 \vec{\one}$, the scalar $\sigma_\mathrm{in}^2$ represents the mean squared tuning strength of all inputs. 
By means of the matrix $\vec{R}$, the distribution of output features will then be perturbed by a component that is normally distributed with isotropic covariance $\vec{C}_\mathrm{ptb} = \sigma_\mathrm{ptb}^2 \vec{\one}$ where 
\[  \sigma_\mathrm{ptb}^2 = \beta^2 \sigma_\mathrm{in}^2 
  \qquad\text{with}\qquad 
  \beta^2 = \sum_j R_{ij}^2. 
\] 
For a random network as described above, the factor $\beta$ is the same for all rows $j$, and it describes now the mean attenuation/amplification of the tuning strength performed by the recurrent network. 

As an example, we compute here the distribution of tuning strengths $\|\vec{\phi}_i^\ast\|$ of neurons that would result in a strongly recurrent network, where the perturbation dominates the result. 
It is given by the probability density 
\[\phi_n(x;\sigma^2) = \eta_n(x) \cdot \psi_n(x;\sigma^2) 
\label{TS_dist} 
\] 
where $\eta_n(x)$ is the surface of the $(n-1)$-sphere with radius $x$, and $\psi_n(x;\sigma^2)$ is the probability density function of the $n$-dimensional normal distribution with zero mean and isotropic covariance matrix $\sigma^2 \vec{\one}$. 
We find 
\begin{align} 
  \phi_n(x)  &= 2 \frac{\pi^{n/2}}{\Gamma(n/2)} x^{n-1}   \cdot \frac{1}{(2\pi)^{n/2}\sigma^n} e^{-x^2/(2\sigma^2)} \nonumber 
  \\ &= \frac{2^{1-n/2}}{\Gamma(n/2)\sigma^n} x^{n-1} e^{-x^2/(2\sigma^2)}. 
\end{align} 
In the case $n=2$, which is particularly relevant for the application described in this paper, we get 
\[ 
  \phi_2(x;\sigma^2) = \frac{x}{\sigma^2} e^{-x^2/(2\sigma^2)} 
  \label{Weibull} 
\] 
which is a Weibull distribution with shape parameter $2$ and scale parameter $\sqrt{2}\sigma$.  

For the scenario, where feedforward and recurrent components of tuning are mixed, one obtains an interpolation between ``purely feedforward'' and ``purely feedback'' operation of the network. 
This is equivalent to computing Eq.~\eqref{TS_dist} with nonzero mean, $\mu$. 
In the specific case of orientation selectivity, with $n = 2$, the final distribution of tuning strength is then obtained as 
\[ 
\phi_2(x;\sigma^2) = \frac{x}{\sigma^2} e^{-(x^2 + \mu^2)/(2\sigma^2)} I_0 (\frac{x \mu}{\sigma^2}), 
  \label{Tot_tuning} 
\] 
where 
\[ 
I_0(z) = \frac{1}{\pi} \int_0^\pi e^{z \cos(\theta)} d\theta \nonumber
\] 
is the modified Bessel function of the first kind and order zero. 

%%%%%%%%%%%%%%%%%%%%%%%%%%%%%%%%%%%%%%%%%%%%%%%%%%%%%%%%%%%%%%%%%%%%%%%%%%%%%%%% 
%%%%%%%%%%%%%%%%%%%%%%%%%%%%%% Acknowledgements
%%%%%%%%%%%%%%%%%%%%%%%%%%%%%%%%%%%%%%%%%%%%%%%%%%%%%%%%%%%%%%%%%%%%%%%%%%%%%%%% 

% Do NOT remove this, even if you are not including acknowledgments
\section*{acknowledgements}

The authors wish to thank A~Aertsen, C~Boucsein, G~Grah, J~Kirsch and A~Kumar for their comments on previous versions of the manuscript. 
We also thank the developers of the simulation software NEST (see http://www.nest-initiative.org) and the maintainers of the BCF computing facilities for their support throughout this study.
Funding by the German Ministry of Education and Research (BCCN Freiburg, grant 01GQ0420 and BFNT Freiburg*T\"ubingen, grant 01GQ0830) is gratefully acknowledged. 

%%\section*{References}
%% The bibtex filename
%\bibliography{refs}

\newpage

%%%%%%%%%%%%%%%%%%%%%%%%%%%%%%%%%%%%%%%%%%%%%%%%%%%%%%%%%%%%%%%%%%%%%%%%%%%%%%%%
%%% Table
%%%%%%%%%%%%%%%%%%%%%%%%%%%%%%%%%%%%%%%%%%%%%%%%%%%%%%%%%%%%%%%%%%%%%%%%%%%%%%%%
\section*{Tables}
\begin{table}[!ht]
\caption{
\bf{Table of notations and parameters.}}
%\begin{tabular}{|c|c|c|}
%table information

\begin{tabular}{lll}   \toprule
  %\multicolumn{2}{|c|}{a} \\
  \textbf{Neuron Model} &~ &~ \\ \hline
  \rowcolor{blue!10} Membrane time constant & $\tau_m$ & $20~\mathrm{ms}$ \\
  Resting potential & $V_\mathrm{rest}$ & $0~\mathrm{mV}$ \\
  \rowcolor{blue!10} Threshold voltage & $V_\mathrm{th}$ & $20~\mathrm{mV}$ \\
  Reset voltage     & $V_\mathrm{reset}$ & $20~\mathrm{mV}$ \\  
  \rowcolor{blue!10} Refractory period & $\tau_\mathrm{ref}$ & $2~\mathrm{ms}$ \\
  
  ~ &~ &~  \\ \hline
  \textbf{Synaptic Model} & & \\ \hline
  \rowcolor{blue!10} Synaptic time constant & $\tau_\mathrm{syn}$ & $0.5~\mathrm{ms}$ \\  
  Peak EPSP          & $\mathrm{EPSP}$ & $0.01, ..., 1~\mathrm{mV}$ \\
  \rowcolor{blue!10} Synaptic efficacy & $J$ & $\tau_\mathrm{syn} e~ \mathrm{EPSP} $ \\
  Inhibition dominance ratio & $g$ & $8$ \\
  \rowcolor{blue!10} Feedforward strength & $\mathrm{EPSP}_\mathrm{ffw}$ & $0.1~\mathrm{mV}$ \\
  Synaptic delay    & $D$ & $1.5~\mathrm{ms}$ \\
    
  ~ &~ &~  \\ \hline
  \textbf{Network Connectivity} & & \\ \hline
  \rowcolor{blue!10} Number of neurons & $N$ & $12\,500$ \\    
  Excitatory fraction & $f$ & $0.8$ \\ 
  \rowcolor{blue!10} Connection probability & $\epsilon$ & $10~\%$ \\
  Weight matrix & $W$ & $w_{ij}$ \\
  \rowcolor{blue!10} Network operator & $A$ & $(\one - W)^{-1}$\\
  
  ~ &~ &~  \\ \hline
  \textbf{Simulation} & & \\ \hline
  \rowcolor{blue!10} Stimulus orientation & $\theta$ & $0^\circ, 15^\circ, ..., 165^\circ$ \\
  Preferred orientation (PO) & $\theta^*$ &  $[0^\circ, 180^\circ)$ \\
  \rowcolor{blue!10} Contrast & $C$ & $9, 39, 99~\%$ \\
  Baseline firing rate & $s_B$ & $12, 16, 20~\mathrm{kHz}$ \\
  \rowcolor{blue!10} Modulation ratio & $m$ & $m=10~\%$ \\
  Simulation time   & $t_\mathrm{sim}$ & $6\,000~\mathrm{ms}$ \\ 
  
    ~ &~ &~  \\ \hline
  \textbf{Analysis} & & \\ \hline
  \rowcolor{blue!10} Orientation selectivity index & OSI & \\
  Orientation selectivity vector & OSV &   \\
  \rowcolor{blue!10} Scatter degree index & SDI &  \\
  Baseline (modulation) gain & $\gamma_B$ ($\gamma_M$) & \\
  \rowcolor{blue!10} Linearized neuronal gain & $\zeta$ & \\
  Tuning curve   & $T_i(\theta)$ & \\ 

\hline
\end{tabular}
\begin{flushleft}%General set of parameters used in the paper.
\end{flushleft}
\label{tab:label}
\end{table}

\end{document}